\documentclass[lineno]{article}
\usepackage{dsfont}
\usepackage{amsmath, amsthm}
\usepackage{color}
\usepackage{float}
\usepackage[left=1.25in, bottom=1.5in, right=1.5in, top=1in]{geometry}

\usepackage{amsfonts}
\usepackage{times}
\usepackage{bm}
\usepackage{natbib}
\usepackage{enumerate}
\usepackage{amsmath,amssymb}
\usepackage{graphicx}
\graphicspath{ {img/} }
\usepackage{subfigure}
\usepackage{rotating} 
\usepackage{authblk}

\usepackage[ruled]{algorithm2e} 

\numberwithin{equation}{section}
\newtheorem{thm}{Theorem}[section]

\newtheorem{lem}[thm]{Lemma}
\newtheorem{assumption}[thm]{Assumption}
\newtheorem{prop}[thm]{Proposition}
\newtheorem*{remark}{Remark}

\newcommand{\var}{\mbox{Var}}
\newcommand{\cov}{\mbox{Cov}}
\newcommand{\rank}{\mbox{rk}}
\newcommand{\argmax}{\operatornamewithlimits{argmax}}
\newcommand{\argmin}{\operatornamewithlimits{argmin}}
\newcommand{\esssup}{\operatornamewithlimits{ess ~sup}}

\newcommand{\iu}{{i\mkern1mu}}

\newcommand{\vertiii}[1]{{\left\vert\kern-0.25ex\left\vert\kern-0.25ex\left\vert #1 
    \right\vert\kern-0.25ex\right\vert\kern-0.25ex\right\vert}}

\begin{document}



\title{Large Spectral Density Matrix Estimation by Thresholding
}
\author[1]{Yiming Sun\thanks{Email: ys784@cornell.edu}}
\author[2]{Yige Li\thanks{Email: yigeli@hsph.harvard.edu}}
\author[3]{Amy Kuceyeski \thanks{Email: amk2012@med.cornell.edu}}
\author[1]{Sumanta Basu\thanks{Corresponding Author. Email: sumbose@cornell.edu}}
\affil[1]{Department of Statistical Science, Cornell University}
\affil[2]{Department of Epidemiology, Harvard T. H. Chan School of Public Health}
\affil[3]{Department of Radiology, Weill Cornell Medical College}

\date{ } 

\maketitle
\begin{abstract}
Spectral density matrix estimation of multivariate time series  is a classical problem in time series and signal processing. In modern neuroscience, spectral density based metrics are commonly used for analyzing functional connectivity among brain regions. In this paper, we develop a non-asymptotic theory for regularized estimation of high-dimensional spectral density matrices of Gaussian and linear processes using thresholded versions of averaged periodograms. Our theoretical analysis ensures that consistent estimation of spectral density matrix of a $p$-dimensional time series using $n$ samples is possible under high-dimensional regime $\log p / n \rightarrow 0$ as long as the true spectral density is approximately sparse. A key technical component of our analysis is a new  concentration inequality of average periodogram around its expectation, which is of independent interest. Our estimation consistency results complement existing results for shrinkage based estimators of multivariate spectral density, which require no assumption on sparsity but only ensure consistent estimation in a regime $p^2/n \rightarrow 0$. In addition, our proposed thresholding based estimators perform consistent and automatic edge selection when learning  coherence networks among the components of a  multivariate time series.  We demonstrate the advantage of our estimators using simulation studies and a real data application on functional connectivity analysis with fMRI data.

\end{abstract}

\section{Introduction}\label{introduction}

Multivariate spectral density estimation is an important problem in time series and signal processing, with applications in many scientific disciplines including economics \citep{granger1969investigating}  and neuroscience \citep{bowyer2016coherence}. Spectral density of a stationary multivariate time series is the frequency domain analogue of covariance and is based on the Fourier transform of autocovariance function. It aggregates information on linear association, both contemporaneous and across different lags,  among the components of a multivariate time series. So it can be used to provide a richer description of cross-sectional dependency than Pearson correlation, which only accounts for contemporaneous association among the time series components. 

In particular, multivariate spectral density and coherence (frequency domain analogue of correlation) are routinely used in neuroscience as metrics of  functional connectivity among brain regions using time series of neurophysiological signals (e.g., fMRI, EEG and MEG) and to construct networks of interactions in a data-driven fashion \citep{bowyer2016coherence}. These connectivity networks, where each node corresponds to a brain region and edge weights correspond to strengths of coherence between regions, are often used to study differential brain connectivity patterns in patients suffering from neurological disorders. More recently, coherence metrics have also been used to construct similarity measures when clustering high-dimensional time series of brain signals \citep{euan2016hierarchical}. With advances in data collection and storage technologies, it is now feasible to analyze time series data on a large number of brain regions. For instance, the freeSurfer brain atlas used in this paper summarizes voxel level data to $p = 86$ brain regions. Consequently, there is an increasing interest among neuroscientists in constructing coherence networks among a large number of brain regions in a principled manner from temporally dependent samples of small to moderate size ($n \ll p^2)$. For instance, we use only $n=200$  samples for our fMRI data analysis in this paper.

This recent interest in learning the cross-sectional dependence from spectral density matrix at different frequencies is complementary to developments in classical time series and signal processing literature, which focused more on studying the \textit{shape} of spectral density function in a low-dimensional asymptotic regime ($p$ fixed, $n \rightarrow \infty$) \citep{brillinger1981time, brockwell2013time}.  In another line of work, \citet{dahlhaus1997identification, dahlhaus2003causality, eichler2007frequency} investigated in depth the issues of inference with coherence and testing of marginal independence between components of multivariate time series using integrated spectral density. Finer and uniform convergence rates of smoothed periodograms were more recently provided by \citet{wu2015uniform}. However, as the dimension of the time series increases, so does the estimation risk of smoothed periodograms. This was first pointed out by \cite{bohm2009shrinkage}, who showed that shrinking smoothed periodogram towards a simpler structure can reduce risk and make the estimates better-conditioned for studying inverse spectral density matrix. 
The authors also proved consistency of their estimates under a double-asymptotic regime $p \rightarrow \infty, n \rightarrow \infty, p^2/n \rightarrow 0$. In a series of papers, \citet{bohm2008structural,  fiecas2016dynamic, fiecas2014datadriven} have made significant progress in this direction by providing a wide variety of shrinkage methods with attractive theoretical and empirical properties.

In this work, we make two additions  to this research direction of learning large spetral density matrices. First, we propose a family of \textit{sparsity regularized estimators} of spectral density matrix based on thresholding averaged periodograms. Our proposed estimators have the added advantage of performing  automatic edge selection and providing sparse, interpretable networks among the component time series. Second, we develop a non-asymptotic theory for estimation of  spectral density and coherence that explicitly connects estimation error bounds to a notion of approximate sparsity of the true spectrum. As a consequence, our theory shows that consistent estimation is possible in a high-dimensional regime $\log p / n \rightarrow 0$ as long as the underlying structure is approximately sparse. 

Our proposal is motivated by recent  developments in covariance matrix estimation literature, where several thresholding based strategies \citep{bickel2008covariance, rothman2009generalized, cai2011adaptive, cai2016rates} have shown to provide good theoretical and empirical properties compared to the shrinkage based estimators proposed in \citet{ledoit2004well}. The thresholding techniques developed in this literature serve as promising candidates for high-dimensional spectral density estimation as well. However, their implementation and theoretical analysis require addressing additional technical challenges. From an implementation consideration, choice of threshold in covariance matrix estimation for i.i.d. data is carried out using multiple sample-splitting \citep{bickel2008covariance} which is not feasible when the data have a temporal ordering. On the theoretical side, non-asymptotic analysis of  periodograms averaged across nearby frequencies requires understanding concentration behavior of a sum of random matrices that are \textit{neither independent nor identically distributed}. Unlike sample covariance estimation with i.i.d. data, the lack of identical distribution results in smoothing bias well-known in nonparametric density estimation. In addition, the additional temporal dependence complicates deriving finite sample deviation of averaged periodogram from its expectation. 

We make three technical contributions in this paper to address the above challenges. First, we select thresholding parameters using a frequency-domain sample-splitting scheme based on the heuristic of approximate independence of periodograms at different Fourier frequencies. Second, we provide upper bounds on the finite sample bias of averaged periodograms and provide insight into how it is affected by temporal dependence in data for some commonly used families of time series. Finally, we develop a non-asymptotic upper bound on the deviation of averaged periodogram using a Hanson-Wright type inequality for complex quadratic forms of temporally dependent random vectors. Building upon these technical ingredients, our main theoretical results include (i) consistency of thresholded averaged periodograms in operator and scaled Frobenius norms in a high-dimensional regime under a weak sparsity assumption on true spectrum, and (ii)  sparsistency results ensuring selection of marginally correlated pairs of time series in a coherence network with high probability. Our analysis  framework accommodates Gaussian time series, and linear processes with subGaussian or generalized subexponential errors, or errors with finite fourth moments. The rates of convergence of thresholded estimators change with the nature of tail distribution of errors.

We demonstrate the merits of our proposed methods using extensive numerical experiments and a real data application on constructing functional connectivity networks from fMRI data. Our numerical experiments show that thresholding methods achieve  estimation accuracy comparable with  the shrinkage method, while simultaneously performing automatic coherence selection. In particular, a  lasso and an adaptive lasso based thresholding strategy show promising performance across different simulation settings. In the real data application, these two methods were able to extract  sparse, interpretable networks that nicely captured known biological patterns in brain networks and distinguished different brain regions from each other.

The rest of the paper is organized as follows. In section \ref{sec:model-methods}, we formally introduce our problem, provide a brief review of shrinkage estimators, and describe our proposed thresholding methods. In section \ref{sec:theory}, we derive non-asymptotic upper bounds on our proposed spectral density estimates for Gaussian time series. In section \ref{sec:heavy-tail} we extend the results for Gaussian time series to general linear processes with different non-Gaussian noise distributions. In section \ref{sec:simulation}, we conduct simulation studies to assess the finite sample properties of our proposed estimators. Section \ref{sec:realdata} contains an empirical application of our proposed method to a functional connectivity analysis with real fMRI data. We defer the proofs of all of our technical results to the Appendix.

\textbf{Notation.} Throughout this paper, $\mathbb{Z}$, $\mathbb{R}$ and $\mathbb{C}$  denote the sets of integers, real numbers and complex numbers, respectively. We use $|c|$ to denote the modulus of a complex number and the absolute value of a real number. We use $\|v\|$ to denote $\ell_2$-norm of a vector $v$. For a matrix $A$, $\|A\|_1$, $\|A\|_{\infty}$, $\|A\|$ and $\|A\|_F$ will denote maximum complex modulus column sum norm, maximum complex modulus row sum norm, { spectral norm} $\sqrt{\Lambda_{\max}(A^\dag A)}$ and Frobenius norm $\sqrt{\text{tr}(A^\dag A)}$, respectively, where $A^\dag$ is conjugate transpose of $A$. We also let $\lambda_{\text{max}}(A)$ denote the spectral radius of a $n \times n$ matrix $A$, i.e., $\lambda_{\text{max}}(A) = \max(|\lambda_1|, \cdots |\lambda_n|)$, where $\lambda_i$ are the eigenvalues of matrix $A$. If $A$ is symmetric or Hermitian, we denote its maximum and minimum eigenvalues by $\Lambda_{\min}(A)$ and $\Lambda_{\max}(A)$. We use $e_i$ to denote the $i^{th}$ unit vector in $\mathbb{R}^p$, for $i = 1, 2, \ldots, p$. For vectors $v_i \in \mathbb{R}^p, i=1,\ldots, n$, we use $[v_1:\ldots:v_n]$ to denote the $p \times n$ matrix formed  by horizontally stacking these column vectors $v_i$, and  $[v_1^\top;\ldots; v_n^\top]$ to denote the $n\times p$ matrix by vertically stacking row vectors $v_i^\top$. Let $vec(A)$ represent the vector got from vectorization of a matrix $A$ by stacking all its columns. We use $rk(A)$ to denote the rank of a matrix $A$. For a complex vector $v\in \mathbb{C}^p$ and any $q > 0$, we define $\|v\|_q:= (\sum_{i=1}^p |v_i|^q)^{1/q}$. We use $\|v\|_0$ to denote the number of non-zero elements in $v$. Note that when $0\le q<1$, it is not really a norm since triangle inequality does not hold, but we keep the notation of a norm for convenience . Then we define the induced matrix norm, $\|A\|_{\alpha, \beta} = \sup_{x\neq 0}\|Ax\|_\alpha/\|x\|_\beta$, for any  $\alpha>0, \beta>0$. We will also use $\|A\|_\alpha$ to denote the induced norm $\|A\|_{\alpha, \alpha}$ for any $\alpha > 0$ and any complex matrix $A \in \mathbb{C}^{p \times p}$. Also, to be succinct, we use $\|A\|_{\rm{max}} :=\max_{r,s}|A_{rs}|$. 
Throughout the paper, we write $A \succsim B$ if there exists a  universal constant $c > 0$, not depending on model dimension or any model parameters, such that $A \ge cB$. We use $A \asymp B$ to denote $A \succsim B$ and $B \succsim A$. 

\section{Background and Methods}\label{sec:model-methods}
Consider a $p$-dimensional weakly stationary real-valued time series $X_t = (X_{t1}, \ldots, X_{tp})^\top, ~ t\in \mathbb{Z}$. Let  $\mathcal{X} = [X_1:\ldots:X_n]^\top$  be the \textit{data matrix} containing $n$ consecutive observations from the time series $\{X_t\}$ in its rows. We assume $\mathbb{E} X_t = 0, ~ t=1,\ldots,n$ for ease of exposition. In practice, multivariate time series are often de-meaned before performing correlation based analysis. Weak stationarity implies that $\cov(X_t, X_{t-\ell}) = \mathbb{E} X_t X_{t-\ell}^\top$ only depends on $\ell$, so we can define autocovariance as function of the lag $\ell$, viz., $\Gamma(\ell) = \cov(X_t, X_{t-\ell})$. Spectral density aggregates information of autocovariance of different lag orders $\ell$  at a specific frequency $\omega \in [-\pi, \pi]$ as 
\begin{equation}
f(\omega) = \frac{1}{2\pi}\sum_{\ell=-\infty}^\infty \Gamma(\ell) e^{-\iu \ell \omega }. 
\end{equation}
Note that the autocovariance functions of different lags can be recovered from the spectral density using the transformation  $\Gamma(\ell) =  \int_{-\pi}^{\pi} f(\omega) e^{i\ell \omega} d\omega$, for any $\ell \in \mathbb{Z}$.


For the matrix-valued spectral density function $f$ over $[-\pi, \pi]$, we define, for $q \ge 0$,
\begin{equation}
\vertiii{f}_q = \esssup_{\omega \in [-\pi, \pi]}\|f(\omega)\|_q. \nonumber
\end{equation}
Following \citet{Basu2015}, we will also use 
$\vertiii{f}: = \vertiii{f}_2 = \esssup_{\omega \in [-\pi, \pi]}\|f(\omega)\| $
as a measure of stability of the time series $X_t$. Larger values of $\vertiii{f}$ are associated with processes having stronger temporal and cross-sectional  dependence and less stability. Since every coordinate of the spectral density matrix is calculated using at most two components of the $p$-dimensional time series $X_t$ and $f(\omega)$ is non-negative definite, a smaller measure of stability, viz. $\max_{1 \le r \le p} \esssup_\omega \|f_{rr}(\omega)\|$ can be also used in our error bound analysis instead, although we present our results in terms of $\vertiii{f}$ for ease of exposition.

In many applications, in particular functional connectivity analyses in neuroscience, it is of interest to estimate standardized spectral density or coherence matrix, an analogue of correlation in the frequency domain,  defined as 
\begin{equation}
\label{def:coherance}
g_{rs}(\omega) = \frac{f_{rs}(\omega)}{\sqrt{f_{rr}(\omega)f_{ss}(\omega)}},
\end{equation}
assuming $f_{rr}(\omega) \neq 0$ for all $1\le r\le p$.

\subsection{Background: Periodogram Smoothing and Shrinkage}\label{sec:model_method_background}
The classical estimate of spectral density is based on the periodogram \citep{brockwell2013time, rosenblatt1985stationary} defined as
\begin{equation}
\label{eq:single_periodogram}
I(\omega)=\sum_{|\ell|<n} \hat{\Gamma}(\ell)e^{-\iu\ell\omega},
\end{equation}
where $\hat{\Gamma}(\ell) = n^{-1}\sum_{t=\ell+1}^{n} X_t X_{t-\ell}^\top$ for $\ell\ge 0$, and 
 $\hat{\Gamma}(\ell) = n^{-1}\sum_{t=1}^{n+\ell} X_t X_{t-\ell}^\top$ for $\ell<0$. 
Note the connection between periodogram and discrete Fourier transformation (DFT)  $d(\omega) = \mathcal{X}^\top(C(\omega)-iS(\omega))$ , where 
\begin{equation}
\label{eq:cos_sin_coef}
\begin{aligned}
C(\omega) = \frac{1}{\sqrt{n}} (1, \cos \omega, \dots, \cos (n-1)\omega)^\top,\\
S(\omega) = \frac{1}{\sqrt{n}} (1, \sin \omega, \dots, \sin (n-1)\omega)^\top.
\end{aligned}
\end{equation}
We can rewrite $I(\omega)$ as $d(\omega) d(\omega)^\dag$. 
In classical asymptotic analysis of time series ($p$ fixed, $n \rightarrow \infty$), it is known that $\frac{1}{2\pi}I(\omega)$ is asymptotically unbiased for $f(\omega)$ but not consistent due to non-diminishing variance.
For instance, for i.i.d Gaussian white noise  
$X_t \overset{i.i.d}{\sim}\mathcal{N}(0,\sigma^2I)$, the variance of $I(\omega)$ is of the order $\sigma^4$ [Proposition 10.3.2,  \citet{brockwell2013time}].  To achieve consistency, it is common to resort to smoothing periodograms over nearby frequencies. In this paper, we focus on the simplest form of smoothing, viz. averaging, of periodograms

\begin{equation}
\label{eq:general_smoothing_estimator}
    \hat{f}(\omega; m) = \frac{1}{2\pi(2m+1)} \sum_{|k|\le m} I(\omega+\omega_k),
\end{equation}
where $\omega_k = 2\pi k/n, ~ k\in F_n$,  the set of Fourier frequencies. To be precise, $F_n$ denotes the set  $\left\{-[\frac{n-1}{2}], \dots, [\frac{n}{2}]\right\}$ where $[x]$ is the integer part of $x$. $F_n$ contains exactly the same frequencies used to calculate discrete Fourier transformation. It is common to evaluate the periodogram at these Fourier frequencies, in which case the smoothing periodogram in \eqref{eq:general_smoothing_estimator} becomes
\begin{equation}
\label{eq:smoothing estimator}
    \hat{f}(\omega_j; m) = \frac{1}{2\pi(2m+1)} \sum_{|k|\le m} I(\omega_{j+k}).
\end{equation}
Note that even though the values of $j+k$ can fall outside $F_n$, it is enough to evaluate periodograms at Fourier frequencies $F_n$ since $I(\omega)$ is $2\pi$-periodic in $\omega$. Theorem 10.4.1 in \citet{brockwell2013time} shows that if $m = o(\sqrt{n})$, \eqref{eq:smoothing estimator} is a consistent estimator. As in general nonparametric function estimation, one can replace the weights $1/(2m+1)$ in \eqref{eq:smoothing estimator} by a more general kernel function. For more details, we refer the  readers to \cite{brockwell2013time}. To make notations simpler, in this paper we will omit the subscript $m$ and use $\hat{f}(\omega_j)$ whenever $m$ is clear from the context. 

\par
This nonparametric smoothing method can be unstable for high-dimensional multivariate spectral density estimation since smoothed periodograms start to become ill-conditioned. Generalizing shrinkage estimation strategy for high-dimensional covariance matrix \citep{ledoit2004well}, \citet{bohm2009shrinkage} proposed shrinking averaged periodogram to estimate spectral density in high-dimension. The idea of shrinkage method is to reduce condition numbers for smoothed periodograms. In particular, the authors changed the estimation target to $f^0(\omega) = \mathbb{E} \hat{f}(\omega)$ and argued that $f^0(\omega)$ is close enough to $f(\omega)$ asymptotically. Subsequently, they considered a Hilbert space for square complex random matrices with inner product defined as  $\mathbb{E}\langle A,B\rangle$ where $A,B$ are two  matrices and 
\begin{equation}
\langle A,B \rangle = \frac{1}{p} \text{tr}(A^\dag B).\nonumber 
\end{equation} 
In this Hilbert space and with the fact that $\hat{f}(\omega)$ is an unbiased estimator for $f^0(\omega)$,  \citet{bohm2009shrinkage} applied the projection argument similar to \citet{ledoit2004well} to build the shrinkage estimator for $f^0(\omega)$. To this end, the authors first projected $f^0(\omega)$ on the space spanned by the identity matrix as $\mu(\omega) I_p$, where $I_p$ is the identity matrix and $ \mu(\omega)= \frac{1}{p} \text{tr}(f(\omega))$. Then the shrinkage estimator is defined as the minimizer of the convex program 
\begin{equation}
\hat{f}^\star(\omega) = \argmin_{\tilde{f}(\omega) \in S(\omega)} \frac 1 p \|f^0(\omega) - \tilde{f}(\omega)\|^2_F, \nonumber
\end{equation}
where 
\begin{equation}
S(\omega) = \rho(\omega) \mu(\omega)I_p + (1-\rho(\omega)) \hat{f}(\omega), ~~~~ 0\le \rho(\omega)\le 1. \nonumber
\end{equation}
\citet{bohm2009shrinkage} derived an explicit formula 
$\rho(\omega) = {\alpha^2(\omega)}/{\delta^2(\omega)}$,
where 
\begin{equation*}
    \alpha^2(\omega) = \frac 1 p \|f^0(\omega) - \mu(\omega)I_p\|^2_F,~~
    \beta^2(\omega) = \frac 1 p \|f^0(\omega)-\hat{f}(\omega)\|^2_F,
\end{equation*}
and $\delta^2(\omega) = \alpha^2(\omega)+\beta^2(\omega)$. 
Then they plugged in estimators of $\alpha(\omega), \beta(\omega), \delta(\omega)$ into the above formula to get the final data-driven estimator of spectral density. \par

\subsection{Method: Thresholding Averaged Periodogram} \label{sec:method_threshold}
In this section, we present our proposed thresholding estimators. We restrict our methodology description and theoretical development on the finite grid of Fourier frequencies for convenience, although all our theoretical results hold for any arbitrary frequency $\omega \in [-\pi, \pi]$. We briefly explain why all theoretical developments still hold for thresholding on smoothed periodograms at a  general frequency defined in \eqref{eq:general_smoothing_estimator}. The key property we used to develop error bound analysis for thresholding estimators is orthogonality of $d(\omega_j), j\in F_n$. For general frequency $\omega$, we can show that $d(\omega+\omega_j), j=-m,\cdots, 0, \cdots, m$, are also orthogonal to each other. Based on this property, we could follow all arguments for Fourier  frequencies to achieve the same theoretical results.

We propose \textit{\underline{hard thresholding}} of averaged periodograms, i.e., 
\begin{equation}
T_{\lambda}(\hat{f}_{rs}(\omega_j))= \begin{cases}
\hat{f}_{rs}(\omega_j) & \mbox{ if } |\hat{f}_{rs}(\omega_j)|\ge \lambda \\
0 & \mbox{ if } |\hat{f}_{rs}(\omega_j)| < \lambda,
\end{cases}
\end{equation}
where $\lambda > 0$ is a threshold chosen by the user, and can potentially be a frequency dependent number $\lambda_j$. $T_\lambda(\cdot)$ is a thresholding operator on spectral density, $T_{\lambda}(\hat{f}_{rs}(\omega_j))$ represents the $(r,s)^{th}$ element of the thresholded matrix, where $1 \le r, s \le p$. For notational convenience, we will often use $\hat{f}_{\lambda, rs}(\cdot)$ instead of $T_{\lambda}(\hat{f}_{rs}(\cdot))$.

Following \citet{rothman2009generalized},  we also propose a variety of \textit{\underline{generalized thresholding operators}} $S_\lambda(.)$ that combine the benefits of shrinkage and thresholding. In particular, we consider elememt-wise shrinkage operator $S_\lambda(.)$ satisfying the following three conditions for any $z \in \mathbb{C}$: 
\begin{enumerate}[(1)]
\item $|S_\lambda(z)|\le |z|$,
\item $S_\lambda(z) = 0$ if $|z|\le \lambda$,
\item $|S_\lambda(z)-z|\le \lambda $.
\end{enumerate}
Similar to hard thresholding $T_\lambda(.)$, we apply this operator to individual elements of averaged periodogram. It turns out conditions (1)-(3) are satisfied by a number of thresholding and shrinkage procedures. 
In particular, the hard thresholding operator $T_\lambda(.)$ satisfies these conditions. In addition, generalizing \citet{rothman2009generalized} to the case of complex variables, we propose a soft thresholding (lasso) operator  
\begin{equation}
S_\lambda^s(z) = \frac{z}{|z|} \left(|z|-\lambda\right)_+, ~~ z \in \mathbb{C}, \nonumber
\end{equation}
and adaptive lasso operator 
\begin{equation}
S_\lambda^{\text{AL}} = \frac{z}{|z|} \left(|z| - \lambda^{(\eta+1)}|z|^{-\eta}\right)_+,  ~~ z \in \mathbb{C}. \nonumber
\end{equation}

Our proposed hard and soft thresholding procedures require selection of two tuning parameters: (i) smoothing span $m$ and (ii) level of threshold $\lambda$. In Section \ref{sec:theory}, we provide a detailed discussion of the theoretical choices of these parameters that ensure consistent estimation in high-dimensional regime. In the next subsection \ref{sec:tuning-parameter}, we discuss how to choose these two parameters in a data-driven fashion. The adaptive lasso based soft thresholding method has a third tuning parameter $\eta$. In our numerical and real data analyses, we set $\eta = 2$ following the suggestion of \cite{rothman2009generalized}, although a more general sample-splitting based choice along the line of Algorithm \ref{al1} can be adopted in practice.

\smallskip

When the thresholded spectral density matrices are sparse, they can be used to construct networks to visualize and analyze marginal dependence relationships among the component time series. However, just like thresholded covariance matrix estimators, thresholding individual entries does not necessarily ensure that the thresholded spectral density matrix estimate is positive definite. Our operator norm consistency results in Section \ref{sec:theory} implies that as long as the true spectral density is positive definite and the sample size is large enough, the thresholded estimate is positive definite with high probability. However, in finite sample, this is a limitation since the estimates cannot be directly used to calculate inverse spectral density and partial coherence. On the other hand, regularization is required to calculate inverse spectral density in high-dimension, and a more principled approach along the line of graphical lasso can be used to directly regularize entries of the inverse spectral density \citep{jung2015learning, jung2015graphical}. We expect that the key concentration inequalities developed in our analysis will be useful in the estimation of inverse spectral density as well.

\subsection{Choice of Tuning Parameters}\label{sec:tuning-parameter}
At any Fourier frequency $\omega_j$, we need to choose two tuning parameters for our method - (i) the smoothing span $2m+1$, and (ii) the threshold level $\lambda$. In this work, we select a single smoothing span for all the frequencies, but choose the threshold level separately for each frequency.

The smoothing span plays the role of ``effective sample size'' in estimating $f(\omega_j)$. Recall that 
\begin{equation}
    \hat{f}(\omega_j; m) = \frac{1}{2\pi (2m+1)} \sum_{|k|\le m} I(\omega_{j+k}). \nonumber
\end{equation}
In classical asymptotics ($p$ fixed, $n \rightarrow \infty$) and the Kolmogorov asymptotics ($p \rightarrow \infty, n \rightarrow \infty, p^2/n \rightarrow 0$) \citep{brockwell2013time, bohm2009shrinkage}, it is shown that for $\hat{f}(\omega_j; m)$ to be consistent, $m$  (depending on $n$) must go to infinity and $m/n \rightarrow 0$ as $n\rightarrow \infty$. Our non-asymptotic analysis in Section \ref{sec:theory} suggests that $m/[n\Omega_n(f)]  \rightarrow 0$, where $\Omega_n(f)$, defined as $\max_{r,s} \sum_{\ell=-n}^n |\ell| |\Gamma_{rs}(\ell)|$ is a measure of temporal dependence in the time series. For our numerical and real data applications, we choose $m$ in the order of  $\sqrt{n}$, with smaller values of $m$ for processes with stronger temporal dependence and larger $\Omega_n(f)$. A more data-driven approach along the line of \cite{ombao2001smoothing} and \citet{fiecas2014datadriven} can be designed with suitable modification to account for high-dimensionality, although we do not pursue this direction in this work.

\begin{algorithm}[t]\small
\DontPrintSemicolon 
\KwIn{$j, m, N$, periodograms at Fourier frequency $\{ I(\omega_k) \}_{k \in F_n}$, finite grid of thresholds $\mathcal{L}$}
\For{$\lambda \in \mathcal{L}$}{
 \For{$\nu \gets 1$ \textbf{to} $N$}{
     Randomly divide $\{j-m, \ldots, j, \ldots, j+m\}$ into two subsets $J_1$ and $J_2$ such that $\left||J_1| - |J_2| \right| \le 1$ and for any $k \in F_n$, $k \in J_1$ iff $-k \in J_1$ \;
     $\hat{f}_{1, \nu} (\omega_j) \gets \sum_{k \in J_1} I(\omega_k)$, ~~ $\hat{f}_{2, \nu} (\omega_j) \gets \sum_{k \in J_2} I(\omega_k)$\;
	 $\hat{R}_{\nu}(\omega_j, \lambda) \gets \left\| T_\lambda (\hat{f}_{1,\nu} (\omega_j)) - \hat{f}_{2,\nu}(\omega_j)\right\|^2_F$\;
	 }
$\hat{R}(\omega_j, \lambda) \gets \sum_{\nu=1}^N \hat{R}_{\nu}(\omega_j, \lambda) / N$
 
}
\KwOut{$\hat{\lambda}_j := \hat{\lambda}(\omega_j) = \argmin_{\lambda \in \mathcal{L}} \hat{R}(\omega_j, \lambda)$}
\caption{Threshold Selection by Frequency Domain Sample-splitting}
\label{alg:sample-split}
\label{al1}
\end{algorithm}

The second tuning parameter is the threshold value. Unlike the shrinkage estimators of spectral density matrices, finding  asymptotically optimal plug-in estimators for threshold level is challenging due to the non-smooth nature of thresholding operators. For covariance estimation from  i.i.d. data using thresholding, a sample-splitting method proposed in \citet{bickel2008covariance} or its variants are normally employed. In this method, the entire sample is split into two sub-samples, and the Frobenius norm difference between thresholded estimation in one sub-sample and regular sample covariance in the other sub-sample is compared for different levels of threshold. The entire exercise is repeated $N$ times and the level of threshold minimizing the average Frobenius norm difference is selected as the threshold.

This approach is not directly amenable to spectral density estimation since for any two given sub-sample sizes, only $N=1$ split is possible maintaining the temporal ordering. However, the periodograms at different positive Fourier frequencies $\omega_j \in F_n, \omega_j \ge 0$, are asymptotically independent. This suggests an analogous sample-splitting algorithm can be designed in the frequency domain. With this heuristic, we propose the following algorithm.

For each frequency $j$ $\in \{1, \cdots, [n/2]\}$, we randomly split the periodograms in $\{j-m ,\cdots,j+m\}$ into two sub-samples $J_1, J_2$ of size $m_1$ and $m_2$, where $|m_1 - m_2 | \le 1$. Since $I(\omega_{-k}) = I(\omega_{k})$, we keep $I(\omega_k)$ and $I(-\omega_k)$ in the same sub-sample. Then, for every $\lambda$ on a finite grid of possible threshold choices $\mathcal{L}$, we calculate the squared Frobenius norm of the difference between thresholded averaged periodogram on $J_1$, viz.,  $\hat{f}_{1}(\omega_j)$,  and averaged periodogram $\hat{f}_{2}(\omega_j)$ on $J_2$. This exercise is repeated $N$ times and the threshold $\lambda \in \mathcal{L}$ minimizing squared Frobenius norm is selected as $\hat{\lambda}_j$ for frequency $\omega_j$. A complete description is provided in Algorithm \ref{al1}.

\section{Theoretical Properties}\label{sec:theory}
In this section, we analyze asymptotic properties of thresholded averaged periodograms under high-dimensional regime. In particular, we derive non-asymptotic upper bound on the estimation error under operator and Frobenius norms and relate them to a notion of weak sparsity of the spectral density matrices. A key technical ingredient of our analysis is a concentration inequality of complex quadratic forms of temporally dependent Gaussian random vectors. In section \ref{sec:heavy-tail} we extend these results to linear processes with more general noise distributions, including subGaussian and   subexponential families.

In contrast with classical asymptotic framework where $p$ is fixed and $n \rightarrow \infty$, a non-asymptotic analysis for high-dimensional time series requires careful quantification of the convergence rates, in particular how they are affected by cross-sectional and temporal dependence inherent in the time series. Therefore, before proceeding with the main theoretical results, we describe parameters of the multivariate time series $X_t$ that appears in our estimation error bounds.

\smallskip
\noindent \textbf{Weak Sparsity of Spectral Density: } 
In order to make meaningful estimation in a high-dimensional regime, we focus on a class of spectral density matrices with suitable low-dimensional structure of \textit{weak sparsity} measured by $\vertiii{f}_q$ for some $0 \le q < 1$. Matrices with small $\vertiii{f}_0$  are \textit{exactly sparse}, while small  $\vertiii{f}_q$ correspond to matrices within a small $\ell_q$ ball in $\mathbb{C}^{p\times p}$. Weak sparsity of regression coefficients and covariance matrices have been proposed earlier in \cite{van2016lecture} and \citet{bickel2008covariance} respectively. 
Weakly sparse covariance matrices have been applied to climate studies according to \cite{cai2016rates} and gene expression array analysis, as mentioned in \cite{cai2012minimax}. \par 

Although the induced norm defined in notation section does not satisfy triangle inequality for $0\le q<1$, $\|A\|_q^q$ satisfies the triangle inequality leading to 
\vspace{-0.1in}
\begin{equation}
\max_{s=1}^p \sum_{s=1}^p |f_{rs}(\omega)|^q = \|f(\omega)\|_q^q  \le \vertiii{f}_q^q, \nonumber
\end{equation}
where $\vertiii{f}_q = \esssup_{\omega\in [-\pi, \pi]} \|f(\omega)\|_q$ as defined before. 
We provide a proof of this statement in lemma   \ref{lemma:q_norm_eq}. 
Since spectral density $f(\omega)$ is a Hermitian matrix, $\vertiii{f}_q^q$ also measures the row weak sparsity. This weakly sparse class covers a  variety of sparse patterns as shown in \citet{bickel2008covariance}. 

\smallskip
\noindent \textbf{Strength of Temporal and Cross-sectional Dependence: } 
The decay rates of the strengths of cross- and autocorrelation between components of $X_t$ capture the strength of temporal and cross-sectional dependence in data, which in turn relates to the effective sample size and appear in our error bounds. For meaningful estimation, we restrict ourselves to the class of short-range dependent time series $X_t$ with  the following summability assumption on its underlying autocovariance function $\Gamma(\ell)$:
\begin{assumption}\label{assumption:finite_auto}
$\sum_{\ell=-\infty}^\infty \|\Gamma(\ell)\|_{\text{max}}<\infty$. 
\end{assumption}

\noindent Under this assumption, we will present our bounds in terms of three quantities. The first one is $\vertiii{f}$ defined before, and will be used to assess the \textit{concentration of averaged periodogram around its expectation}. Note that $\vertiii{f}$ is finite since  
\begin{equation}
\|f(\omega)\| = \left\|\sum_{\ell = -\infty}^\infty \Gamma(\ell) e^{-\iu \omega \ell }\right\| \le \sum_{\ell = -\infty}^\infty \|\Gamma(\ell)\| \le \sum_{\ell=-\infty}^\infty p\|\Gamma(\ell)\|_{\rm{max}}.
\end{equation}
The other two quantities that capture the strength of temporal and contemporaneous dependence in the multivariate time series $\{X_t\}_{t \in \mathbb{Z}}$ are 
\begin{eqnarray}
\Omega_{n}(f) = \max_{1 \le r,s \le p} \sum_{\ell=-n}^n |\ell| |\Gamma_{rs}(\ell)|, ~~~~ L_n(f) = \max_{1 \le r,s \le p}\sum_{|\ell|>n} |\Gamma_{rs}(\ell)|.
\end{eqnarray}
Together, these two quantities help assess  how the \textit{bias of averaged periodogram} depends on the the degree of decay of the autocovariance function with increasing lag order $\ell$. Under Assumption \ref{assumption:finite_auto}, both of these quantities are finite. In Proposition \ref{prop:order_bias}, we show how these quantities grow for some common classes of multivariate time series.

\subsection{Estimation Consistency: Stable Gaussian Time Series} 

We start with a key technical ingredient of our analysis, a  Hanson-Wright type inequality \citep{rudelson2013hanson} for quadratic forms of random vectors generated by a multivariate Gaussian time series. This result generalizes Proposition 2.4 in \cite{Basu2015} by allowing an arbitrary matrix $A$ in the quadratic form. In Section \ref{sec:heavy-tail}, we extend this inequality to accommodate more general non-Gaussian time series.

Our modified Hanson-Wright inequality is crucial for understanding the concentration behaviour of averaged periodograms around the true spectral density $\left| \hat{f}_{rs}(\omega_j) - f_{rs}(\omega_j) \right|$, for a fixed coordinate $(r,s)$ of the $p\times p$ spectral density matrix. This deviation is required for selecting threshold $\lambda$ that ensures consistency in high-dimension. Unlike high-dimensional covariance estimation problem where sample covariance is an unbiased estimator of population covariance, the averaged periodogram at frequency $\omega_j$ is a biased estimator of $f(\omega_j)$. This requires developing upper bounds on both the ``bias'' and ``variance'' terms in the deviation of $\hat{f}_{rs}$ around $f_{rs}$: 
\begin{equation}
\left|\hat{f}_{rs}(\omega_j) - f_{rs}(\omega_j)\right| \le \left|\mathbb{E}\hat{f}_{rs}(\omega_j) - f_{rs}(\omega_j)\right| + \left| \hat{f}_{rs}(\omega_j) - \mathbb{E}\hat{f}_{rs}(\omega_j) \right|. \nonumber
\end{equation}
Note that while the first term above is indeed capturing bias of $\hat{f}_{rs}(\omega_j)$, the second term is not technically ``variance'' since this is the centered version of $\hat{f}_{rs}(\omega_j)$ and not its $L_2$ norm. Nevertheless, we continue to use the term 'variance' in this context since it captures the fluctuation of $\hat{f}_{rs}(\omega_j)$ around its expectation. The upper bounds on bias and variance terms are obtained in Propositions \ref{prop:bias_bound} and \ref{prop:variance_bound}, respectively. Finally, in Proposition \ref{prop: gauss_prop} we extend the deviation bound on a single $(r,s)$ to all $p^2$ elements of $f(\omega_j)$ and provide a non-asymptotic upper bound on the estimation error of the hard-thresholded averaged periodogram.

\begin{lem}
\label{lemma: hason_bound_time_gauss}
Suppose $\mathcal{\mathcal{X}}_{n\times p} = [X_1:\ldots:X_n]^\top$ is a data matrix from a stable  Gaussian time series $X_t$ satisfying Assumption \ref{assumption:finite_auto}. Then there exists a universal constant $c>0$ such that for any $\eta > 0$ and any $p \times p$ real matrix $A$, 
\begin{equation}
\begin{aligned}
&\mathbb{P}\left(\left|vec(\mathcal{\mathcal{X}}^\top)^\top A ~vec(\mathcal{\mathcal{X}}^\top) - \mathbb{E} \left[vec(\mathcal{\mathcal{X}}^\top)^\top A ~vec(\mathcal{\mathcal{X}}^\top)\right]\right| >2\pi \eta\vertiii{f} \right) \nonumber \\
&\le 2\exp\left[-c\min\left\{\cfrac{\eta}{\|A\|}, \cfrac{\eta^2}{\rank(A)\|A\|^2}\right\}\right]. \nonumber
\end{aligned}
\end{equation}
\end{lem}

For Gaussian $\mathcal{X}$, the above lemma generalizes Hanson-Wright inequality by allowing dependence among the entries of $\mathcal{X}$, and controlling the effect of dependence in the tail bound using $\vertiii{f}$, $\|A\|$ and $\rank(A)$. As will be evident from our analysis, this simple generalization will be immensely useful for studying concentration behaviour of averaged periodogram around the true spectral density in appropriate norms. Note that we replace  $\|A\|_F^2$ in standard Hanson-Wright inequality by a larger quantity $rk(A) \|A\|^2$, which makes the presentation easier in the asymptotic regime of our interest. In a lower dimensional regime, it is possible to get sharper rate using $\|A\|_F^2$ and $\int_{[-\pi,\pi]} \|f(\omega)\|^2 d\omega$ instead of $\vertiii{f}$, as discussed in \citet{Basu2015}. 
\smallskip
\par
\noindent \textbf{Bound on Bias Term: } In low-dimensional asymptotic regime ($p$ fixed, $n \rightarrow \infty$) the bias term is asymptotically negligible. In the double-asymptotic analysis of \citep{bohm2009shrinkage} as well, the  authors claim the bias of the estimator i.e., $|\mathbb{E}\hat{f}(\omega_j)-f(\omega)| = o(\frac{m}{n})$ which is negligible. In our non-asymptotic analysis, we need to derive an upper bound for this bias term in terms of $\{ \Gamma(\ell)\}_{\ell \in \mathbb{Z}}$, since the choice of threshold $\lambda$ depends crucially on this. The following proposition establishes such an  upper bound in terms of the temporal dependence present in the multivariate time series $X_t$. 

\begin{prop}
\label{prop:bias_bound}
For any coordinate $(r,s)$ with $1\le r,s\le p$ and any Fourier frequency $\omega_j$, $j\in F_n$, the estimation bias of averaged periodogram with a smoothing span $2m + 1$ satisfies
\begin{equation}
\begin{aligned}
\left|\mathbb{E}\hat{f}_{rs}(\omega_j) - f_{rs}(\omega_j)\right| \le \frac{m+1/2\pi}{n}\Omega_n(f) + \frac{1}{2\pi}L_n(f). \nonumber
\end{aligned}
\end{equation}
\end{prop}

A consequence of this proposition is that it shows $m/[n /(\Omega_n(f)] \rightarrow 0$ is sufficient to ensure bias vanishes asymptotically. In particular, for two $p$-dimensional time series and same sample size $n$, it suggests choosing a smaller $m$ for the series with stronger temporal dependence (larger $\Omega_n(f)$) since the effective sample size after accounting for dependence ($n/\Omega_n(f)$) is smaller.

We defer its proof to Appendix A. The upper bound on the bias depends on two terms: $\Omega_n(f)$ and $L_n(f)$. In previous works \cite{bohm2009shrinkage, bohm2008structural}, authors  argue that this upper bound on bias is of the order $\mathcal{O}(m/n)$. But since we focus on non-asymptotic analysis, these two terms $\Omega_n(f)$ and $L_n(f)$ appear in the choices of our two tuning parameters: threshold $\lambda$ and the smoothing span $2m+1$. To ensure we choose these parameters appropriately so that the bias vanishes asymptotically under a high-dimensional regime, it is important to understand how the above quantities grow with sample size $n$. Our next proposition provides some upper bounds on these quantities under  three different conditions. The first one is assuming a geometric decay rate on $\|\Gamma(\ell)\|_{\max}$, second one is 
about $\rho$-mixing condition (equivalent to strongly mixing for stationary Gaussian processes  \citep{bradley2005basic}) and VAR processes. Before that, we briefly review definition of $\rho$ mixing for condition 2 in Proposition \ref{prop:order_bias} and VAR process for condition 3 in Proposition \ref{prop:order_bias}. 

\cite{bradley2005basic} provides a good summary of various mixing conditions. Here we introduce the definition for $\rho$ mixing: for two $\sigma$-algebras $\mathcal{A}$ and $\mathcal{B}$,  we define 
\begin{equation}
\rho(\mathcal{A}, \mathcal{B}) = \sup  |\text{Corr}(f,g)|,~~ f\in L^2(\mathcal{A}), g\in L^2(\mathcal{B}), \nonumber
\end{equation}
where $f,g$ are two measurable functions with respect to $\sigma$-algebras $\mathcal{A}$ and $\mathcal{B}$ respectively. For stationary multivariate time series $X_t$, we define the $\rho$-mixing coefficient for gap $\ell$ as 
\begin{equation}
\label{eq:ts_rho_mixing}
\rho(\ell) = \rho(\sigma(X_t, t\le 0), \sigma(X_t, t\ge \ell)).
\end{equation}
The two characteristics $\|\Gamma(\ell)\|_{\max}$ and $\rho(\ell)$ are usually easy to describe for finite order VMA and  $\rm{VAR}(1)$ model. For $\rm{VAR}(d)$ with $d > 1$, however, it is more complicated. It is well known that we can rewrite a VAR(d)  model  
\begin{equation}
X_t = \sum_{\ell=1}^d A_{\ell} X_{t-\ell}+\varepsilon_t, \nonumber
\end{equation}
as a VAR(1) model $\tilde{X}_t = \tilde{A}_1 \, \tilde{X}_{t-1} + \tilde{\varepsilon}_t$, where 
\begin{equation}\label{eqn:var_dto1}
\tilde{X}_t = \left[ \begin{array}{c} X_t \\ X_{t-1} \\ \vdots \\ X_{t-d+1} \end{array} \right]_{dp \times 1}
\tilde{A}_1 = \left[ \begin{array}{ccccc} A_1 & A_2 & \cdots & A_{d-1} & A_d \\ 
										I_p & \mathbf{0} & \cdots & \mathbf{0} & \mathbf{0} \\
										\mathbf{0} & I_p & \cdots & \mathbf{0} & \mathbf{0} \\
										\vdots & \vdots & \ddots & \vdots & \vdots \\
										\mathbf{0} & \mathbf{0} & \cdots & I_p & \mathbf{0} \end{array}\right]_{dp \times dp}
\tilde{\varepsilon}_t = \left[ \begin{array}{c} \varepsilon_t \\ \mathbf{0} \\ \vdots \\ \mathbf{0} \end{array} \right]_{dp \times 1}. \nonumber
\end{equation} 
The sufficient and necessary condition for $X_t$ being stationary is that $\lambda_{\text{max}}( \tilde{A}_1)<1$. As we will discuss later that first two conditions in Proposition \ref{prop:order_bias} could be achieved by assuming coefficients has operator norm less than 1 for VAR(1) model.  But for VAR(d) with $d > 1$, it is known that $\|\tilde{A}_1\| \ge 1$ \citep{Basu2015}. So we cannot directly verify the geometric decay conditions \ref{geo_decay} and \ref{rho-mixing} in Proposition \ref{prop:order_bias}. But we can still get some compact bound by assuming $\tilde{A}_1$ is diagonalizable. Note that the assumption of diagonalizability is not stringent since  we can add a sufficiently small perturbation to the entries of $\tilde{A}_1$  so that its eigenvalues are distinct and we still have  $\lambda_{\text{max}}( \tilde{A}_1)<1$. We make this statement precise in Lemma \ref{lemma:spectral_simple} in the Appendix. 

\begin{prop}
\label{prop:order_bias}
Consider a weakly stationary,  centered time series $X_t$.  
\begin{enumerate}
    \item \label{geo_decay} Suppose $X_t$ satisfies $\|\Gamma(\ell) \|_{\textup{max}} \le \sigma_X \rho_X^{|\ell|}$ for all $\ell \in \mathbb{Z}$ for some $\sigma_X > 0$ and  $\rho_X \in (0, 1)$. Then 
    \begin{equation}
    \begin{aligned}
     \Omega_n \le 2\sigma_X\rho_X\left[\frac{1-(n+1)\rho_X^n +n\rho_X^{n+1}}{(1-\rho_X)^2}\right],  ~~~ L_n\le \frac{2\sigma_X\rho_X^{n+1}}{1-\rho_X}. \nonumber
    \end{aligned}
    \end{equation}
    \item \label{rho-mixing} Suppose $X_t$ satisfies $\rho(\ell) \le \sigma_X \rho_X^{|\ell|}$ where $\rho({\ell})$ is the $\rho$-mixing coefficient defined in \eqref{eq:ts_rho_mixing}. Then 
     \begin{equation}
    \begin{aligned}
     \Omega_n \le 2\|\Gamma(0)\|_{\textup{max}}\sigma_X\rho_X\left[\frac{1-(n+1)\rho_X^n +n\rho_X^{n+1}}{(1-\rho_X)^2}\right],  ~~~ L_n\le \frac{2\sigma_X\|\Gamma(0)\|_{\textup{max}}\rho_X^{n+1}}{1-\rho_X}. \nonumber
    \end{aligned}
    \end{equation}
    \item \label{var-condition} Suppose $X_t$ is a stable VAR(d) process $X_t = \sum_{\ell=1}^d A_d  X_{t-d} + \varepsilon_t$, where $\varepsilon_t \stackrel{i.i.d.}{\sim} N(0, \sigma^2 I)$. Set $\tilde{A}_1$ as in \eqref{eqn:var_dto1}, and assume $\tilde{A}_1$ is diagonalizable with an eigendecomposition $\tilde{A}_1 = SDS^{-1}$. Then 
    \begin{equation}
    \begin{aligned}
    &\Omega_n \le 2\kappa^2\frac{\lambda_{\textup{max}}(\tilde{A}_1)(1+n\lambda_{\textup{max}}^{n+1}(\tilde{A}_1)-(n+1)\lambda_{\textup{max}}(\tilde{A}_1))}{(1-\lambda_{\textup{max}}(\tilde{A}_1))^2(1-\lambda^2_{\textup{max}}(\tilde{A}_1))}, \nonumber \\
    &L_n \le 2\kappa^2\frac{\lambda_{\textup{max}}^{n+1}(\tilde{A}_1)}{(1-\lambda_{\textup{max}}(\tilde{A}_1))(1-\lambda^2_{\textup{max}}(\tilde{A}_1))}, \nonumber
    \end{aligned}
    \end{equation}   
 where $\kappa = \|S\|\|S^{-1}\|$. 
\end{enumerate}
\end{prop}

\begin{remark}
These bounds show that for a large class of stationary processes $X_t$, $\Omega_n(f)/n \rightarrow 0$ and $L_n(f) \rightarrow 0$ as $n \rightarrow \infty$. This implies it is possible to choose a large smoothing span $m \rightarrow \infty$ (required for asymptotically vanishing variance) that also ensures bias vanishing at a rate $O(m \Omega_n(f)/n)$. \end{remark}

\noindent \textbf{Bound on Variance term:} Unlike the bias term, the variance term $|\hat{f}_{rs}(\omega_j) - \mathbb{E} \hat{f}_{rs} (\omega_j)|$ is non-deterministic, so we need to establish high probability upper bound on this quantity. Compared to analogous bounds derived in covariance estimation for i.i.d. \citep{bickel2008covariance} or time series \citep{shu2014estimation} data, concentration of sample average of periodograms over nearby frequencies requires additional care since the summands are neither independent nor identically distributed to each other. However, the following proposition shows that the deviation bounds are the same order as i.i.d. data modulo a \textit{price of dependence} captured by $\vertiii{f}$. From a purely technical perspective, this Proposition forms the core of all our subsequent theoretical developments, and we believe this deviation bound will potentially be useful in other problems involving high-dimensional spectral density, e.g., estimation of partial coherence using graphical lasso type algorithms  \citep{jung2015graphical}. 
\begin{prop}
\label{prop:variance_bound}
There exist universal positive constants $c_1, c_2$ such that for any $\eta > 0$, 
\begin{equation}\label{eqn:conc-entrywise}
\mathbb{P}\left(\left|\hat{f}_{rs}(\omega_j) - \mathbb{E}\hat{f}_{rs}(\omega_j)\right| \ge \vertiii{f}\eta\right)\le c_1\exp\left[-c_2(2m+1)\min\{\eta, \eta^2\}\right].
\end{equation} 
\end{prop}

A complete proof is provided in Appendix \ref{appendix:proof_gaussian}. It is worth noting that the effective sample size in this bound is $(2m+1)$, a function of the smoothing span. The proof proceeds by separating the real and imaginary parts of $\hat{f}_{rs}(\omega_j) - \mathbb{E}\hat{f}_{rs}(\omega_j)$ into two quadratic forms involving random vectors $\{X_t\}_{t=1}^n$, subsequently applying Lemma \ref{lemma: hason_bound_time_gauss} to each part and deriving upper bounds on the spectral norm and ranks of the resulting $A$ matrices.

With the aforementioned bounds on bias and variance parts, we are now ready to present our main result that provides non-asymptotic upper bounds on the estimation error of the high-dimensional thresholded averaged periodogram in operator norm and Frobenius norm for Gaussian time series. The proof adapts techniques of \citet{bickel2008covariance} and \citet{rothman2009generalized} to combine the individual, entry-wise bounds on bias and variance terms across all the entries of the high-dimensional matrix.

\begin{prop}
\label{prop: gauss_prop}
Assume ${X}_t, t=1,\ldots,n$, are $n$ consecutive observations from a stable Gaussian time series satisfying Assumption \ref{assumption:finite_auto}, and consider a single Fourier frequency $\omega_j \in [-\pi, \pi]$. Assume $n \succsim \Omega_n(f) \vertiii{f}^2 \log p$. Then for any $m $ satisfying $m \precsim n/ \Omega_n(f)$ and $m \succsim \vertiii{f}^2 \log p$, and any $R > 0$,  
there exist universal constants $c_1, c_2 > 0$ such that choosing a threshold 
\begin{equation}
\label{eq:threshold_value}
\lambda = 2R \vertiii{f}\sqrt{\frac{\log p}{m}} +2\left[ \frac{m+1/2\pi}{n}\Omega_n(f)+\frac{1}{2\pi}L_n(f)\right], 
\end{equation}
the estimation error of thresholded averaged periodogram satisfies 
\begin{equation}
\begin{aligned}
\mathbb{P}\left(\left\|T_{\lambda}(\hat{f}(\omega_j)) - f(\omega_j)\right\|\ge 7\vertiii{f}_q^q \lambda^{(1-q)} \right)
\le c_1 \exp\left[-(c_2 R^2-2)\log p\right]. \nonumber
\end{aligned}
\end{equation}
Similarly, there exist universal positive constants $c_1, c_2$ such that for any $R>0$, with the same choice of threshold in \eqref{eq:threshold_value}, we have 
\begin{equation}
\mathbb{P}\left(\frac{1}{p}\left\|T_{\lambda}(\hat{f}(\omega_j)) - f(\omega_j)\right\|_F^2 \ge 13\vertiii{f}_q^q\lambda^{2-q} \right)\le c_1 \exp\left[-(c_2 R^2-2)\log p\right]. \nonumber
\end{equation}
\end{prop}

\begin{remark}
The estimation errors of our thresholded averaged periodogram in both operator norm and Frobenius norm depend on three factors: (i) the weak sparsity level of the true spectral density matrix $\vertiii{f}_q$; (ii) measure of stability of the process $\vertiii{f}$ to control variance of our estimate; (iii) rate of decay of autocovariances $\Omega_n$ and $L_n$ to control bias of our estimates. For any process satisfying $\Omega_n/n \rightarrow 0$ faster than $1/\vertiii{f}^2 \log p$, it is possible to find a sequence of smoothing span $m$ such that $\lambda \rightarrow 0$ as $n\rightarrow \infty$. The two appears in the threshold is only for an easy writing for technical proof. 
\end{remark}

The above result is non-asymptotic in nature, and our choice of threshold includes an upper bound on the bias. This is in contrast with existing works in the regime $p^2/n \rightarrow 0$, where this bias term is asymptotically negligible. 
Our choices of tuning parameters $m$ and $\lambda$ then ensure that both bias and variance decrease as $n,p$ grow, which is necessary for meaningful estimation, i.e.,   
\begin{equation}
\label{eq:var_bias_zero}
\max\left\{R\vertiii{f}\sqrt{\frac{\log p}{m}}, \frac{m}{n}\Omega_n(f)\right\} = o(1).
\end{equation}

\noindent \textbf{Generalized Thresholding of Averaged Periodogram:} Building up on the bounds on bias and variance terms of the individual entries of averaged periodogram, we are now ready to present our results for the generalized thresholding operator $S_\lambda(.)$. Suppose we have a generalized thresholding operator $S_\lambda(.)$ satisfying conditions (1)-(3) in Section \ref{sec:model-methods}.
The following proposition generalizes our previous estimation guarantees of hard thresholding to this more generalized family of estimates that includes lasso and adaptive lasso thresholds.
\begin{prop}
Suppose $S_{\lambda}(.)$ satisfies conditions (1) - (3) above. Then, for any Fourier frequency $\omega_j, j\in F_n$, and the same choices of tuning parameters $m$ and $\lambda$ as in Proposition \ref{prop: gauss_prop}, there exist universal constants $c_i > 0$ such that 


\begin{equation}
\mathbb{P} \left( \|S_{\lambda}(\hat{f}(\omega_j))-f(\omega_j)\| >  7\vertiii{f}_q^q \lambda^{(1-q)}\right) \le c_1 \exp \left[ -(c_2R^2-2)\log p \right]. \nonumber
\end{equation}
\end{prop}

As pointed out in \citet{rothman2009generalized}, the key is to build concentration inequality for each element of  $\hat{f}(\omega_j)-f(\omega_j)$ which is provided by proof in  Proposition \ref{prop:bias_bound} and \ref{prop:variance_bound}. After building the concentration inequality, all the proof left is exactly same as in Proposition \ref{prop: gauss_prop} and \citet{rothman2009generalized}. We omit this proof for sake of brevity.

\smallskip

\noindent \textbf{Sparsistency of Thresholded Averaged Periodograms}: A key motivation for using thresholded averaged periodogram for estimating high-dimensional spectral density matrix is the automatic selection of marginal independence graph among the $p$ times series. Our next result provides a support recovery guarantee at each frequency, justifying usage of these estimates to build weighted networks for downstream functional connectivity analysis in neuroscience problems (see Section \ref{sec:realdata}). In particular, the results show that with an appropriate choice of threshold, the support of estimated spectral density matrix is contained in the true support of $f(\omega)$ with high probability. In addition, if the spectral density is exactly sparse and minimum strength of cross-spectral density is sufficiently large, the entire support is recovered with high probability. For general weakly sparse spectral densities, our proposed thresholding procedures can still recover the strong connections with high probability.


\begin{prop}
\label{prop:consistency}
Assume ${X}_t, t=1,\ldots,n$, are $n$ consecutive observations from a stable Gaussian time series satisfying Assumption \ref{assumption:finite_auto}, and consider a single Fourier frequency $\omega_j, j\in F_n$. Assume $n \succsim  \Omega_n(f) \vertiii{f}^2 \log p.$ Then for any $m $ satisfying $m \precsim n/ \Omega_n(f)$ and $m \succsim \vertiii{f}^2\log p$, and any $R > 0$,
if we set threshold value $\lambda$ as \eqref{eq:threshold_value} , then there exists universal constant $c_1, c_2$  s.t.
\begin{equation}
\mathbb{P}\left(\exists ~r,s : T_\lambda(\hat{f}_{rs}(\omega_j)) \neq 0, f_{rs}(\omega_j)=0\right)\le  c_1\exp[-(c_2R^2-2) \log p]. \nonumber
\end{equation}

Define $\mathcal{S}(\gamma) = \left\{(r,s):|f_{rs}(\omega_j)|\ge \gamma \lambda \right\}$ with some $\gamma>3/2$, then
\begin{equation}
\mathbb{P}\left(\exists ~(r,s) \in \mathcal{S}(\gamma) : T_\lambda(\hat{f}_{rs}(\omega_j))=0, f_{rs}(\omega_j) \neq 0\right)\le  c_1\exp[-(c_2 (\gamma-1)^2R^2-2) \log p]. \nonumber
\end{equation}
\end{prop}

\begin{remark}
The first probabilistic bound claims that probability of false positive selection goes to zero if $\lambda = o(1)$ with R large enough and the second probabilistic bound claims that we could recover the signal with strength larger than the threshold we choose ($\gamma>3/2$). 
\end{remark}

\noindent \textbf{Coherence Matrix Estimation}:
Our next proposition provides an error bound for each element of this plug-in estimator of coherence matrix defined in \eqref{def:coherance}, 
\begin{equation}
\hat{g}_{rs}(\omega_j) = \frac{\hat{f}_{rs}(\omega_j)}{\sqrt{\hat{f}_{rr}(\omega_j)\hat{f}_{ss}(\omega_j)}}. \nonumber
\end{equation}
Note that $\hat{f}_{rr} \neq 0$ ($\hat{f}_{rr}$ is a real number) almost surely for Gaussian time series $X_t$. The sparsistency results can be generalized along the line of Proposition \ref{prop:consistency} to ensure coherence graph selection consistency.
\begin{prop}
\label{prop:coherance}
Assume ${X}_t, t=1,\ldots,n$, are $n$ consecutive observations from a stable Gaussian time series $X_t$ satisfying Assumption \ref{assumption:finite_auto}, and $\tau := \min_{r=1}^p f_{rr}(\omega_j)>0$. Consider a single Fourier frequency $\omega_j, j\in F_n$. Assume $n \succsim \Omega_n(f) \vertiii{f}^2 \log p$. Then for any $m $ satisfying $m \precsim n/ \Omega_n(f)$ and $m \succsim \vertiii{f}^2\log p$ and $\lambda$ as in  \eqref{eq:threshold_value}, there exist universal positive constants $c_1, c_2$ such that for any $R>0$, \begin{equation}
\mathbb{P}\left(\exists ~r,s : |T_{2\lambda/\tau}(\hat{g}_{rs}(\omega_j))|>0, g_{rs}(\omega_j)=0\right)\le  c_1\exp[-(c_2R^2-2) \log p]. \nonumber
\end{equation}
Define $\mathcal{S}(\gamma) := \left\{(r,s):|g_{rs}(\omega_j)|\ge \gamma\lambda /\tau \right\}$ with some $\gamma>3/2$. Then we have 
\begin{equation}
\mathbb{P}\left(\exists ~(r,s) \in \mathcal{S}(\gamma) : T_{2\lambda/\tau}(\hat{g}_{rs}(\omega_j))=0, |g_{rs}(\omega_j)| >0\right)\le  c_1\exp[-(c_2 (\gamma-1)^2R^2-2) \log p]. \nonumber
\end{equation}

\end{prop}

\section{Spectral Density Estimation of Linear Processes}\label{sec:heavy-tail} 

In this section, we extend the estimation consistency results of our thresholding based spectral density estimators beyond Gaussian time series. The proof of the Hanson-Wright type inequality for temporally dependent data in Lemma \ref{lemma: hason_bound_time_gauss} crucially relies on the fact that uncorrelated Gaussian random  variables are also independent with each other. This does not apply for non-Gaussian time series in general. However, we show in this section that for some linear processes with error tail heavier than Gaussian distribution, it is possible to derive similar concentration inequalities. Using these concentration inequalities, we then extend the theoretical results of previous section to a larger class of non-Gaussian linear time series.

We focus on linear processes with absolutely summable MA($\infty$) coefficients:
\begin{equation}
\label{eq:infinite_ma}
    X_t = \sum_{\ell = 0}^\infty B_\ell \varepsilon_{t-\ell},
\end{equation}
where $B_\ell\in \mathbb{R}^{p\times p}$ and $\varepsilon_t\in \mathbb{R}^p$ have i.i.d. centered distribution with possibly heavier tails than Gaussian. \citet{rosenblatt1985stationary} shows that stationarity of $X_t$ is ensured under element-wise absolute summability of MA coefficients
\begin{equation}
\label{eq:infinity_moving_average}
    \sum_{\ell=0}^\infty |B_{\ell, (r,s)}| < \infty 
\end{equation}
for any $r,s$, $1\le r, s\le p$. Under this condition, the autocovariance $\Gamma(\ell) = \sum_{t=0}^\infty B_{t}B_{t+\ell}^\top$ is well-defined for every $\ell \in \mathbb{Z}$, and Assumption \ref{assumption:finite_auto} holds. A proof is given in Lemma \ref{lemma:linear_assumption} for completeness.\par 
We assume that each component $\varepsilon_{tr}, \, 1 \le r \le p$, of the random vector $\varepsilon_t$ is from one of the following three types of distributions. 
\begin{enumerate}[(C1)]
    \item sub-Gaussian: there exists some $\sigma>0$ such that for all $\eta > 0$,   $\mathbb{P}[|\varepsilon_{tr}|>\eta]\le 2\exp\left(-\frac{\eta^2}{2\sigma^2}\right)$; 
    \label{C1} 
    \item generalized sub-exponential with parameter $\alpha>0$: there exist positive constants $a, b$ such that for all $\eta > 0$,  
    $\mathbb{P}[|\varepsilon_{tr}|\ge \eta^\alpha]\le a\exp(-b\eta)$ \citep{erdHos2012bulk};
    \label{C2}
    \item $\varepsilon_{tr}$ has finite $4^{th}$ moment: $\mathbb{E} \varepsilon_{tr}^4 \le K <\infty$.  \label{C3}
\end{enumerate}
\begin{remark}
$\varepsilon_{tr}$ has generalized sub-exponential distribution defined in \citet{erdHos2012bulk}, which is more general than the usual definition of sub-exponential used in the literature with $\alpha = 1$. In some recent works  \citep{FiniteTimeIdentification2017, wong2017lasso}, such distributions were also referred to as sub-Weibull distributions.
\end{remark}
Next we establish concentration inequalities similar to Lemma \ref{lemma: hason_bound_time_gauss} for linear processes where the distribution of each coordinate of  noise terms comes from one of the  families C\ref{C1}, C\ref{C2} and C\ref{C3}. 

For i.i.d. data, existing works have generalized Hanson-Wright type inequality  for distributions in C\ref{C1} and  C\ref{C2} \citep{rudelson2013hanson, erdHos2012bulk}. We can use Markov inequality to get an upper bound for C\ref{C3} as well. We summarize these results in the following lemma. Its proof is defered to Appendix \ref{Appendix:proof_heavytail}.

\begin{lem}
\label{lemma:heavy_tail_hanson}
Consider a random vector $\varepsilon \in \mathbb{R}^p$ with i.i.d. coordinates following one of the three distributions C\ref{C1} - C\ref{C3}, and a deterministic $p \times p$ matrix $A$. For simplicity, let us assume $A$ is a real matrix, and  $\mathbb{E} \varepsilon_r = 0$ and $\mathbb{E} \varepsilon^2_r = 1$ for every $r, \, 1 \le r \le p$. Then 
\begin{equation}
\mathbb{P}\left(|\varepsilon^\top A\varepsilon - \mathbb{E} \varepsilon^\top A\varepsilon|\ge \eta\right) \le \mathcal{T}_j(\eta, A), \nonumber
\end{equation}
where $\mathcal{T}_j(\eta, A), j=1,2,3$, are tail decay functions for the three families, given by 
\begin{equation}
\begin{aligned}
&\mathcal{T}_1(\eta, A) =  2\exp\left[-c\min\left\{\cfrac{\eta}{\|A\|}, \cfrac{\eta^2}{\rank(A)\|A\|^2}\right\}\right], \nonumber \\
&\mathcal{T}_2(\eta, A) = c_1\exp\left[-c_2\left(\frac{\eta}{\sqrt{\rank(A)}\|A\|}\right)^{\frac{1}{2+2\alpha}}\right], \nonumber \\
& \mathcal{T}_3(\eta, A) =  \frac{c_3\rank(A)\|A\|^2}{\eta^2}. \nonumber
\end{aligned}
\end{equation}
Here $c$ only depends on $\sigma$ in C\ref{C1}, $c_1, c_2$ only depend on $a, b$ in C\ref{C2} and $c_3$ only depends on K in C\ref{C3}, and none of them depends on the MA coefficients $B_\ell$, $\ell \ge 0$.
\end{lem}

\noindent Now we extend these three inequalities by replacing $\varepsilon$ with $n$ random variables of the form $X_t = \sum_{\ell \ge 0} B_\ell \varepsilon_{t-\ell}$. The main technical difficulty stems from handling the sum of infinitely many terms $\varepsilon_t$. We apply a truncation argument to overcome this.

\begin{prop}
\label{lemma:heavy_tail_time_hanson} 
Suppose $\mathcal{X} = [X_1:X_2:\ldots:X_n]^\top$ is a data matrix with $n$ consecutive observations from  a stationary linear process  $\{X_t\}$ in  \eqref{eq:infinity_moving_average}  with each coordinate of $\varepsilon_t$ is i.i.d. from one of the families C\ref{C1}, C\ref{C2} or  C\ref{C3}, and consider a deterministic $np \times np$ matrix $A$. Then 
\begin{equation}
\begin{aligned}
\mathbb{P}\left(\left|vec(\mathcal{X}^\top)^\top A ~vec(\mathcal{X}^\top) - \mathbb{E} \left[~vec(\mathcal{X}^\top)^\top A ~vec(\mathcal{X}^\top)\right]\right| >2\pi
\eta \vertiii{f} \right) \le \mathcal{T}_j(\eta, A), \nonumber
\end{aligned}
\end{equation}
where $\mathcal{T}_j(\eta, A), j=1,2,3$, are tail decay functions for the three families, as defined in Lemma \ref{lemma:heavy_tail_hanson}.
\end{prop}
\begin{remark}
The main difference between the  concentration inequalities in Lemma \ref{lemma:heavy_tail_hanson} and Proposition  \ref{lemma:heavy_tail_time_hanson} is that $\vertiii{f}$ appears in the right side of the inequality. As pointed by \cite{Basu2015}, $\vertiii{f}$ can be viewed as a ``price of dependence'' present in time series data. For instance, if $B_\ell = 0$ for all $\ell > 0$, $B_0 = I$, and $\var(\varepsilon_{tr}) = 1$ for all $r, t$, we have  $\vertiii{f} = \frac{1}{2\pi}$ which coincides with the result in Lemma \ref{lemma:heavy_tail_hanson} applied to a $np$-dimensional random vector. 
\end{remark}

This result generalizes the Hanson-Wright type concentration inequality in Lemma \ref{lemma: hason_bound_time_gauss} to the case of three non-Gaussian families with potentially heavier tails.
After building concentration inequalities for these three cases, we could bound the variance term as Proposition \ref{prop:variance_bound} which we listed as following Proposition. 
The proof follows the same line as the proof of Proposition \ref{prop:variance_bound}, by replacing Gaussian Hanson-Wright type inequality with those in Proposition \ref{lemma:heavy_tail_time_hanson}. We omit this for sake of brevity.

\begin{prop}
\label{prop:heavy_tail_bound_variance}
Suppose $\mathcal{X} = [X_1:X_2:\ldots:X_n]^\top$ is a data matrix with $n$ consecutive observations from  a stationary linear process  $\{X_t\}$ in  \eqref{eq:infinity_moving_average} , each coordinate of $\varepsilon_t$ is i.i.d. from one of the families C\ref{C1}, C\ref{C2} or  C\ref{C3}. Then there exist  general constants $c_i > 0$ (depending only on the error distribution but not on the coefficients $B_\ell$ of the linear process) 
such that  for any $r, s$, $1 \le r, s \le p$, and any Fourier frequency $\omega_j \in F_n$, we have 
\begin{equation}\label{eqn:heavy-tail-conc-entrywise}
\mathbb{P}\left(\left|\hat{f}_{rs}(\omega_j) - \mathbb{E}\hat{f}_{rs}(\omega_j)\right| \ge \vertiii{f}\eta\right)\le \mathcal{B}_k(\eta, m),
\end{equation} 
where $\mathcal{B}_k$, $k=1, 2, 3$, are defined as 
\begin{equation}
\begin{aligned}
& \mathcal{B}_1(\eta, m) = c_1\exp\left[-c_2\min\{\eta, \eta^2\}\right],\\
& \mathcal{B}_2(\eta, m) = c_3\exp(\left[-c_4\left(\sqrt{m}\eta\right)^{\frac{1}{2+2\alpha}}\right]),\\
& \mathcal{B}_3(\eta, m) = \frac{c_5}{m\eta^2}.\nonumber
\end{aligned}
\end{equation}
\end{prop}
After showing the bound for variance term for linear process, we can derive estimation consistency of hard-thresholding estimators similar to Proposition \ref{prop: gauss_prop} for linear processes with any of the three different types of noise distributions. 

\begin{prop}
\label{prop: linear_prop}
Suppose $\{X_t\}$ is a linear process defined in \eqref{eq:infinite_ma}, with $\varepsilon_t$ from one of the three distributions C\ref{C1}, C\ref{C2} and C\ref{C3}, and consider a Fourier frequency $\omega_j \in F_n$.
Assume $n \succsim \Omega_n(f) \mathcal{N}_k$, where $\mathcal{N}_1 = \vertiii{f}^2 \log p$, $\mathcal{N}_2 = \vertiii{f}^2 (\log p)^{4+4\alpha}$, and $\mathcal{N}_3 = p^{2}$ for the three families C\ref{C1}, C\ref{C2} and C\ref{C3}. Then for any $m $ satisfying $m \precsim n/ \Omega_n(f)$ and $m \succsim \vertiii{f}^2 \mathcal{N}_k$, and any $R > 0$, 
if we choose threshold for the three different distributions as 
\begin{enumerate}[(C1)]
    \item $\lambda = 2R\vertiii{f}\sqrt{\frac{\log p}{m}} + 2\left[\frac{m+1/2\pi}{n}\Omega_n(f)+\frac{1}{2\pi}L_n(f)\right]$,
    \item $\lambda = 2\vertiii{f}\frac{(R\log p)^{2+2\alpha}}{\sqrt{m}} + 2\left[\frac{m+1/2\pi}{n}\Omega_n(f)+\frac{1}{2\pi}L_n(f)\right]$,
    \item $\lambda = 2\vertiii{f}\frac{p^{1+R}}{\sqrt{m}} + 2\left[\frac{m+1/2\pi}{n}\Omega_n(f)+\frac{1}{2\pi}L_n(f)\right]$,
\end{enumerate}
then 
\begin{equation}
\mathbb{P} \left( \|T_{\lambda}(\hat{f}(\omega_j))-f(\omega_j)\| >  7\vertiii{f}_q^q \lambda^{(1-q)}\right) \le \mathcal{B}_k, \nonumber
\end{equation}
where the tail probability $\mathcal{B}_k$ are given as 
\begin{equation}
\begin{aligned}
& \mathcal{B}_1 =  c_1 \exp\left[-(c_2 R^2-2)\log p\right], \\
& \mathcal{B}_2 =  c_3 \exp\left[-(c_4 R-2)\log p\right], \\
& \mathcal{B}_3 = c_5 \exp \left[ - 2R \log p \right],
\end{aligned}
\end{equation}
where $c_i > 0$ are some general constants depending only on the error distribution but not on the coefficients $B_\ell$ of the linear process. 
\end{prop}

The proof follows the same line as the proof of Proposition \ref{prop: gauss_prop}, by replacing Gaussian variance bound in Proposition \ref{prop:variance_bound} with Proposition \ref{prop:heavy_tail_bound_variance}. We omit this for sake of brevity.
\begin{remark}
The heavier is the tail of the noise distribution, the wider bandwidth of periodogram averaging ($2m+1$ in our notation) is required for consistent estimation. For generalized sub-exponential, we can ensure consistency in high-dimensional regime  $p=O(n^\alpha), \alpha>1$, while if we only assume existence of fourth moment, we will require $p=o(\sqrt{n})$ for consistency. 
\end{remark}


\section{Simulation Studies}\label{sec:simulation}
We assess the finite sample properties of our proposed spectral density estimators through numerical experiments on simulated data sets. To this end, we compare the performance of smoothed periodogram, shrinkage estimator from \cite{bohm2009shrinkage},  hard thresholding, soft thresholding (lasso) and adaptive lasso thresholding. In particular, we simulate data from vector moving average (VMA) and autoregressive (VAR) processes with block-diagonal transition matrices and evaluate estimation and model selection performance of these methods for different values of $n$ and $p$. Overall, the results demonstrate that thresholding methods provide substantial improvements in estimation accuracy over smoothed periodograms and shrinkage methods when $p$ is large and the true spectral density is approximately sparse. In addition, thresholding methods accurately recovers the edges in coherence networks, as measured by their precision, recall and area under receiver operating characteristic (ROC)  curves. 

\textit{Generative models: } { We consider VAR(1) models $X_t = A X_{t-1} + \varepsilon_t$  of three different dimensions: $p =12, 48, 96$. Each element in $\varepsilon_t$ is independent and identically distributed as $\mathcal{N}(0,1)$, and the transition matrix $A$ is composed of $3 \times 3$ block matrices on the diagonal. Each block matrix $A^0$ has $0.5$ on the diagonal and $0.9$ on the first upper off-diagonal. We also consider VMA(1) models $X_t = B \varepsilon_{t-1}+\varepsilon_t$ of the same dimensions as the VAR models. These transition matrix structures are adopted from \citet{fiecas2014datadriven}, where a data-driven shrinkage method was shown to improve upon smoothed periodograms in high-dimensional settings. For each model, we generate $n =100, 200, 400, 600$ consecutive observations from the multivariate time series.}


The transition matrix $A$ of VAR is a block diagonal composed of identical blocks consisting of a $3 \times 3$ upper triangular matrix $A^0$. 
Similarly, the VMA transition matrix $B$ is a block diagonal matrix composed of identical $3 \times 3$ upper triangular matrix $B^0$.

\begin{equation}
A^0 = B^0 = \left[ \begin{array}{ccc} 
0.5 & 0.9 & 0  \\ 
0 & 0.5 & 0.9  \\
0 & 0 & 0.5  \\
\end{array}\right].
\end{equation}



The estimated spectral density matrices are compared to the true spectral densities. For stable, invertible VARMA(1,1) processes $X_t = AX_{t-1} + 
  \varepsilon_t 
  + B\varepsilon_{t-1}
$, true spectral densities take the form 
\begin{equation*}
f(\omega) = \frac{1}{2\pi} (\mathcal{A}^{-1}(e^{-i\omega}))\mathcal{B}(e^{-i\omega})\Sigma_{\varepsilon}\mathcal{B}^{\dag}(e^{-i\omega})(\mathcal{A}^{-1}(e^{-i\omega}))^{\dag},
\end{equation*}

where $\mathcal{A}(z) = I_p - A z$ and $\mathcal{B}(z) = I_p +  B z$.

\smallskip
\textit{Performance Metrics: } We compare the estimation performances of different estimators of  $f(\omega_j)$ using Relative Mean Integrated Squared Error (RMISE) in Frobenius norm, defined as
 \begin{equation*}
{RMISE}(\hat{f}) := \frac{\sum_{j\in F_n} \|\hat{f}(\omega_j) - f(\omega_j)\|_F^2}{\sum_{j\in F_n} \|f(\omega_j)\|_F^2}. 
\end{equation*}

In order to capture how well the three thresholding methods recover the non-zero coordinates in a spectral density matrix under exactly sparse generative VMA and VAR models, we also record their precision, recall and F1 measures over all Fourier frequencies 
\begin{eqnarray}
    && \text{precision}(\omega_j) = 
    \frac{\# \{(r,s): ~ |\hat{f}_{rs}(\omega_j)|\neq 0, ~ |f_{rs}(\omega_j)|\neq 0 \}}{\# \{(r,s): ~ |\hat{f}_{rs}(\omega_j)|\neq 0\}} \nonumber \\
    && \text{recall}(\omega_j) = \frac{\# \{(r,s): ~ |\hat{f}_{rs}(\omega_j)|\neq 0, ~ |f_{rs}(\omega_j)|\neq 0 \}}{\# \{(r,s): ~ |f_{rs}(\omega_j)|\neq 0\}} \nonumber \\
    && \text{F1}(\omega_j) =  2\times (\text{precision}(\omega_j) \cdot \text{recall}(\omega_j))/(\text{precision}(\omega_j) + \text{recall}(\omega_j)). \nonumber 
\end{eqnarray}
We calculate each of the three criteria averaged across all Fourier frequencies $j\in F_n$.  All the experiments are replicated  $50$ times, and mean and standard deviation of the performance metrics are reported.

We also evaluate the accuracy of thresholding methods in selecting the graph $G = \{ (r,s) \in V \times V: \hat{f}_{rs}(\omega_j) \neq 0 \mbox{ for some } \omega_j \in F_n \}$. For this purpose, we use averaged absolute coherence (across all Fourier frequencies) to construct a single $p \times p$ weighted adjacency matrix $\hat{G}$, and then measure its accuracy in selecting edges of the true graph $G$. \par 

\textit{Tuning parameter selection: } For each of the three thresholding methods, we use the sample-splitting algorithm \ref{alg:sample-split} with $N = 1$ to determine the value of threshold for individual frequencies. We choose a  grid $\mathcal{L}$ of equispaced values between the minimum and maximum moduli of off-diagonal entries in smoothed periodogram.  Based on the theoretical considerations in Section \ref{sec:theory}, the smoothing spans for VMA models are chosen by setting $m = \sqrt{n}$. Since $\Omega_n(f)$ is  larger for VAR than VMA models considered here, a smaller  smoothing span is chosen by setting $m = 2/3\sqrt{n}$. The results are qualitatively similar in our sensitivity analysis with different values of $m$ of this order. 

\textit{Results: } The RMISE of smoothed (averaged) periodograms, shrinkage and thrsholding methods are reported in Table \ref{table:rmise-homogeneous-final}. The results show that both shrinkage and thresholding outperform smoothed periodogram, and the improvement is more prominent for larger $p$. Further, thresholding procedures show some improvement over shrinkage methods in these approximately sparse data generative models. Amongst the three thresholding methods,  lasso and adaptive lasso tend to have lower error than hard thresholding in most settings. 

Precision, recall and F1 scores of the three thresholding methods are reported 
in Appendix \ref{appendix:more_tables}. In most of the simulation settings, the methods have high precision but low recall, indicating higher true negative in general. This matches with our theoretical predictions for weakly sparse spectral densities in Proposition \ref{prop:consistency}. The F1 scores are in the range of $50-60\%$ in most simulation settings. As in the RMISE results, lasso and adaptive lasso thresholds perform significantly better than hard thresholding in most simulation settings.

\begin{figure}[!t]
    \centering
    \includegraphics[width=0.48\textwidth]{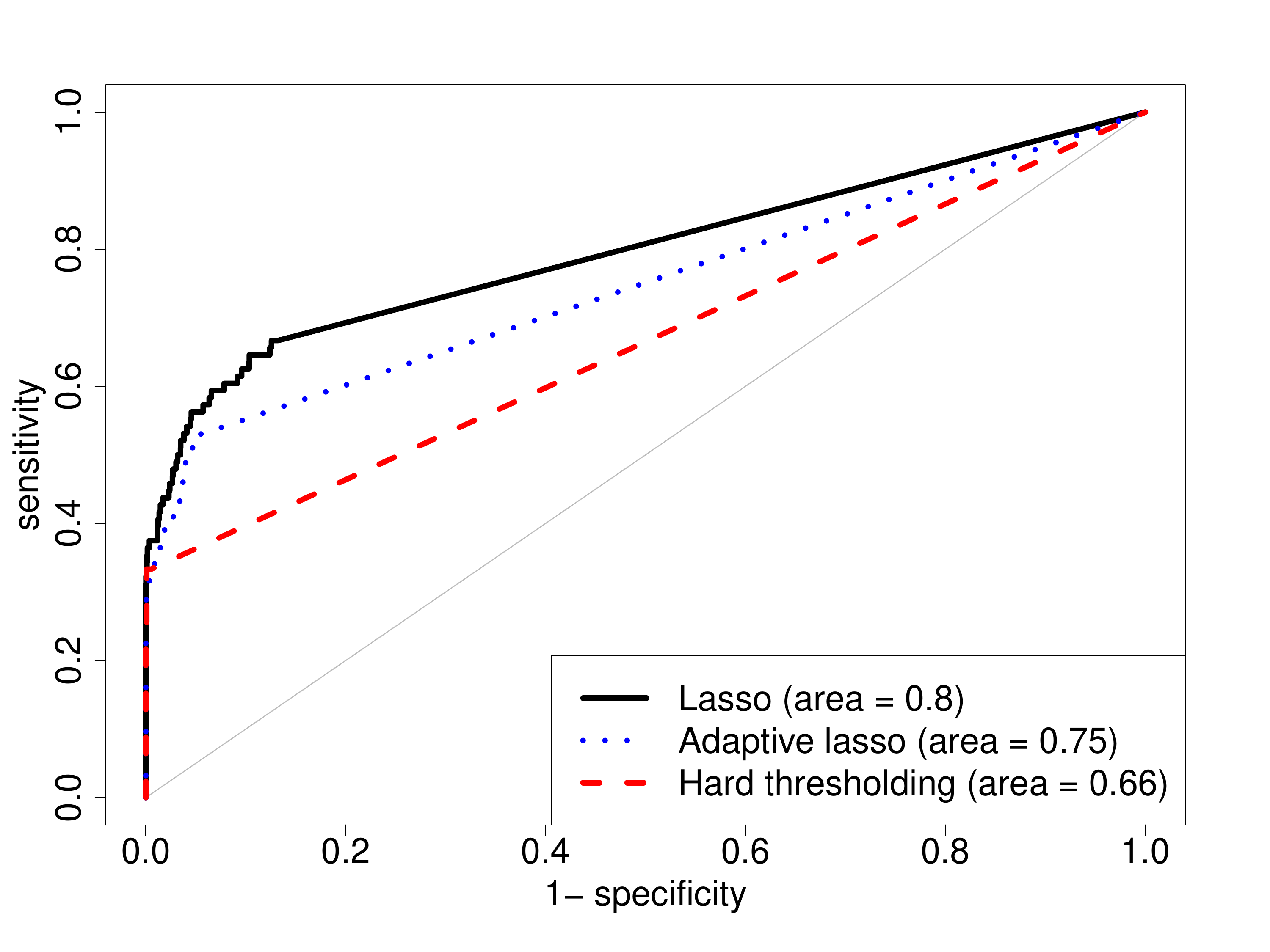}
    \includegraphics[width=0.48\textwidth]{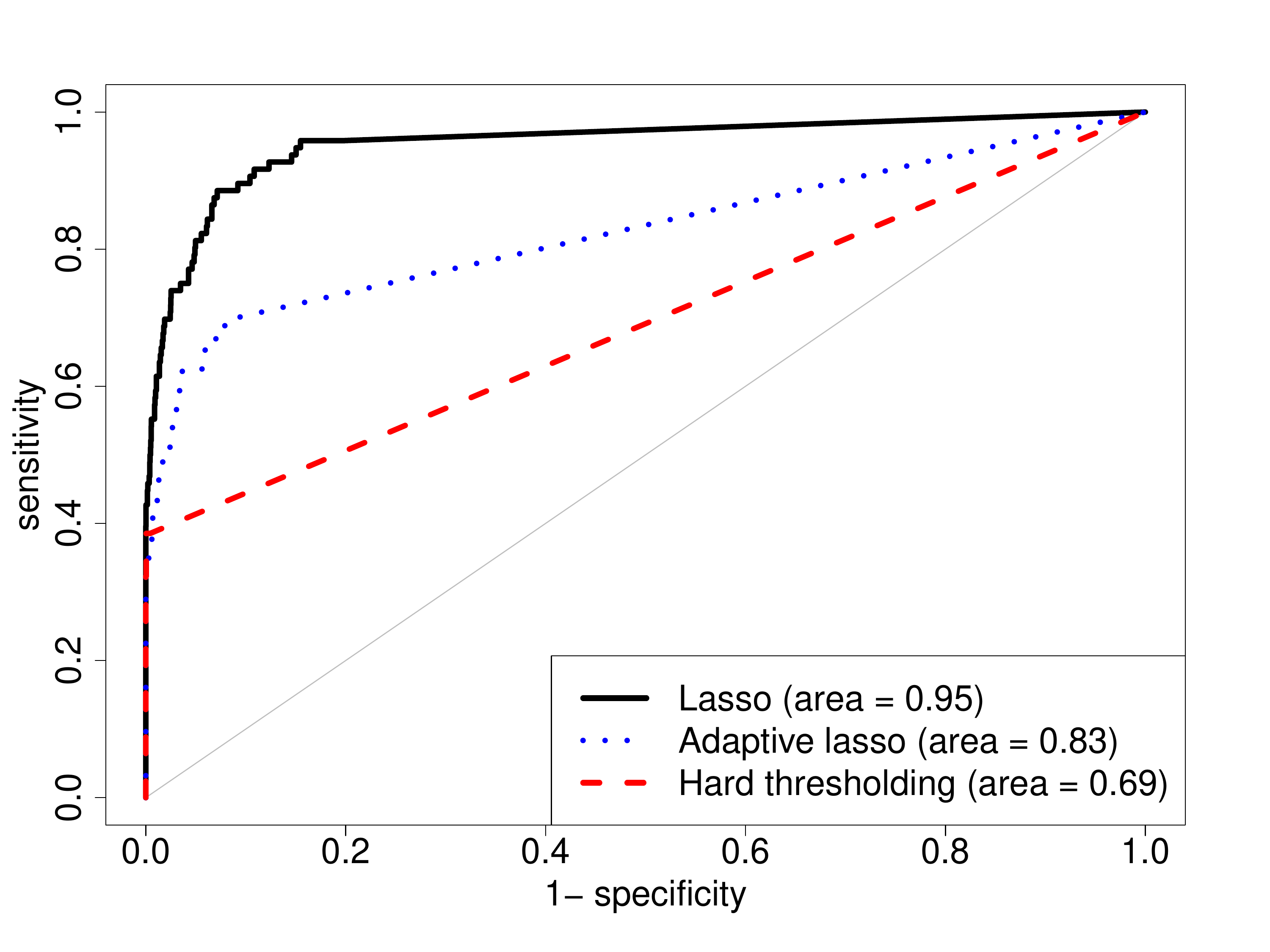}
    \includegraphics[width=0.48\textwidth]{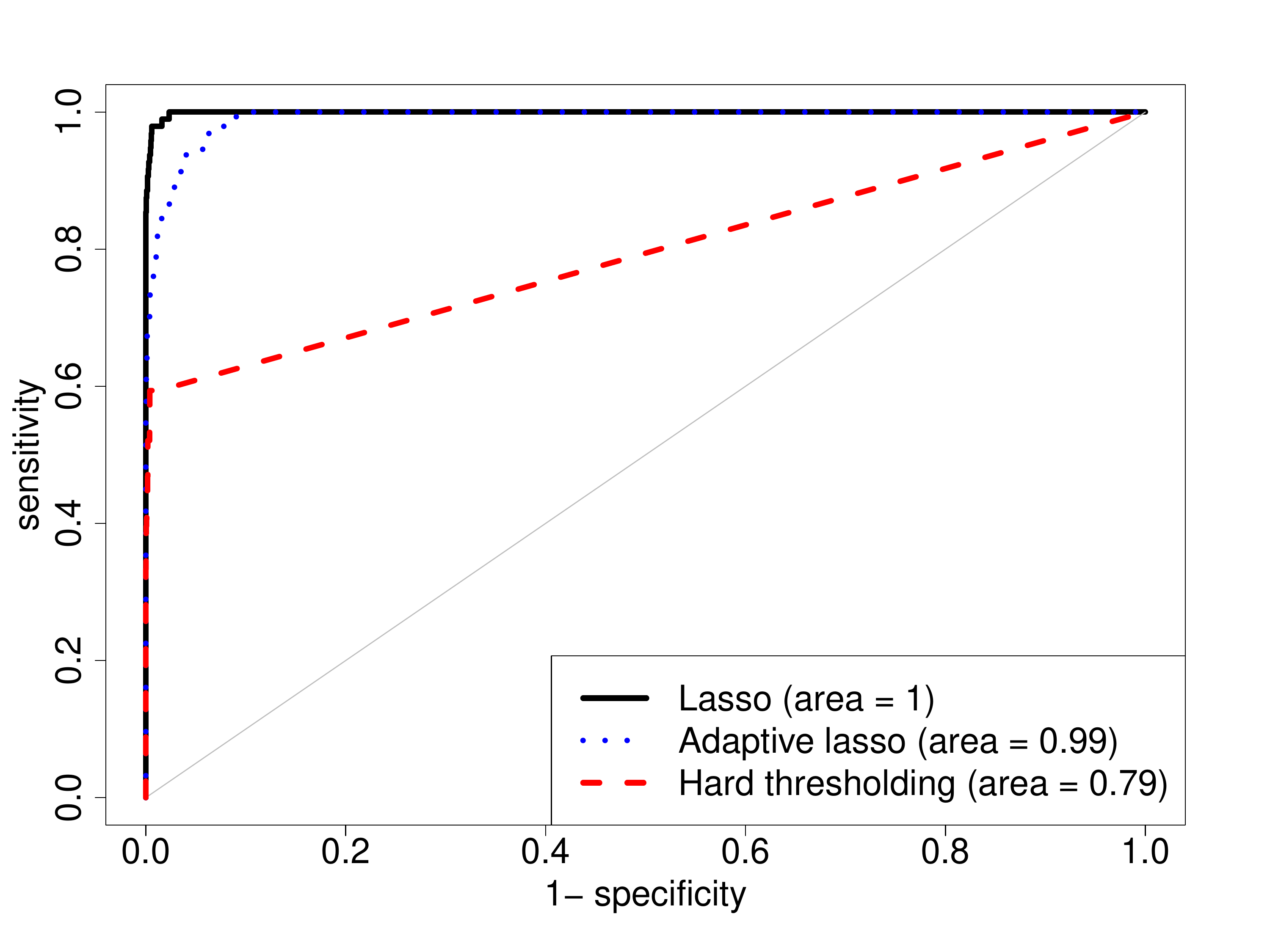} 
    \includegraphics[width=0.48\textwidth]{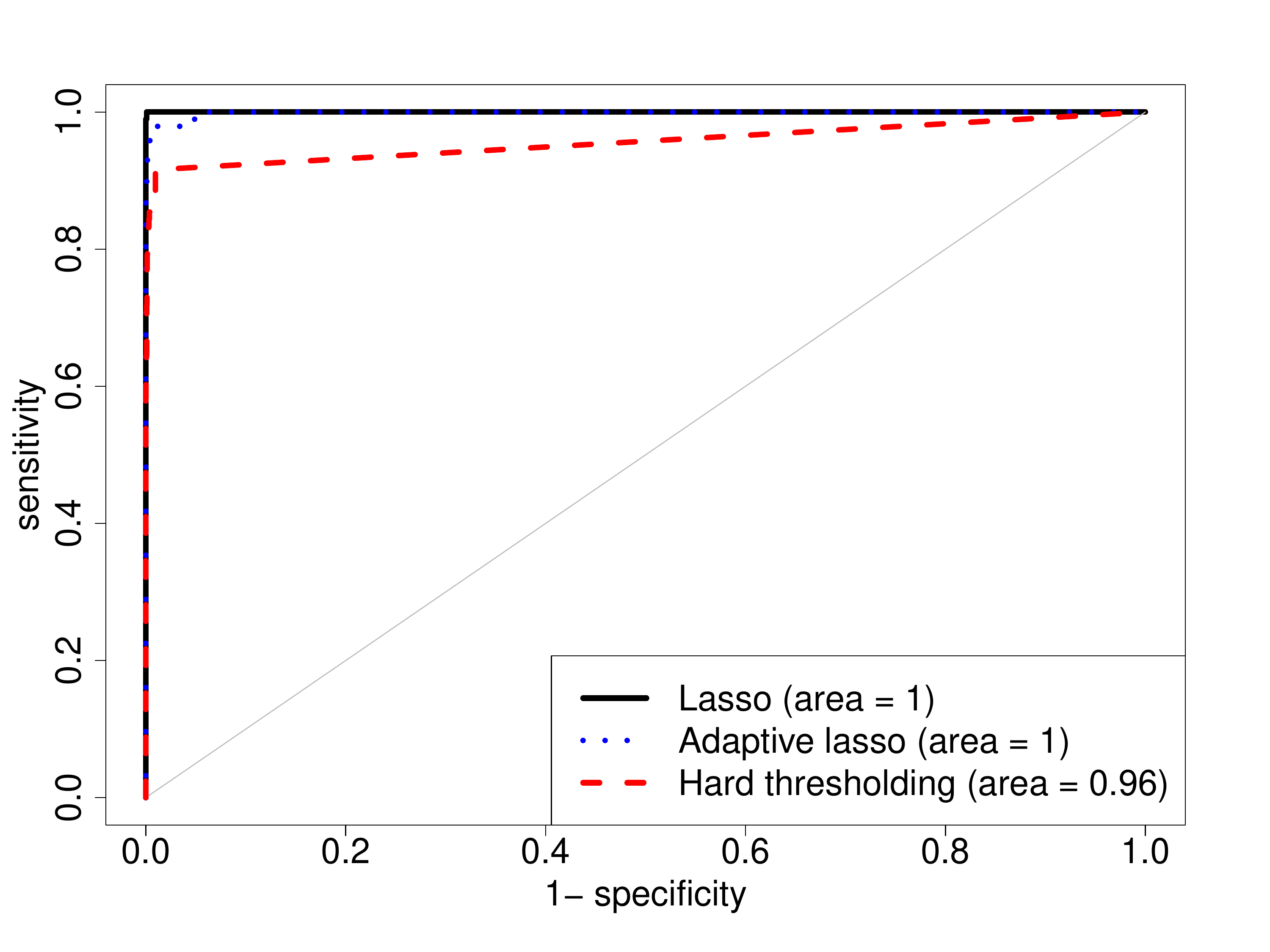}    
    \caption{Receiver Operating Characeristic (ROC) curves of hard thresholding, lasso and adaptive lasso for recovering coherence network of a $p = 96$ dimensional VAR(1) model using $n = 100$ (top left), $n = 200$ (top right), $n = 400$ (bottom left) and $n = 600$ (bottom right) time series observations.}
    \label{fig:roc}
\end{figure}

The ROC curves for the three thresholding methods in selecting coherence graph of a VAR(1) model with $p=48$ and $n \in \{100, 200, 400, 600\}$ are provided in Figure \ref{fig:roc}. Consistent with the frequency-specific precision and recall results, lasso and adaptive lasso thresholding methods perform better than hard thresholding. 

Overall, our numerical experiments confirm that thresholding procedures can be successfully used to estimate large spectral density matrices with same order of accuracy as shrinkage methods, and with an additional advantage of performing automatic edge selection in coherence networks.

\begin{table}
\begin{tabular}{l@{\hskip 0.4in}ccccc}
\\
& Smoothed   & Shrinkage & Hard Threshold  & Lasso & Adaptive Lasso\\
\\
VMA & & & & &\\
p = 12 & & & & &\\
\multicolumn{1}{r}{n = 100}&43.21(6.86)&22.15(1.77)&27.43(2.11)&22.89(2.05)&25.54(1.91)\\
\multicolumn{1}{r}{n = 200}&29.95(2.93)&17.67(1.01)&20.5(1.45)&16.18(1.33)&18.74(1.32)\\
\multicolumn{1}{r}{n = 400}&21.11(1.75)&14.24(0.67)&12.39(1.52)&10.84(1.01)&11.33(1.27)\\
\multicolumn{1}{r}{n = 600}&17.28(1.39)&12.58(0.59)&8.73(1.16)&8.84(0.71)&8.45(0.92)\\
p = 24 & & & & &\\
\multicolumn{1}{r}{n = 100}&80.49(7.63)&26.28(1.62)&29.86(1.21)&26.36(1.5)&28.38(1.31)\\
\multicolumn{1}{r}{n = 200}&59.79(4.7)&22.92(0.81)&25.62(0.86)&19.29(1.35)&22.09(1.21)\\
\multicolumn{1}{r}{n = 400}&41.83(1.98)&19.54(0.45)&17.16(1.26)&13.0(0.99)&13.71(1.25)\\
\multicolumn{1}{r}{n = 600}&35.86(1.6)&17.83(0.36)&12.27(1.01)&10.36(0.65)&9.89(0.84)\\
p = 48 & & & & &\\
\multicolumn{1}{r}{n = 100}&162.79(9.94)&29.58(1.24)&30.62(0.91)&28.78(0.83)&29.9(0.8)\\
\multicolumn{1}{r}{n = 200}&119.58(4.21)&27.0(0.57)&28.29(0.37)&22.68(0.76)&25.48(0.72)\\
\multicolumn{1}{r}{n = 400}&83.48(2.67)&24.09(0.37)&22.35(0.65)&15.86(0.54)&17.21(0.74)\\
\multicolumn{1}{r}{n = 600}&69.83(1.77)&22.58(0.28)&16.95(0.81)&12.88(0.48)&12.73(0.7)\\
p = 96 & & & & &\\
\multicolumn{1}{r}{n = 100}&324.57(14.7)&32.34(1.15)&30.3(0.46)&29.71(0.43)&30.11(0.44)\\
\multicolumn{1}{r}{n = 200}&235.78(7.75)&29.58(0.67)&28.83(0.28)&25.28(0.43)&27.31(0.38)\\
\multicolumn{1}{r}{n = 400}&167.89(4.28)&27.44(0.37)&25.67(0.33)&18.58(0.5)&20.34(0.55)\\
\multicolumn{1}{r}{n = 600}&139.4(2.02)&26.26(0.24)&21.25(0.48)&15.35(0.37)&15.72(0.51)\\
VAR & & & & &\\
p = 12 & & & & &\\
\multicolumn{1}{r}{n = 100}&39.11(10.1)&37.49(5.27)&41.09(6.36)&38.46(5.25)&41.81(5.18)\\
\multicolumn{1}{r}{n = 200}&28.06(8.4)&25.2(4.15)&30.52(5.83)&27.6(4.19)&30.69(5.21)\\
\multicolumn{1}{r}{n = 400}&17.31(4.63)&16.51(2.93)&19.37(3.74)&16.84(2.61)&19.5(3.41)\\
\multicolumn{1}{r}{n = 600}&25.0(5.86)&19.23(3.95)&23.07(4.62)&18.55(2.85)&21.65(3.92)\\
p = 24 & & & & &\\
\multicolumn{1}{r}{n = 100}&73.83(15.52)&49.25(4.16)&49.18(4.78)&44.64(3.8)&47.59(3.54)\\
\multicolumn{1}{r}{n = 200}&54.77(9.83)&36.84(2.97)&40.95(3.46)&34.29(3.52)&38.46(3.47)\\
\multicolumn{1}{r}{n = 400}&35.53(6.01)&27.34(2.05)&27.43(2.86)&22.32(1.76)&25.05(2.67)\\
\multicolumn{1}{r}{n = 600}&28.53(2.24)&21.82(0.74)&17.25(0.97)&15.17(0.11)&16.64(0.98)\\
p = 48 & & & & &\\
\multicolumn{1}{r}{n = 100}&131.88(20.49)&61.75(4.11)&49.3(3.35)&47.12(1.89)&48.1(2.25)\\
\multicolumn{1}{r}{n = 200}&99.46(12.68)&48.3(2.17)&44.24(1.53)&39.31(1.77)&42.63(1.41)\\
\multicolumn{1}{r}{n = 400}&69.19(7.07)&38.38(1.44)&35.52(1.55)&26.69(1.35)&30.5(1.75)\\
\multicolumn{1}{r}{n = 600}&53.08(1.38)&32.58(0.4)&25.23(0.41)&20.38(0.3)&21.16(0.52)\\
p = 96 & & & & &\\
\multicolumn{1}{r}{n = 100}&259.85(31.63)&75.46(5.47)&48.6(1.69)&47.96(1.41)&48.15(1.59)\\
\multicolumn{1}{r}{n = 200}&200.12(16.87)&59.45(1.88)&45.18(1.23)&43.34(1.2)&44.63(1.06)\\
\multicolumn{1}{r}{n = 400}&135.52(8.76)&50.08(1.25)&41.41(0.7)&32.53(1.11)&37.13(1.14)\\
\multicolumn{1}{r}{n = 600}&97.13(1.32)&42.62(0.08)&31.65(0.45)&24.6(0.34)&24.56(0.69)\\
\end{tabular}
\label{table:rmise-homogeneous-final}
\caption{Relative Mean Integrated Squared Error (RMISE, in \%) of smoothed periodogram, shrinkage towards a diagonal target and three different thresholding methods - hard thresholding, lasso and adaptive lasso. Results are averaged over $20$ replicates. Standard deviations (also in \%) are reported in parentheses.}
\end{table}

\section{Functional Connectivity Analysis with fMRI Data}\label{sec:realdata}

We demonstrate the advantage of thresholding based spectral density estimators for visualization and interpretation in functional connectivity analysis among different brain regions of a human subject using resting state fMRI data. This data is part of a study involving 51 subjects (29.6 $\pm$ 8.6 years of age, 35 males) that suffered from mild traumatic brain injury (TBI). Magnetic resonance imaging (MRI) data and neuropsychological data were collected at 1 week, 1 month, 6 months and 12 months post-injury. TBI is defined as Glasgow Coma Scale of 13-15 at injury, loss of consciousness less than 30 minutes and post-traumatic amnesia less than 24 hours. More  details are available in  \cite{Kuceyeski2018functional}. 


\begin{figure}
    \centering
    \includegraphics[width=0.48\textwidth]{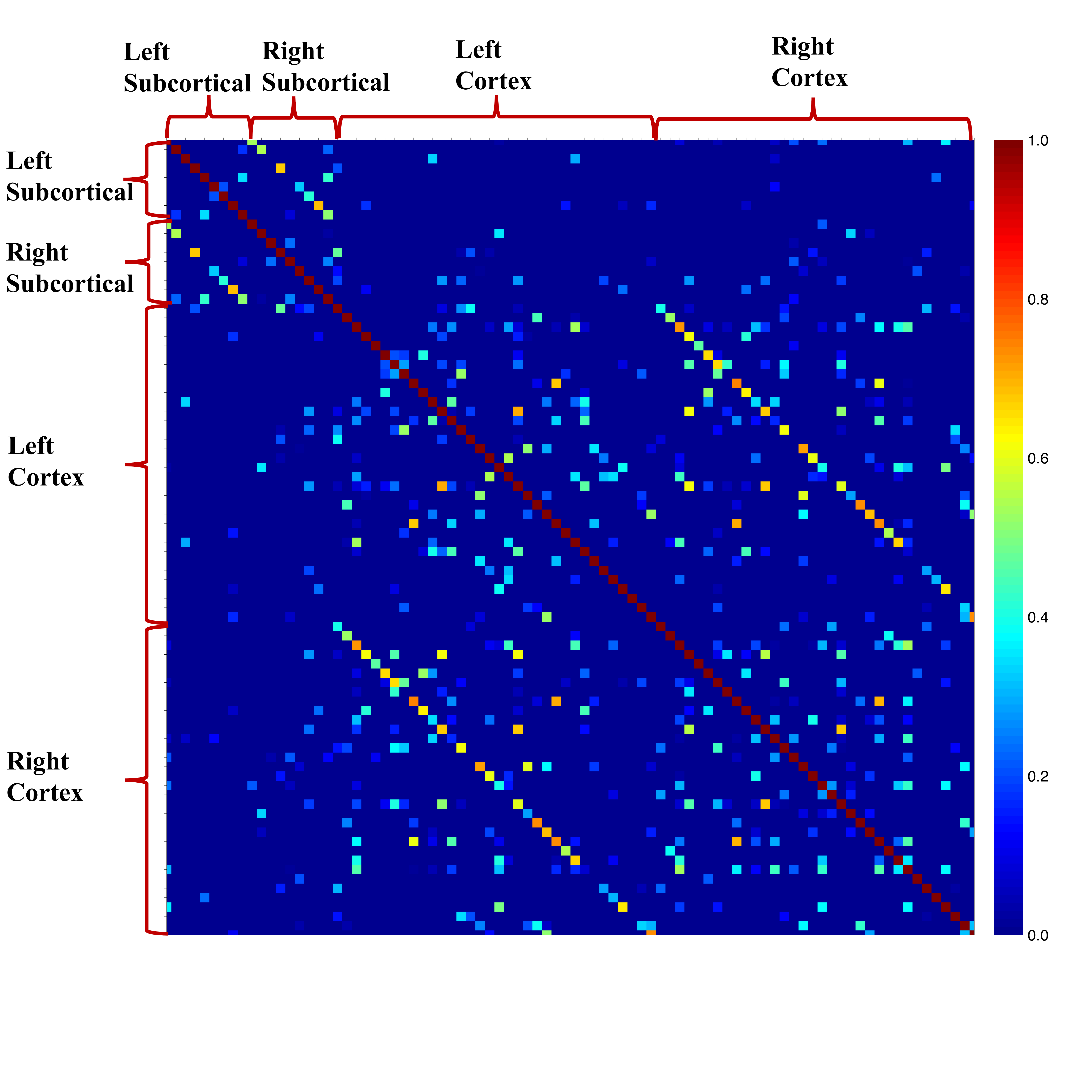}
    \includegraphics[width=0.48\textwidth]{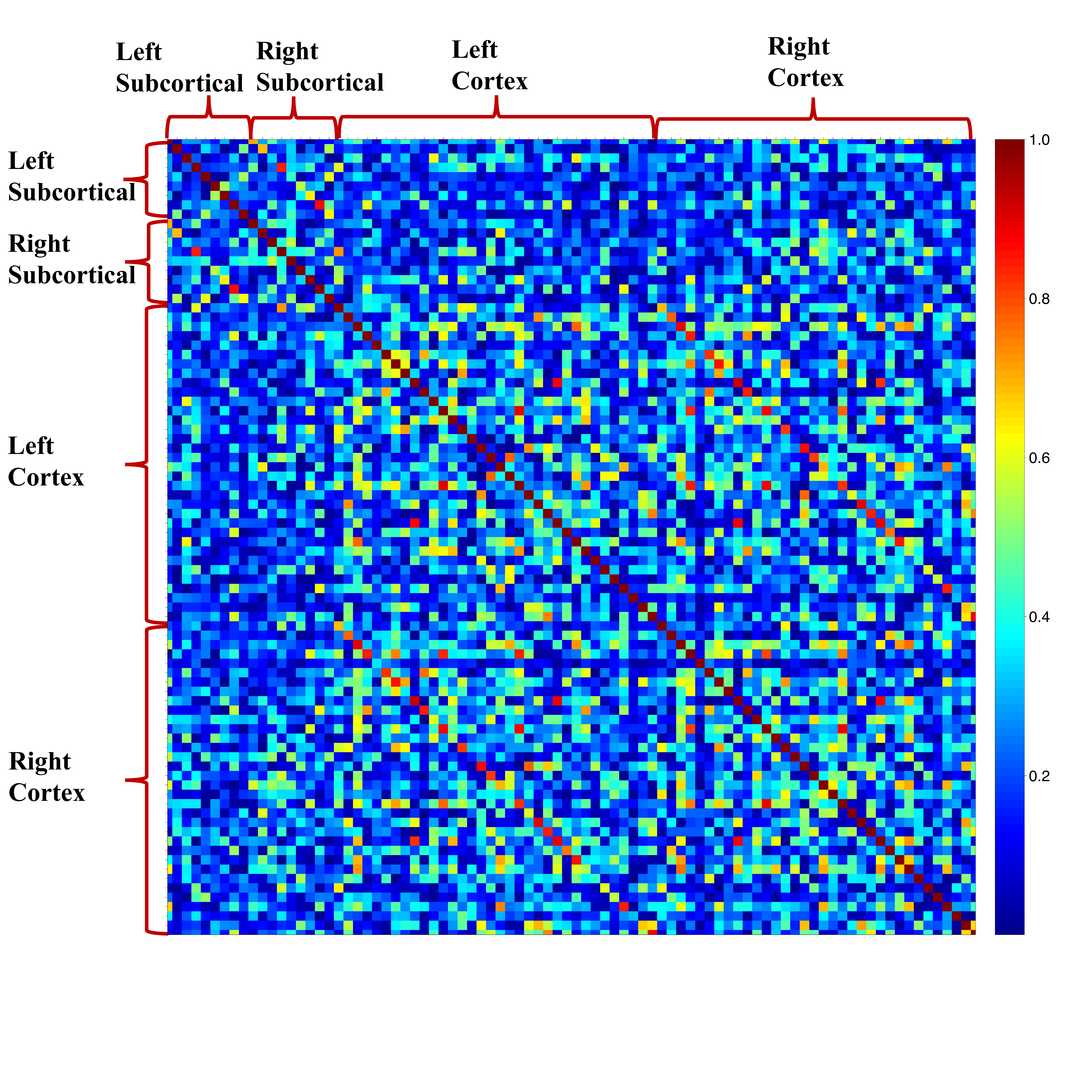}
    \includegraphics[width=0.95\textwidth]{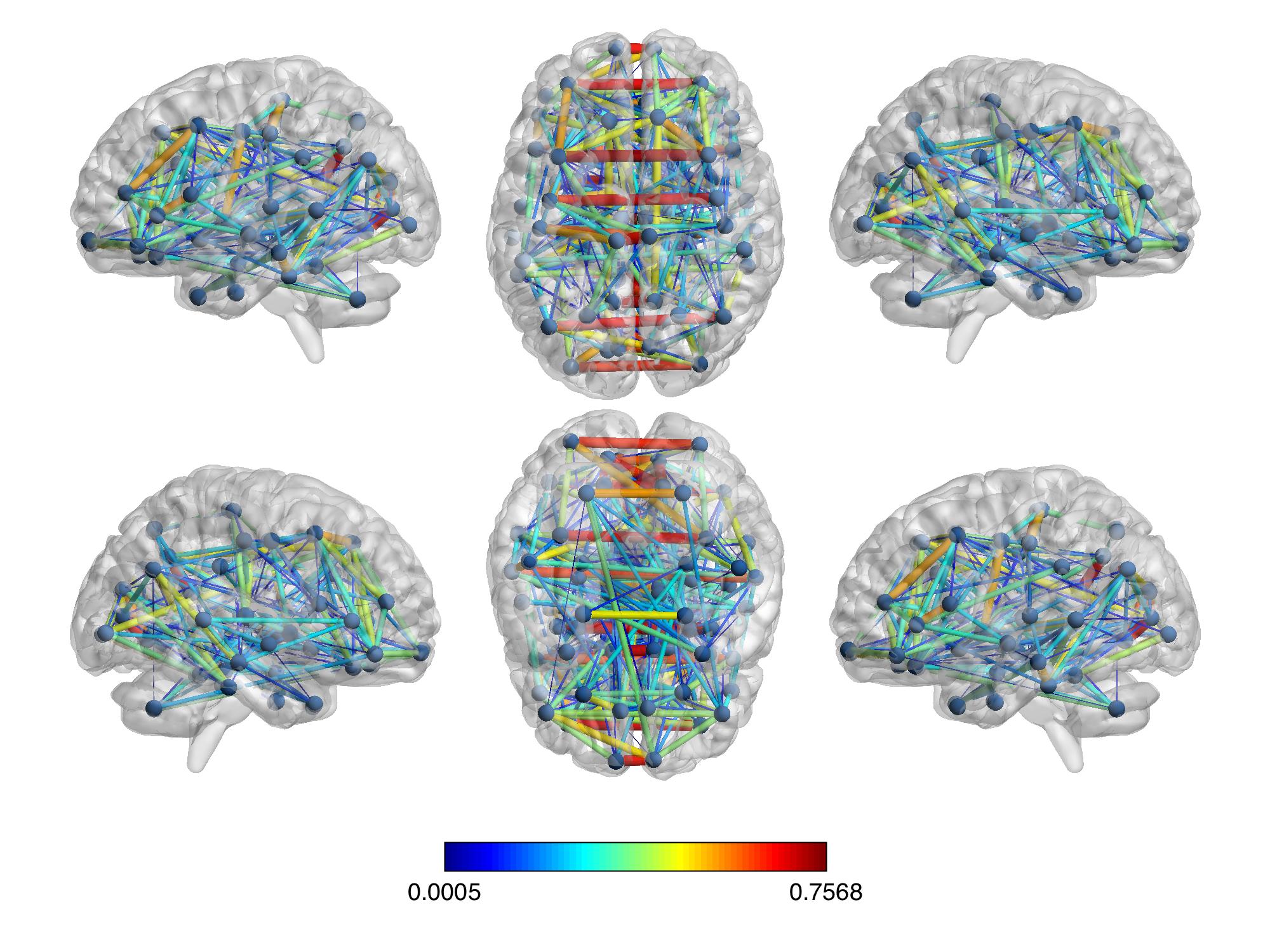}
    \caption{[top]: Heat maps of absolute coherence matrices (at frequency $0$) obtained from  spectral density estimated using [top left] adaptive lasso thresholding  and [top right] a shrinkage method. [bottom]: Absolute coherence network among brain regions obtained using adaptive lasso and visualized using BrainNet Viewer. The coherence network estimated by adaptive lasso retains known biological patterns, including presence of bilateral homologues, i.e. strong connectivity between same ROIs in the left and right parts of brain.}
    \label{fig:realdata}
\end{figure}

A 3T GE Signa EXCITE scanner was used to  acquire the MRIs, which included structural scans (FSPGR T1, $1\times 1 \times 1$ mm$^3$ voxels) and resting-state functional magnetic resonance imaging (fMRI) (7 min, $3.4 \times 3.4 \times 4.0$ mm$^3$ voxels, 2 sec sampling rate). The MRIs were processed by parcellating the gray matter into $p=86$ anatomical regions of interest (ROIs) using the semi-automated FreeSurfer software \citep{fischl2000measuring}. Cortical and subcortical parcellations and the fMRI time series data were then used in the construction of coherence based functional connectivity (FC) networks. The adjacency matrix of FC network captures the similarity of the neuronal activation over time between pairs of ROIs. 

We calculated coherence matrices at frequency $0$ using adaptive lasso thresholding (with $\eta = 2$) and shrinkage of averaged periodograms. The smoothing span was chosen by setting $m=\sqrt{n}$, and the tuning parameters in our sample-splitting algorithm were selected as in our simulation studies. 

\textit{Results}: In Figure \ref{fig:realdata}, we show an example of the FC coherence network for a particular TBI patient using our proposed adaptive lasso thresholding (top left) and the same patient's FC network estimated using the shrinkage method (top right) of \citep{bohm2009shrinkage} that does not perform automatic coherence selection. One of the many issues with using fMRI data is the spurious functional connections that arise from the method's abundant noise (due to instrumentation and physiology). It is often preferable in a clinical context to filter out this noise, but it is not currently done in a universally accepted and statistically principled way. As shown in the top panel of Figure \ref{fig:realdata}, the coherence matrix estimated by adaptive lasso thresholding obviously is more sparse in nature compared to the one from shrinkage method, while maintaining known physiological connections. For example, we see strong FC in the bilateral homologues (the same ROI in the left versus right hemisphere), which are known to have strong functional connections \citep{zuo2010growing}. This is even more readily apparent in the bottom panel of Figure \ref{fig:realdata} where we see strong connections between the same ROI in the left and right sides of the brain. Other than the bilateral homologues, the left and right precuneus, isthmus cingulate, lingual gyrus and pericalcarine have prominent connections to many regions (see Figures \ref{fig:realdatafullalasso} and \ref{fig:realdatafullshrinkage} in Appendix \ref{appendix:more_tables}). The precuneus, which plays a role in visual, sensorimotor, and attentional information processing, is central to resting-state (task negative) fMRI networks detected using correlation analysis \citep{Utevsky14}. Additionally, the isthmus cingulate, part of the posterior cingulate cortex, is known to be highly functionally connected to many regions across the brain at rest \citep{FRANSSON20081178}. In addition, we see a stronger FC between the left and right homologues in the subcortical ROIs (upper left corner) than between subcortical and cortical ROIs. It is interesting to note that while some of these connections are also strong in the shrinkage based coherence matrix estimate, it is not easy to separate them from other moderately strong coherences between brain regions.

\section{Discussion}\label{sec:discussion}
We proposed hard thresholding and generalized thresholding of averaged periodogram for estimation of high-dimensional spectral density matrices of stable Gaussian time series and linear processes with errors having potentially heavier tails than Gaussian. Under high-dimensional regime $\log p /n \rightarrow 0$, we established consistency of the above estimation procedures when the true spectral densities are weakly sparse. At the core of our technical results lie concentration inequalities of  complex quadratic forms of temporally dependent, high-dimensional random vectors, which were used to derive finite sample deviation of averaged periodograms around their expectation. These results  are of independent interest and are potentially useful in other problems involving high-dimensional spectral density. In our next steps, we plan to extend the theoretical analyses to more general adaptive thresholding methods \citep{cai2011adaptive}, which will explicitly account for heterogeneity in the strengths of cross-spectral association across different pairs of time series and different frequency bands. We also plan to develop estimation and inference procedures for high-dimensional partial coherence at different frequencies.

Another direction of potential interest is to develop thresholding strategies that incorporate  information on different brain regions and prior biological knowledge on brain networks. Dynamic functional connectivity of brain networks is known to play important roles behind progression of neurodegenerative diseases. A common approach to build such networks is using coherence measures of Fourier or wavelet transform of multi-channel fMRI/EEG/MEG signals and thresholding small entries of zero. Selection of threshold level that represents heterogeneous modular structure of human brain has been a topic of active research \citep{bordier2017graph}. We expect that more sophisticated thresholding methods, building up on universal and adaptive thresholds and incorporating prior neuroscientific knowledge, will be potentially useful in  data-driven discovery of scientifically and clinically relevant connectivity patterns in human brain.

\section*{Acknowledgements} The authors wish to thank Pratik Mukherjee for providing and Keith Jamison for pre-processing the TBI patient MRI data.  SB was supported by NSF award (DMS-1812128) and AK was supported by a Kellen Foundation Fellowship and the NIH (R21 NS104634-01 and R01 NS102646-01A1).
\newpage 

\appendix
\section{Appendix: Proofs for Gaussian Time Series}\label{appendix:proof_gaussian}
\subsection{Proof of Lemma \ref{lemma: hason_bound_time_gauss}}
\begin{proof}
We can write $vec(\mathcal{\mathcal{X}}^\top)\stackrel{d}{=}\Sigma^{1/2}Z$, where $\Sigma$ is the covariance matrix of the $np$-dimensional random vector $vec(\mathcal{\mathcal{X}}^\top)$ and $Z\sim N(0,I)$. Then using Hanson-Wright inequality [Theorem 1.1,  \cite{rudelson2013hanson}] and the fact that the subGaussian norm of $Z$ is $1$, we conclude that there exists a universal constant $c>0$ satisfying 
\begin{equation}
\label{eq: hanson1}
\begin{aligned}
&\mathbb{P}\left(\left|vec(\mathcal{\mathcal{X}}^\top)^\top A ~vec(\mathcal{\mathcal{X}}^\top) - \mathbb{E} \left[vec(\mathcal{\mathcal{X}}^\top)^\top A~ vec(\mathcal{\mathcal{X}}^\top)~\right]\right| >2\pi\eta\vertiii{f}\right) \\
&= \mathbb{P}\left(\left|Z^\top \Sigma^{1/2}A\Sigma^{1/2}Z - \mathbb{E}\left[ Z^\top \Sigma^{1/2}A\Sigma^{1/2}Z \right]\right|>2\pi \eta\vertiii{f}\right)\\
&\le 2\exp\left[-c\min\left\{\cfrac{2\pi \eta\vertiii{f}}{\|\Sigma^{1/2}A\Sigma^{1/2}\|},\cfrac{4\pi^2\eta^2\vertiii{f}^2}{\|\Sigma^{1/2}A\Sigma^{1/2}\|_F^2}\right\}\right].
\end{aligned}
\end{equation}

Using Lemma \ref{lemma:max-L2-norm}, $\|\Sigma^{1/2}A\Sigma^{1/2}\|\le \|\Sigma\|\|A\|\le 2\pi \vertiii{f} \|A\|$. It follows from \citet{golub2012matrix},
\begin{equation}
\begin{aligned}
&\|\Sigma^{1/2}A\Sigma^{1/2}\|_F \le \sqrt{\rank(\Sigma^{1/2}A\Sigma^{1/2})} \|\Sigma^{1/2}A\Sigma^{1/2}\| \nonumber \\
&\le \sqrt{\rank(A)} \|\Sigma^{1/2}A\Sigma^{1/2}\| \le 2\pi \sqrt{\rank(A)} \|A\| \vertiii{f}. \nonumber
\end{aligned}
\end{equation}
Then plugging in the bound for $\|\Sigma^{1/2}A\Sigma^{1/2}\|$ and 
$\|\Sigma^{1/2}A\Sigma^{1/2}\|_F$ into \eqref{eq: hanson1} completes the proof.
\end{proof}

\subsection{Proof of Proposition \ref{prop:bias_bound}}
\begin{proof}
It suffices to show that for any two unit vectors $e_r,e_s$, 
\begin{equation}
\begin{aligned}
\left|e_r^\top \left[\mathbb{E}\hat{f}(\omega_j) - f(\omega_j)\right]e_s\right| \le \frac{m}{n}\Omega_n(f) + \frac{1}{2\pi}\left(\frac{\Omega_n(f)}{n}+L_n(f)\right). \nonumber
\end{aligned}
\end{equation}
Since  
\begin{equation}
\hat{f}(\omega_j) = \frac{1}{2\pi(2m+1)} \sum_{\ell =-m}^m I(\omega_{j+\ell}),  \nonumber
\end{equation}
we have 
\begin{equation}
\label{eq:mul_dev}
\begin{aligned}
\left|e_r^\top \left[\mathbb{E}\hat{f}(\omega_j) - f(\omega_j)\right]e_s\right| 
&\le \frac{1}{2\pi(2m+1)} \sum_{\ell = -m}^m |e_r^\top\left[\mathbb{E}I(\omega_{j+\ell}) - \mathbb{E} I(\omega_{j})\right]e_s|\\
&+\left|e_r^\top \left[\frac{1}{2\pi}\mathbb{E}I(\omega_j) - f(\omega_j)\right]e_s\right|. 
\end{aligned}
\end{equation}
By definition of $I(\omega_j)$ in \eqref{eq:single_periodogram}, we have $\mathbb{E} I(\omega_j) = \sum_{|k| \le n} \Gamma(k) \frac{(n-|k|)}{n} e^{-ik \omega_j}$.  Therefore, the second term above takes the form 
\begin{equation}
\label{eq:mul_dev1}
\begin{aligned}
\left|\frac{1}{2\pi} e_r^\top \left[\mathbb{E} I(\omega_j) - 2\pi f(\omega_j)\right]e_s\right| &= \frac{1}{2\pi}\left|\sum_{|k|\le n} \frac{|k|}{n}  \Gamma_{rs}(k) e^{-ik\omega_j}+\sum_{|k|>n} \Gamma_{rs}(k) e^{-ik\omega_j}\right|\\
&\le \frac{1}{2\pi} \left [\sum_{|k|\le n} \frac{|k|}{n}  |\Gamma_{rs}(k)|+ \sum_{|k|>n} |\Gamma_{rs}(k)|\right]\\
&= \frac{1}{2\pi}\left(\frac{\Omega_n(f)}{n} + L_n(f)\right). 
\end{aligned}
\end{equation}
For the first term, note that $|e^{ix} - e^{iy}|\le |x-y|$ and $|\omega_j-\omega_{j+\ell}| = 2\pi \frac{|\ell|}{n}$. This implies 
\begin{equation}
\label{eq:mul_dev2}
\begin{aligned}
\left|\frac{1}{2\pi} e_r^\top \left[\mathbb{E} I(\omega_j) - \mathbb{E} I(\omega_{j+\ell}) \right]e_s\right| &= \frac{1}{2\pi}\left|\sum_{|k|\le n} \left(1-\frac{|k|}{n}\right) |\Gamma_{rs}(k)| (e^{-ik\omega_j} - e^{-i k\omega_{j+\ell}})\right|\\
&\le \frac{1}{2\pi} \sum_{|k|\le n} |\Gamma_{rs}(k)| |k||\omega_j - \omega_{j+\ell}| =  |\ell| \Omega_n(f)/n. 
\end{aligned}
\end{equation}
Plugging in \eqref{eq:mul_dev1} and \eqref{eq:mul_dev2}  into \eqref{eq:mul_dev},  
\begin{equation*}
\begin{aligned}
\left|e_r^\top \left[\mathbb{E}\hat{f}(\omega_j) - f(\omega_j)\right]e_s\right| &\le \frac{1}{2\pi}\left(\frac{\Omega_n(f)}{n} + L_n(f)\right)+ \left(\frac{\sum_{|\ell|\le m} |\ell|}{2m+1}\right)\frac{\Omega_n(f)}{n}\\
&\le \frac{m}{n}\Omega_n(f) + \frac{1}{2\pi}\left(\frac{\Omega_n(f)}{n}+L_n(f)\right).\nonumber
\end{aligned}
\end{equation*}
\end{proof}

\subsection{Proof for Proposition \ref{prop:order_bias}}
\begin{proof}
We will prove the proposition one by one for its three conditions. Proof for all three conditions uses the simple fact that, for $0<x<1$
\begin{equation}
\label{eq:sum-help}
\sum_{\ell=1}^n \ell x^\ell = \frac{x(1+nx^{n+1}-(n+1)x^n)}{(1-x)^2}.
\end{equation}

\paragraph{Condition 1:} Directly plug the bound on $\|\Gamma(\ell)\|_{\text{max}}$ and with $|\Gamma_{r,s}(\ell)|\le \|\Gamma(\ell)\|_{max}$, we have 

\begin{equation}
\Omega_n \le 2 \sum_{\ell=1}^n \ell \|\Gamma(\ell)\|_{\text{max}}\le  2 \sigma_X \sum_{\ell=1}^n \ell \rho_X^{\ell} = \frac{2\sigma_X\rho_X (1+n\rho_X^{n+1}-(n+1)\rho_X^n)}{(1-\rho_X)^2}. \nonumber
\end{equation}

For $L_n$, 
\begin{equation}
L_n \le 2\sum_{\ell> n} \|\Gamma_{\ell}\|_{\text{max}} \le 2 \sigma_X \sum_{\ell>n} \rho_X^\ell = \frac{2 \sigma_X \rho_X^{n+1}}{1-\rho_X}. \nonumber
\end{equation}

\paragraph{Condition 2:} Note that condition of geometrically decaying $\rho$-mixing coefficient leads to condition 1 as 
\begin{equation}
\begin{aligned}
|\Gamma_{rs}(\ell)| &= \left|\frac{\mathbb{E} e_r^\prime X_{\ell}X_{0}^\top e_s}{\sqrt{|\Gamma_{rr}(0)||\Gamma_{ss}(0)|}}\right|\sqrt{|\Gamma_{rr}(0)||\Gamma_{ss}(0)|}\\
& \le \|\Gamma(0)\|_{\text{max}}\sigma_X \rho_X^{|\ell|} . \nonumber
\end{aligned}
\end{equation}
Then follow the argument for condition 1, we finish the proof.

\paragraph{Condition 3:} 
Let $\tilde{\Omega}_n$, $\tilde{L}_n$ be the $\Omega_n, L_n$ defined before for time series $\tilde{X}_t$ and $\tilde{\Gamma}(\ell)$ be the auto-covariance for $\tilde{X}$. We first show that $\tilde{\Omega}_n$, $\tilde{L}_n$
are upper bounds for $\Omega_n$, $L_n$. Then we present upper bounds for $\tilde{\Omega}_n$ and $\tilde{L}_n$ although it may lose the tightness in controlling of growth rates of these two. To see this, we partition $\tilde{\Gamma}(\ell)$ into blocks as follows.
\begin{equation}
\tilde{\Gamma}(\ell) = 
\begin{bmatrix} 
\Gamma(\ell) & \Gamma(\ell+1) & \cdots & \Gamma(\ell+d-1)\\ 
\vdots & \vdots & \ddots & \vdots \\
\Gamma(\ell-d+1) & \Gamma(\ell-d) & \cdots & \Gamma(\ell)
\end{bmatrix}.\nonumber
\end{equation}
Since $\Gamma(\ell)$ appears as diagonal block of $\tilde{\Gamma}(\ell)$, based on the definition of $\Omega_n$ and $L_n$, we can claim that $\tilde{\Omega}_n$, $\tilde{L}_n$ are upper bounds for $\Omega_n$ and $L_n$ respectively. Next, we focus on gettting upper bound for $\tilde{\Omega}_n$ and $\tilde{L}_n$. 
Consider the infinite moving average representation of $\tilde{X}_t$ 
\begin{equation}
\tilde{X}_t = \sum_{\ell=0}^\infty \tilde{B}_\ell \tilde{\varepsilon}_{t-\ell}, \nonumber
\end{equation}
where $\tilde{B}_\ell = (\tilde{A}_1)^\ell$ and autocovariance becomes
\begin{equation}
\tilde{\Gamma}(\ell) = \sum_{t=0}^\infty \tilde{B}_{t+\ell} \begin{bmatrix}
I_{p} & \mathbf{0} &\ldots & \mathbf{0} \\
\mathbf{0} &  \mathbf{0} & \ldots & \mathbf{0} \\
\vdots  & \ddots  &  \mathbf{0} & \mathbf{0} \\
\mathbf{0} &\ldots   &  \mathbf{0} & \mathbf{0}
\end{bmatrix}
\tilde{B}_{t}^\top.  \nonumber
\end{equation}


Since $\|\tilde{A}^\ell\| = \|SD^\ell S^{-1}\| =  \kappa\lambda_{\textup{max}}^\ell(\tilde{A}_1)$,

\begin{equation}
\begin{aligned}
\|\tilde{\Gamma}(\ell)\| &\le  \sum_{t=0}^\infty  \|\tilde{B}_{t+\ell}\|\|\tilde{B}_{t}\|   \\
&\le \kappa^2 \lambda^\ell_{\textup{max}}(\tilde{A}_1)\sum_{k=0}^{\infty} \lambda^2_{\textup{max}}(\tilde{A}_1)= \kappa^2 \frac{\lambda_{\textup{max}}^\ell(\tilde{A}_1)}{1-\lambda^2_{\text{max}}(\tilde{A}_1)}.\nonumber
\end{aligned}
\end{equation}

Then noticing $\|\tilde{\Gamma}(\ell)\|_{\text{max}}\le \|\tilde{\Gamma}(\ell)\|$, using \eqref{eq:sum-help}
\begin{equation}
\begin{aligned}
\tilde{\Omega}_n &\le 2\sum_{\ell=1}^n |\ell|\|\tilde{\Gamma}(\ell)\| \\
&\le 2\kappa^2 \sum_{\ell=1}^n  \frac{\ell\lambda_{\textup{max}}^\ell(\tilde{A}_1)}{(1-\lambda_{\textup{max}}(\tilde{A}_1))(1-\lambda^2_{\textup{max}}(\tilde{A}_1))} = 2\kappa^2\frac{\lambda_{\textup{max}}(\tilde{A}_1)(1+n\lambda_{\textup{max}}^{n+1}(\tilde{A}_1)-(n+1)\lambda_{\textup{max}}(\tilde{A}_1))}{(1-\lambda_{\textup{max}}(\tilde{A}_1))^2(1-\lambda^2_{\textup{max}}(\tilde{A}_1))}. \nonumber
\end{aligned}
\end{equation}

For $\tilde{L}_n$, 
\begin{equation}
\tilde{L}_n \le 2\sum_{\ell> n} \|\tilde{\Gamma}(\ell)\| = 2\kappa^2\sum_{\ell > n}\frac{\lambda_{\textup{max}}^\ell(\tilde{A}_1)}{1-\lambda^2_{\textup{max}}(\tilde{A}_1)} = 2\kappa^2\frac{\lambda_{\textup{max}}^{n+1}(\tilde{A})}{(1-\lambda_{\textup{max}}(\tilde{A}_1))(1-\lambda^2_{\textup{max}}(\tilde{A}_1))}.\nonumber
\end{equation}
\end{proof}

\subsection{Proof of Proposition \ref{prop:variance_bound}}
\begin{proof}
We focus on bounding the tail probability of the variance term
\begin{equation}
\mathbb{P}\left(\left|\hat{f}_{rs}(\omega_j) - \mathbb{E}\hat{f}_{rs}(\omega_j)\right| \ge \vertiii{f}\eta\right). \nonumber
\end{equation}
First, note that $\mathbb{P}\left(\left|\hat{f}_{rs}(\omega) - \mathbb{E}\hat{f}_{rs}(\omega)\right|\ge  \vertiii{f}\eta \right)$ is at most 
\begin{equation}
\label{eq:bound_with_real_img}
\begin{aligned}
\mathbb{P}\left(\left|\mathbf{Re}\left(\hat{f}_{rs}(\omega) - \mathbb{E}\hat{f}_{rs}(\omega)\right)\right|\ge \frac{\vertiii{f}\eta}{2} \right)  + \mathbb{P}\left(\left|\mathbf{Im}\left(\hat{f}_{rs}(\omega) - \mathbb{E}\hat{f}_{rs}(\omega)\right)\right|\ge \frac{\vertiii{f}\eta}{2} \right),
\nonumber 
\end{aligned}
\end{equation}
so it is sufficient to derive upper bounds for the real and imaginary parts separately.\par 
The main idea of our proof is to express the real and imaginary parts of $\hat{f}(\omega_j)$ as quadratic forms in $vec(\mathcal{X}^\top)$, and apply Lemma \ref{lemma: hason_bound_time_gauss} on each part. First, we express the periodogram $I(\omega_j)$ in terms of  trigonometric series. 
As pointed out before, $I(\omega_j)$ defined in \eqref{eq:single_periodogram} can be written as 
\begin{equation}
\label{eq:realImaginaryParts}
\begin{aligned}
I(\omega_j) = & \left(\mathcal{X}^\top C_j-  \iu \mathcal{X}^\top S_j\right) \left(\mathcal{X}^\top C_j-\iu \mathcal{X}^\top S_j\right)^\dag \\
= & \mathcal{X}^\top (C_jC_j^\top + S_jS_j^\top)\mathcal{X} + \iu \mathcal{X}^\top(C_jS_j^\top-S_jC_j^\top)\mathcal{X}. 
\end{aligned}
\end{equation}
Note that in the univariate case ($p=1$) the imaginary part becomes zero and we only need to bound the real part. However, for multivariate case, we need to understand the concentration behaviour of both parts.
\smallskip
\par 
\noindent \textbf{Concentration Inequality for Real Part: } We claim that  there exists a universal constant $c>0$ s.t. for any two unit vectors $u$ and $v$, 
\begin{equation*}
\begin{aligned}
&\mathbb{P}\left(\left|u^T\mathrm{\bf{Re}}\left(\hat{f}(\omega_j) - \mathbb{E}\hat{f}(\omega_j)\right)v\right|\ge \vertiii{f} \eta/2 \right)\\
\leq & 6\exp \left( - c\min \left\{(2m+1)\eta^2,(2m+1)\eta \right\} \right).
\end{aligned}
\end{equation*}
We notice for any symmetric matrix $A$ and unit vectors $u$ and $v$:
\begin{equation}
\label{eq: sym_matrix_ine}
2|u^\top Av| \le |u^\top Au| + |v^\top Av| + |(u+v)^\top A(u+v)|.
\end{equation}
Now, $\mathrm{\bf{Re}}\left(\hat{f}(\omega_j)\right) = \mathcal{X}^\top \sum_{|\ell|\le m}(C_{j+\ell}C_{j+\ell}^\top + S_{j+\ell}S_{j+\ell}^\top)\mathcal{X}$ is a symmetric matrix, and 
so is $\mathbb{E}\left[ {\bf Re}\left(\hat{f}(\omega_j)\right)\right]$. Thus, $\mathrm{\bf{Re}}\left(\hat{f}(\omega_j) - \mathbb{E}\hat{f}(\omega_j)\right)$ is a symmetric matrix. Then applying \eqref{eq:realImaginaryParts}, we get 
\begin{equation}
\label{eq:mul_three_parts}
\begin{aligned}
&\mathbb{P}\left(\left|u^\top \mathrm{\bf{Re}}\left(\hat{f}(\omega_j) - \mathbb{E}\hat{f}(\omega_j)\right)v\right|\ge 1/2\vertiii{f} \eta \right) \\
&\leq  \mathbb{P}\left(\left|u^\top \mathrm{\bf{Re}}\left(\hat{f}(\omega_j) - \mathbb{E}\hat{f}(\omega_j)\right)u\right|\ge 1/4\vertiii{f} \eta \right)\\
&+\mathbb{P}\left(\left|v^\top \mathrm{\bf{Re}}\left(\hat{f}(\omega_j) - \mathbb{E}\hat{f}(\omega_j)\right)v\right|\ge 1/4\vertiii{f} \eta \right)\\
&+ \mathbb{P}\left(\left|(u+v)^\top\mathrm{\bf{Re}}\left(\hat{f}(\omega_j) - \mathbb{E}\hat{f}(\omega_j)\right)(u+v)\right|\ge 1/2\vertiii{f} \eta \right).
\end{aligned}
\end{equation}
Next, we note that 
\begin{equation}
\label{eq:qudratic representation}
\mathbf{\bf{Re}}(\hat{f}(\omega_j)) = \frac{1}{2\pi(2m+1)} \|Q_j\mathcal{X}\|^2,\nonumber
\end{equation}
where 
\begin{equation}
Q_j := \left[ 
\begin{array}{llll}
C_{j-m}^\top\\
S_{j-m}^\top \\
\vdots\\
C_j^\top\\
S_j^\top\\
\vdots\\
C_{j+m}^\top\\
S_{j+m}^\top\\ 
\end{array}
\right]_{(4m+2)\times n}.\nonumber
\end{equation}
Then for any unit vector $v$, 
\begin{equation}
\left|v^\top \mathrm{\bf{Re}}\left(\hat{f}(\omega_j) - \mathbb{E}\hat{f}(\omega_j)\right)v \right| = \frac{1}{2\pi(2m+1)}\left|v^\top \mathcal{X}^\top Q_j^\top Q_j \mathcal{X}v - \mathbb{E} v^\top \mathcal{X}^\top Q_j^\top Q_j \mathcal{X}v\right|.\nonumber
\end{equation}
Let $Y_t = v^\top X_t$, and let $\mathcal{Y} = [Y_1: \ldots: Y_n]^\top$ be a data matrix with $n$ consecutive observations.  Using Lemma  \ref{lemma:max-L2-norm-Y}, $\vertiii{f_Y} \le \|v\|^2 \vertiii{f} = \vertiii{f}$. Now note that  $\rank(Q^{\top}_jQ_j) \le 4m+2$, and $\|Q_j\|\le \|Q_{F_n}\|=1$, where $Q_{F_n}$ expands the rows of $Q_j$ to include all the Fourier frequencies (see Lemma \ref{lemma:maximum_L2_Q} for definition). Since all the rows of $Q_j$ are partially selected from those in $Q_{F_n}$ and Lemma \ref{lemma:maximum_L2_Q} states that $\|Q_{F_n}\|=1$, using this bound and applying Lemma \ref{lemma: hason_bound_time_gauss}, we get 
\begin{equation}
\label{eq:mul_real_vv}
\begin{aligned}
&\mathbb{P}\left(\left|v^\top \mathrm{\bf{Re}}\left(\hat{f}(\omega_j) - \mathbb{E}\hat{f}(\omega_j)\right)v\right| \ge 1/4\vertiii{f}\eta\right) \\
&\le P\left(\frac{1}{2\pi}\left|\mathcal{Y}^\top Q^{\top}_jQ_j\mathcal{Y} - \mathbb{E}\mathcal{Y}^\top Q^{\top}_jQ_j\mathcal{Y}\right|\ge 1/4  \vertiii{f_Y}(2m+1)\eta\right)\\
&\le 2\exp\left[-c_1\min\left\{\cfrac{(2m+1)\eta}{\|Q_j\|^2}, \cfrac{(2m+1)^2\eta^2}{\rank(Q_j)\|Q_j\|^4}\right\}\right] \\
&\le 2\exp\left[-c_1\min\left\{(2m+1)\eta, \cfrac{(2m+1)^2\eta^2}{(4m+2)}\right\}\right] \\
& \le 2\exp\left[-c\min\left((2m+1)\eta^2, (2m+1)\eta\right)\right],
\end{aligned}
\end{equation}
where $c_1, c$ are universal constants not depending on $n,p$ { or any other model parameters}. 
We can write 
\begin{equation}
\begin{aligned}
&\mathbb{P}\left(|(u+v)^\top \mathrm{\bf{Re}}\left(\hat{f}(\omega_j) - \mathbb{E}\hat{f}(\omega_j)\right)(u+v)|\ge 1/2 \vertiii{f} \eta \right) \\
=& \mathbb{P}\left(\left|\frac{(u+v)}{\sqrt{2}}^\top \mathrm{\bf{Re}}\left(\hat{f}(\omega_j) - \mathbb{E} \hat{f}(\omega_j)\right)\frac{(u+v)}{\sqrt{2}}\right|\ge 1/4 \vertiii{f} \eta \right),\nonumber
\end{aligned}
\end{equation}
with $\frac{u+v}{\sqrt{2}}$ as a unit vector. Thus three terms appearing in right hand side of inequality \eqref{eq:mul_three_parts} can all be bounded by \eqref{eq:mul_real_vv}, which completes our proof.  Note that when $u, v$ are canonical vectors $e_r$, $e_s$ respectively, then $(u+v)$ has at most two non-zero entries. Further, since $f$ is non-negative definite, the quantity $\vertiii{f_Y}$ can be upper bounded by a smaller quantity $\max_{1 \le r \le p} \vertiii{f_{r}}$, where $f_r$ denotes the spectral density of the $r^{th}$ component of $X_t$.
\smallskip

\noindent \textbf{Concentration Inequality for Imaginary Part: } We claim that there exists a universal positive constant $c$ such that for any two  unit vectors $u$ and $v$ and any $\eta > 0$,
\begin{equation}
\begin{aligned}
& \mathbb{P}\left(\left|u^\top \mathrm{\bf{Im}}\left(\hat{f}(\omega_j) - \mathbb{E}\hat{f}(\omega_j)\right)v\right|\ge 1/2\vertiii{f}\eta \right) \\
& \le 4\exp \left[ - c\min \left\{(2m+1)\eta^2,(2m+1)\eta \right\} \right].\nonumber
\end{aligned}
\end{equation}
To prove this claim, note that \eqref{eq:realImaginaryParts} implies
\begin{equation}
\mathrm{\bf{Im}}\left(\hat{f}(\omega_j)\right) = \mathcal{X}^\top\sum_{|\ell|\le m}(C_{j+\ell}S_{j+\ell}^\top - S_{j+\ell}C_{j+\ell}^\top)\mathcal{X} .\nonumber
\end{equation}
Therefore, for any $\eta > 0$, we have 
\begin{equation}
\label{eq:mul_im_two}
\begin{aligned}
&~ \mathbb{P}\left(\left|u^\top \mathrm{\bf{Im}}\left(\hat{f}(\omega_j) - \mathbb{E}\hat{f}(\omega_j)\right)v\right|\ge 2\vertiii{f} \eta \right)  \\
& \le \mathbb{P}\left(\frac{1}{2\pi(2m+1)}\left|u^\top \mathcal{X}^\top\sum_{|\ell|\le m}(S_{j+\ell}C_{j+\ell}^\top)\mathcal{X}  v - \mathbb{E}\left[u^\top \mathcal{X}^\top\sum_{|\ell|\le m}(S_{j+\ell}C_{j+\ell}^\top)\mathcal{X}  v\right]\right|\ge \vertiii{f} \eta \right) \\
&+\mathbb{P}\left(\frac{1}{2\pi(2m+1)}\left|u^\top \mathcal{X}^\top\sum_{|\ell|\le m}(C_{j+\ell}S_{j+\ell}^\top)\mathcal{X}  v - \mathbb{E}\left[ u^\top \mathcal{X}^\top\sum_{|\ell|\le m}(C_{j+\ell}S_{j+\ell}^\top)\mathcal{X}  v\right]\right|\ge \vertiii{f} \eta \right).
\end{aligned}
\end{equation}
It takes the same technique to get upper bound for two parts in the right hand side of inequality  \eqref{eq:mul_im_two}. So we will only show the proof for getting upper bound for the first part. \par 
Let $Y_t = [v^\top ; u^\top ]X_t$ be a 2-dimensional time series. It follows from Lemma \ref{lemma:max-L2-norm-Y} that $\vertiii{f_Y}\le \|[v^\top; u^\top]\|^2 \vertiii{f} = 2\vertiii{f}$. \par 
Define
\begin{equation}
P_j = \left[ 
\begin{array}{ll}
M_j & 0 \\
0 & N_j 
\end{array}
\right]_{(4m+2)\times{2n}}, 
\end{equation}
where 
\begin{align}
M_j=\begin{bmatrix}
S_{j-m}^\top\\
\vdots \\
S_j^\top\\
\vdots \\
S_{j+m}^\top
\end{bmatrix}_{(2m+1)\times n}  \qquad
N_j =\begin{bmatrix}
C_{j-m}^\top\\
\vdots \\
C_j^\top\\
\vdots \\
C_{j+m}^\top
\end{bmatrix}_{(2m+1)\times n}. \nonumber 
\end{align}
We can express the first part in \eqref{eq:mul_im_two} as 

\begin{equation}
\label{eq:mul_img_final}
\begin{aligned}
&~\mathbb{P}\left(\frac{1}{2\pi(2m+1)}\left|u^\top \mathcal{\mathcal{X}}^\top\sum_{|\ell|\le m}(S_{j+\ell}C_{j+\ell}^\top)\mathcal{\mathcal{X}}  v - \mathbb{E}\left[u^\top \mathcal{X}^\top\sum_{|\ell|\le m}(S_{j+\ell}C_{j+\ell}^\top)\mathcal{X}  v\right]\right|\ge 1/2\vertiii{f} \eta \right)\\
& = \mathbb{P}\left(\frac{1}{2\pi(2m+1)}\left|vec(\mathcal{Y}^\top)^\top P_j^\top MP_j vec(\mathcal{Y}^\top)\ - \mathbb{E} \left[vec(\mathcal{Y}^\top)^\top P_j^\top MP_j vec(\mathcal{Y}^\top)\right]\right|\ge 1/2\vertiii{f}\eta\right),
\end{aligned}
\end{equation}
where 
\begin{equation}
M = \left[
\begin{array}{ll}
0_{2m+1,2m+1} & I_{2m+1,2m+1}\\
0_{2m+1,2m+1} & 0_{2m+1,2m+1}
\end{array}
\right]. \nonumber 
\end{equation}
Since $M_j$ and $N_j$ are both composed with rows from $Q_{F_n}$, $\|M_j\|\le \|Q_{F_n}\|=1$ and $\|N_j\|\le \|Q_{F_n}\|=1$. Furthermore, as block-wise diagonal matrix, $\|P_j\|=\max\{\|M_j\|, \|N_j\|\}=1$. 
Now $\|P_j^\top M P_j\|\le \|P_j\|^2\|M\| \le 1$ and $\rank(P_j^\top M P_j) \le \rank(M) = 2m+1$. Since $\vertiii{f_Y}\le 2\vertiii{f}$, we can apply lemma \ref{lemma: hason_bound_time_gauss} to show that the probability in  \eqref{eq:mul_img_final} is at most 
\begin{equation}
\begin{aligned}
&2\exp\left[-c\min\left\{\cfrac{(2m+1)\eta}{\|P_jMP_j^\top\|}, \cfrac{(2m+1)^2\eta^2}{\rank(P_j)\|P_jMP_j^\top\|}\right\}\right] \\
&\le 2\exp\left[-c\min\left\{(2m+1)\eta, \cfrac{(2m+1)^2\eta^2}{(4m+2)}\right\}\right] \\
& \le 2\exp\left[-c\min\left\{(2m+1)\eta^2, (2m+1)\eta\right\}\right], \nonumber 
\end{aligned}
\end{equation}
where $c$ is an universal constant. 

Combining bounds for real and imaginary parts, and plugging these two bounds in  \eqref{eq:bound_with_real_img}, we can show that there exist universal positive constants $c_1, c_2$ such that for any $\eta > 0$, 
\begin{equation}
\mathbb{P}\left(\left|\hat{f}_{rs}(\omega_j) - \mathbb{E}\hat{f}_{rs}(\omega_j)\right| \ge \vertiii{f}\eta\right)\le c_1\exp\left[-c_2(2m+1)\min\{\eta, \eta^2\}\right].
\end{equation}
\end{proof}

\subsection{Proof of Proposition \ref{prop: gauss_prop}}
\begin{proof}
For any Hermitian matrix $M$ \cite{golub2012matrix}, we have
\begin{equation}
\|M\|\le \sqrt{\|M\|_1 \|M\|_\infty} = \|M\|_1.
\end{equation}
Since both $f(\omega_j)$ and $\hat{f}_{\lambda}(\omega_j)$ are Hermitian, we can bound { spectral norm} of estimation error matrix with its { maximum absolute column sum norm}, i.e.
\begin{equation}
\label{eq:L1_error}
\|\hat{f}_{\lambda}(\omega_j)-f(\omega_j)\| \le \|\hat{f}_{\lambda}(\omega_j)-f(\omega_j)\|_1.
\end{equation}
Following the proof technique of Theorem 1 in \citet{bickel2008covariance}, the first step is to bound probability of event 
\begin{equation}
A_0 = \left\{\max_{1\le r,s \le p}|\hat{f}_{rs}(\omega_j) - f_{rs}(\omega_j)| \ge \lambda/2 \right\}. \nonumber
\end{equation}
Our goal is to prove that there exist universal constants $c_1, ~ c_2$ such that for any $r,s \in\{1,\cdots, p\}$, 
\begin{equation}
\mathbb{P}\left(\left|\hat{f}_{rs}(\omega_j) - f_{rs}(\omega_j)\right|\ge \frac{\lambda}{2} \right) \le c_1 \exp\left[-c_2\min\left\{(2m+1)\eta^2, (2m+1)\eta\right\}\right], \nonumber
\end{equation}
where 
\begin{equation}
\lambda = 2\left[ R\vertiii{f}\sqrt{\log p/m} + \frac{m+1/2\pi}{n}\Omega_n(f) + \frac{1}{2\pi}L_n(f)\right]. \nonumber
\end{equation}
Then, with union bound, we could get probability bound for $\mathcal{A}_0$. 

To accomplish this, we first divide the error into two terms along the line of a bias-variance decomposition.
\begin{equation}
\left|\hat{f}_{rs}(\omega_j) - f_{rs}(\omega_j)\right| \le \left|\mathbb{E}\hat{f}_{rs}(\omega_j) - f_{rs}(\omega_j)\right| + \left| \hat{f}_{rs}(\omega_j) - \mathbb{E}\hat{f}_{rs}(\omega_j) \right|. \nonumber
\end{equation}
Proposition \ref{prop:bias_bound} provides an upper bound on the bias term 
\begin{equation}
\left|\mathbb{E}\hat{f}_{rs}(\omega_j) - f_{rs}(\omega)\right| \le \frac{m+1/2\pi}{n}\Omega_n(f) + \frac{1}{2\pi}L_n(f).. \nonumber
\end{equation}
This bound in bias shows that 
\begin{equation}
\label{eq:change_bound_to_variance}
    \mathbb{P}\left(\left|\hat{f}_{rs}(\omega_j)-f_{rs}(\omega_j)\right|\ge \lambda/2\right)\le 
    \mathbb{P}\left(\left|\hat{f}_{rs}(\omega_j)-\mathbb{E}\hat{f}_{rs}(\omega_j)\right|\ge R\vertiii{f}\sqrt{\frac{\log p}{m}}\right).
\end{equation}
Next, proposition \ref{prop:variance_bound} shows that there exists general constants $c_1, c_2$ s.t. 
such that for any $\eta > 0$, 
\begin{equation}
\mathbb{P}\left(\left|\hat{f}_{rs}(\omega_j) - \mathbb{E}\hat{f}_{rs}(\omega_j)\right| \ge \vertiii{f}\eta\right)\le c_1\exp\left[-c_2(2m+1)\min\{\eta, \eta^2\}\right]. \nonumber 
\end{equation}
We set $\eta = R\sqrt{\frac{\log p}{m}}$. Combined with   \eqref{eq:change_bound_to_variance}, and noting that we are working in the  regime $m\succsim \log p$, we conclude  $\eta^2 = R^2\frac{\log p}{m} \le \eta = R\sqrt{\frac{\log p}{m}}$. This implies 
\begin{equation}
\begin{aligned}
P({A}_0) = \mathbb{P}(\max_{r, s}|\hat{f}_{rs}(\omega_j)-f_{rs}(\omega_j)|\ge \lambda/2) \le c_1 p^2\exp\left[-c_2 (2m+1)R^2\frac{\log p}{m}\right]. \nonumber
 \end{aligned}
\end{equation}
This concentration playes essential role in the proof of Theorem 1 as equation (12) in \citet{bickel2008covariance}. Theorem 1 in \cite{bickel2008covariance} provides the techniques to complete the asymptotic analysis, while here we do some modification to achieve non-asymptotic analysis. \par 
\smallskip 
\noindent \textbf{$L_2$ norm bound: } We separate our target into two terms 
\begin{equation}
\|T_\lambda(\hat{f}(\omega_j))-f(\omega_j)\| \le \|T_\lambda(f(\omega_j))-f(\omega_j)\|+\|T_\lambda(f(\omega_j))-T_\lambda(\hat{f}(\omega_j))\| \nonumber
\end{equation}
The first term can be bounded by its $L_1$ norm 
\begin{equation}
\label{eq:two_parts_L2_norm}
\begin{aligned}
& \|T_\lambda(f(\omega_j))-f(\omega_j)\|\le \|T_\lambda(f(\omega_j))-f(\omega_j)\|_1\\
& \le \max_{r=1}^p \sum_{s=1}^p |f_{rs}(\omega_j)|\mathds{1}(|f_{rs}(\omega_j)|<\lambda) \le  \lambda^{1-q}\vertiii{f}_q^q,
\end{aligned}
\end{equation}
for any $0\le q<1$. 

Then we can upper bound the second term in \eqref{eq:two_parts_L2_norm} by three terms as follows:
\begin{equation*}
\begin{aligned}
&\|T_\lambda(f(\omega_j))-T_\lambda(\hat{f}(\omega_j))\| \\
&\le \max_{r=1}^p \sum_{s=1}^p |\hat{f}_{rs}(\omega_j)|\mathds{1}(|\hat{f}_{rs}(\omega_j)|\ge \lambda, |f_{rs}(\omega_j)|\le \lambda)\\
&+\max_{r=1}^p \sum_{s=1}^p |f_{rs}(\omega_j)|\mathds{1}(|\hat{f}_{rs}(\omega_j)|\le \lambda, |f_{rs}(\omega_j)|\ge \lambda)\\
&+\max_{r=1}^p \sum_{s=1}^p |\hat{f}_{rs}(\omega_j)-f_{rs}(\omega_j)|\mathds{1}(|\hat{f}_{rs}(\omega_j)|\ge \lambda, |f_{rs}(\omega_j)|\ge \lambda)\\
& = \rm{I}+\rm{II}+\rm{III}
\end{aligned}
\end{equation*}
Define three events:
\begin{equation*}
\begin{aligned}
&A_1 = \left\{\rm{I} \ge 3\vertiii{f}_q^q \lambda^{(1-q)}\right\}\\
&A_2 = \left\{\rm{II}\ge 2\vertiii{f}_q^q \lambda^{(1-q)}\right\}\\
&A_3 = \left\{\rm{III}\ge \vertiii{f}_q^q \lambda^{(1-q)}\right\}
\end{aligned}
\end{equation*}
We will show that on $A_0^\complement$, none of these three events can happen, i.e., \begin{equation}
A_1\cup A_2 \cup A_3 \subset A_0. \nonumber
\end{equation}
To this end, note that on $A_0^\complement$, 
\begin{equation}
\label{eq:bound_III}
\begin{aligned}
\rm{III} &\le \max_r \left|\hat{f}_{rs}(\omega_j)-f_{rs}(\omega_j)\right|\sum_{s=1}^p\mathds{1}(|f_{rs}(\omega_j)|\ge \lambda)\\
& \le \lambda \sum_{s=1}^p \frac{|f_{rs}(\omega_j)|^q}{\lambda^q} \le \vertiii{f}_q^q \lambda^{1-q}. \nonumber 
\end{aligned}
\end{equation}
Here we use the fact that on event $A_0^\complement$, 
$|\hat{f}_{rs}(\omega_j) - f_{rs}(\omega_j)|\le \frac{\lambda}{2}<\lambda$. Similarly, on $A_0^\complement$,
\begin{equation*}
\begin{aligned}
\rm{II} &\le \max_{r=1}^p |\hat{f}_{rs}(\omega_j)-f_{rs}(\omega_j)|\sum_{s=1}^p\mathds{1}(|f_{rs}(\omega_j)|\ge \lambda)+|\hat{f}_{rs}(\omega_j)|\sum_{s=1}^p \mathds{1}(|\hat{f}_{rs}(\omega_j)|\le \lambda, |f_{rs}(\omega_j)|\ge \lambda)\\
& \le \max_{r=1}^p \left[\lambda \sum_{s=1}^p\mathds{1}(|f_{rs}(\omega_j)|\ge \lambda)+\lambda  \sum_{s=1}^p \mathds{1}(|f_{rs}(\omega_j)|\ge \lambda)\right]\le 2\vertiii{f}_q^q \lambda^{1-q},
\end{aligned}
\end{equation*}
where the last inequality follows from the same argument as in \eqref{eq:bound_III}. Next, we focus on $A_1$. 
\begin{equation*}
\begin{aligned}
&\rm{I} = \max_{r=1}^p \sum_{s=1}^p |\hat{f}_{rs}(\omega_j)|\mathds{1}(|\hat{f}_{rs}(\omega_j)|\ge \lambda, |f_{rs}(\omega_j)|\le \lambda)\\
& \le \max_{r=1}^p \sum_{s=1}^p |\hat{f}_{rs}(\omega_j)-f_{rs}(\omega_j)|\mathds{1}(|\hat{f}_{rs}(\omega_j)|\ge \lambda, |f_{rs}(\omega_j)|\le \lambda)\\
&+\max_{r=1}^p \sum_{s=1}^p |f_{rs}(\omega_j)|\mathds{1}(|\hat{f}_{rs}(\omega_j)|\ge \lambda, |f_{rs}(\omega_j)|\le \lambda)\\
& = \rm{IV}+\rm{V}.
\end{aligned}
\end{equation*}
A similar argument as above can show that 
\begin{equation}
\rm{V}\le \vertiii{f}_q^q \lambda^{1-q}. \nonumber
\end{equation}
For $\rm{IV}$, on $A_0^\complement$, 
\begin{equation*}
\begin{aligned}
&\rm{IV} = \max_{r=1}^p \sum_{s=1}^p |\hat{f}_{rs}(\omega_j)-f_{rs}(\omega_j)|\mathds{1}(|\hat{f}_{rs}(\omega_j)|\ge \lambda, |f_{rs}(\omega_j)|\le \lambda)\\
&= \max_{r=1}^p \sum_{s=1}^p |\hat{f}_{rs}(\omega)-f_{rs}(\omega_j)|
\mathds{1}(|\hat{f}_{rs}(\omega_j)|\ge \lambda, ~\lambda/2<|f_{rs}(\omega_j)|\le \lambda)\\
& \le \max_{r=1}^p \sum_{s=1}^p \lambda \mathds{1}(|f_{rs}(\omega_j)|\ge \lambda/2) \le \max_r \sum_{s=1}^p \lambda \sum_{s=1}^p \frac{|f_{rs}(\omega_j)|^q}{(\lambda/2)^q}\\
&\le 2\lambda^{1-q} \vertiii{f}_q^q. 
\end{aligned}
\end{equation*}
Combining these two parts, we have $\rm{I}\le 3\lambda^{1-q}\vertiii{f}_q^q$. Also, since 
\begin{equation*}
\left\{\|T_\lambda(\hat{f}(\omega_j))-f(\omega_j)\|
\ge 7\lambda^{1-q}\vertiii{f}_q^q\right\} \subset A_1\cup A_2 \cup A_3 \subset A_0,
\end{equation*}
we have 
\begin{equation*}
\begin{aligned}
&\mathbb{P}(\|\hat{f}(\omega_j)-f(\omega_j)\| 
\ge 7\lambda^{1-q}\vertiii{f}_q^q\})
\le \mathbb{P}(A_0) \le c_1p^2\exp\left[-c_2(2m+1)\min\{\eta, \eta^2\}\right]. \end{aligned}
\end{equation*}
\par
\smallskip
\noindent\textbf{Proof of upper bound on Frobenius norm: } Like the proof for operator norm, we decompose the error term as 
\begin{equation}
\|T_\lambda(\hat{f})(\omega_j)-f(\omega_j)\|_F^2 \le \|T_\lambda(f(\omega_j))-f(\omega_j)\|_F^2+\|T_\lambda(f(\omega_j))-T_\lambda(\hat{f}(\omega_j))\|_F^2. \nonumber
\end{equation}
The same argument for opertator norm then ensures that on $A_0^c$
\begin{equation*}
\begin{aligned}
\|T_\lambda(f)-f\|_F^2 &= \sum_{r,s} |f_{rs}(\omega_j)|^2\mathds{1}(|f_{rs}(\omega_j)|\le \lambda)\\
& \le \sum_{r,s} \lambda^{2-q} |f_{rs}(\omega_j)|^q \le \lambda^{2-q}\vertiii{f}_q^2. 
\end{aligned}
\end{equation*}
As before, we decompose the second term in the next step as follows:
\begin{equation*}
\begin{aligned}
&\|T_\lambda(f(\omega_j))-T_\lambda(\hat{f}(\omega_j))\|_F^2 \\
&\le \sum_{r,s} |\hat{f}_{rs}(\omega_j)|^2\mathds{1}(|\hat{f}_{rs}(\omega_j)|\ge \lambda, |f_{rs}(\omega_j)|\le \lambda)\\
&+\sum_{r,s} |f_{rs}(\omega_j)|^2 \mathds{1}(|\hat{f}_{rs}(\omega_j)|\le \lambda, |f_{rs}(\omega_j)|\ge \lambda)\\
&+ \sum_{r,s} |\hat{f}_{rs}(\omega_j)-f_{rs}(\omega_j)|^2\mathds{1}(|\hat{f}_{rs}(\omega_j)|\ge \lambda, |f_{rs}(\omega_j)|\ge \lambda)\\
&= \rm{I}+\rm{II}+\rm{III},
\end{aligned}
\end{equation*}
and we define following events:
\begin{equation*}
\begin{aligned}
&A_1 = \left\{\rm{I} \ge 7p\vertiii{f}_q^q \lambda^{2-q}\right\}\\
&A_2 = \left\{\rm{II}\ge 4p\vertiii{f}_q^q \lambda^{2-q}\right\}\\
&A_3 = \left\{\rm{III}\ge p\vertiii{f}_q^q \lambda^{2-q}\right\}.
\end{aligned}
\end{equation*}
We will show that $A_1\cup A_2 \cup A_3 \subset A_0$ by showing on $A_0^\complement$, none of these three events can happen.  $\rm{III}\le p\lambda^{2-q}\vertiii{f}_q^q$ is obvious with same techniques before. For $\rm{II}$, on $A_0$, 
\begin{equation*}
\begin{aligned}
\rm{II} &\le \left[|\hat{f}_{rs}(\omega_j)-f_{rs}(\omega_j)|^2 + |\hat{f}_{rs}(\omega_j)|^2 +2 |\hat{f}_{rs}(\omega_j)||\hat{f}_{rs}(\omega_j)-f_{rs}(\omega_j)|\right]\\
&  \mathds{1}(|\hat{f}_{rs}(\omega_j)|\le \lambda, |f_{rs}(\omega_j)|\ge \lambda)\\
& \le \sum_{r,s} \lambda^2\mathds{1}(|f_{rs}(\omega_j)|\ge \lambda) + \lambda^2\mathds{1}(|f_{rs}(\omega_j)|\ge \lambda)+2\lambda^2\mathds{1}(|f_{rs}(\omega_j)|\ge \lambda) \\
& \le 4p\lambda^{2-q}\vertiii{f}_q^q. 
\end{aligned}
\end{equation*}
For $\rm{I}$, on $A_0$, we have
\begin{equation*}
\begin{aligned}
\rm{I}\le& \sum_{r,s}\left[|\hat{f}_{rs}(\omega_j)-f_{rs}(\omega_j)|^2 + |f_{rs}(\omega_j)|^2+ 2|f_{rs}(\omega_j)||\hat{f}_{rs}(\omega_j)-f_{rs}(\omega_j)|\right] \\ 
& \mathds{1}(|f_{rs}(\omega_j)|\le \lambda, |\hat{f}_{rs}(\omega_j)|\ge \lambda)\\
& = \rm{V}+\rm{VI}+\rm{VII}. 
\end{aligned}
\end{equation*}
Note that on $A_0^\complement$, $\mathds{1}(|f_{rs}(\omega_j)|\le \lambda, |\hat{f}_{rs}(\omega_j)|\ge \lambda)=\mathds{1}(\lambda/2<|f_{rs}(\omega_j)|\le \lambda, |\hat{f}_{rs}(\omega_j)|\ge \lambda)$. Using this, we can show that 
\begin{equation*}
\begin{aligned}
& \rm{V} = \lambda^2 \sum_{r,s}\mathds{1}(|f_{rs}(\omega_j)|\ge \lambda/2)=\le \lambda^2 \sum_{r,s}\left(\frac{|f_{rs}(\omega_j)|^q}{(\lambda/2)^q}\right) \le 2p\lambda^{2-q}\vertiii{f}_q^q\\
& \rm{VI}\le  \sum_{r,s}|f_{rs}(\omega_j)|^2 1(|f_{rs}(\omega_j)|\le \lambda)\le \sum_{r,s} \left(\frac{\lambda}{|f_{rs}(\omega_j)|}\right)^{2-q}|f_{rs}(\omega_j)|^2 = p\vertiii{f}_q^q \lambda^{2-q}\\
& \rm{VII}\le 2\lambda^2 \sum_{r,s}1(|f_{rs}(\omega_j)|\ge \lambda/2)=\le 2\lambda^2 \sum_{r,s}\left(\frac{|f_{rs}(\omega_j)|^q}{(\lambda/2)^q}\right) \le 4p\lambda^{2-q}\vertiii{f}_q^q.
\end{aligned}
\end{equation*}
Thus, we have shown that $\rm{I}\le 7p\lambda^{2-q}\vertiii{f}_q^q$. Putting all these pieces together, we obtain 
\begin{equation}
\left\{\|T_\lambda(\hat{f}(\omega_j))-f(\omega_j)\|_F^2\ge 13p\lambda^{2-q}\vertiii{f}_q^q\right\} \subset  A_1\cup A_2 \cup A_3 \subset A_0, \nonumber
\end{equation}
which completes the proof. 
\end{proof}

\subsection{Proof of Proposition \ref{prop:consistency}}
\begin{proof}
In order to prove the first bound, we note that 
\begin{equation*}
\begin{aligned}
&\mathbb{P}\left(\exists ~r,s : |T_\lambda(\hat{f}_{rs}(\omega_j))|>0, f_{rs}(\omega_j)=0\right)\\
& \le \mathbb{P}\left(\exists ~r,s : |T_\lambda(\hat{f}_{rs}(\omega_j))-f_{rs}(\omega_j)|>\lambda \right)\\
&\le p^2 c_1\exp[-c_2R^2 \log p].
\end{aligned}
\end{equation*}
where the last inequality comes from proposition \ref{prop: gauss_prop}. 

Now we turn to the second part. Since $\mathcal{S}(\gamma) = \left\{(r,s):|f_{rs}(\omega_j)|\ge \gamma \lambda \right\}$ with some $\gamma>1$. 
\begin{equation*}
\begin{aligned}
& \mathbb{P}\left(\exists ~(r,s) \in \mathcal{S}(\gamma) : T_\lambda(\hat{f}_{rs}(\omega_j))=0, |f_{rs}(\omega_j)| >0\right)\\
& \mathbb{P}\left(\exists ~(r,s) \in S(\gamma), |\hat{f}_{rs}(\omega_j)-f_{rs}(\omega_j)| >(\gamma-1)\lambda\right)\\
&\le p^2 c_1\exp[-c_2 (\gamma-1)^2R^2 \log p].
\end{aligned}
\end{equation*}
The last inequality comes from the following decomposition  
\begin{equation}
(\gamma-1)\lambda = 2(\gamma-1)R \vertiii{f}\sqrt{\frac{\log p}{m}} +2(\gamma-1)\left[ \frac{m+1/2\pi}{n}\Omega_n(f)+\frac{1}{2\pi}L_n(f)\right], \nonumber
\end{equation}
where the second part serves as an upper bound for bias because $\gamma>1.5$.
\end{proof}

\subsection{Proof of Proposition \ref{prop:coherance}}
We first build the concentration bound for error terms under asymptotic  region stated in the proposition \ref{prop:coherance}, i.e.,  there exist universal positive constants $c_1, c_2$ s.t.
\begin{equation}
\mathbb{P}\left(\max_{r,s} |\hat{g}_{rs}(\omega_j) - g_{rs}(\omega_j)|\ge \frac{2\lambda}{\tau}\right)\le c_1 p^2 \exp[-c_2 R\log p].
\end{equation}
Define the events 
\begin{equation}
A_0 = \left\{\max_{r, s}|\hat{f}_{rs}(\omega_j)-f_{rs}(\omega_j)|\ge \lambda \right\} \nonumber
\end{equation}
and 
\begin{equation}
A_1 = \left\{\max_{r, s}|\hat{g}_{rs}(\omega_j)-g_{rs}(\omega_j)|\ge 2\lambda/\tau \right\}. \nonumber
\end{equation}
We will show that 
$A_1\subset A_0$. Since 
\begin{equation}
|\hat{g}_{rs}(\omega_j)-g_{rs}(\omega_j)| \le    |\hat{g}_{rs}(\omega_j)-\tilde{g}_{rs}(\omega_j)|+|\tilde{g}_{rs}(\omega_j)-g_{rs}(\omega_j)| \nonumber
\end{equation}
with $\tilde{g}_{rs}(\omega_j) = \frac{\hat{f}_{rs}(\omega_j)}{\sqrt{f_{rr}(\omega_j)f_{ss}(\omega_j)}}$, it suffices to show that for any $r,s$, 
\begin{equation*}
\begin{aligned}
& \left\{|\tilde{g}_{rs}(\omega_j)-g_{rs}(\omega_j)| \ge \lambda/\tau\right\}\subset A_0\\
&\left\{|\hat{g}_{rs}(\omega_j)-\tilde{g}_{rs}(\omega_j)| \ge \lambda/\tau\right\}\subset A_0 
\end{aligned}
\end{equation*}
For the first inclusion, note that with $|f_{rr}(\omega_j)|\ge \tau$ for $1\le r\le p$, 
\begin{equation*}
\begin{aligned}
&\left\{|g_{rs}(\omega_j)-\tilde{g}_{rs}(\omega_j)| \ge \lambda/\tau \right\}\\
&= \left\{\left|\frac{\hat{f}_{rs}(\omega_j) - f_{rs}(\omega_j)}{\sqrt{f_{rr}(\omega_j)f_{ss}(\omega_j)}}\right| \ge \lambda/\tau\right\}\\
&\subset \left\{\left|\frac{\hat{f}_{rs}(\omega_j) - f_{rs}(\omega_j)}{\tau}\right| \ge \lambda/\tau\right\} = A_0.\\
\end{aligned}
\end{equation*}
Similarly, for the second one,  
\begin{equation*}
\begin{aligned}
&\left\{|\hat{g}_{rs}(\omega_j)-\tilde{g}_{rs}(\omega_j)| \ge \lambda/\tau \right\}\\
&= \left\{\left|\hat{g}_{rs}(\omega_j) \right| \left|\sqrt{\frac{\hat{f}_{rr}(\omega_j)\hat{f}_{ss}(\omega_j)}{f_{rr}(\omega_j)f_{ss}(\omega_j)}}-1\right| \ge \lambda/\tau\right\}.
\end{aligned}
\end{equation*}
Since the averaged periodogram  ($\hat{f}(\omega_j)$) is positive semi-definite with positive diagonal elements(almost surely), we have $|\hat{g}_{rs}(\omega_j)|\le 1$. This implies that the above event is a subset of 
\begin{equation}
\left\{\left|\sqrt{\frac{\hat{f}_{rr}(\omega_j)\hat{f}_{ss}(\omega_j)}{f_{rr}(\omega_j)f_{ss}(\omega_j)}}-1\right| \ge \lambda/\tau \right\}. \nonumber
\end{equation}
For all $1\le r\le p$, 
\begin{equation}
\label{eq:single-ratio-bound}
\begin{aligned}
&\left\{\left|\frac{\hat{f}_{rr}(\omega_j)}{f_{rr}(\omega_j)}-1\right|\ge \frac{\lambda}{\tau}\right\}  \\
& = \left\{\left|\frac{f_{rr}(\omega_j) - f_{rr}(\omega_j)}{f_{rr}(\omega_j)}\right|\ge \frac{\lambda}{\tau}\right\}\\
& \subset \left\{\left|f_{rr}(\omega_j) - f_{rr}(\omega_j)\right|\ge \lambda\right\} = A_0. \nonumber 
\end{aligned}
\end{equation}
This indicates that 
\begin{equation}
\begin{aligned}
\left\{\max_{r=1}^p\left|\frac{\hat{f}_{rr}(\omega_j)}{f_{rr}(\omega_j)}-1\right|\ge \frac{\lambda}{\tau}\right\} \subset A_0. \nonumber 
\end{aligned}
\end{equation}

Noticing on the event $A_0^\complement$, for all $1\le r\le p$, 
\begin{equation}
\left\{1-\frac{\lambda}{\tau}\le \left|\frac{\hat{f}_{rr}(\omega_j)}{f_{rr}(\omega_j)}\right|\le 1+\frac{\lambda}{\tau}\right\}, \nonumber
\end{equation}
with $\lambda/\tau<1$ (since $\lambda = o(1)$), 
\begin{equation}
1-\frac{\lambda}{\tau}\le \sqrt{\frac{\hat{f}_{rr}(\omega_j)\hat{f}_{ss}(\omega_j)}{f_{rr}(\omega_j)f_{ss}(\omega_j)}}\le 1+\frac{\lambda}{\tau}, 
\end{equation}
indicating
\begin{equation}
\left\{\left|\sqrt{\frac{\hat{f}_{rr}(\omega_j)\hat{f}_{ss}(\omega_j)}{f_{rr}(\omega_j)f_{ss}(\omega_j)}}-1\right| \ge \lambda/\tau \right\}\subset A_0. \nonumber
\end{equation}
This in turn implies 
\begin{equation}
\begin{aligned}
&\left\{|\hat{g}_{rs}(\omega_j)-\tilde{g}_{rs}(\omega_j)| \ge \lambda/\tau \right\}\subset A_0. \nonumber
\end{aligned}
\end{equation}
Combining two inclusion relations, we can claim that 
\begin{equation}
\left\{\left|\hat{g}_{rs}(\omega_j)-g_{rs}(\omega_j)\right|\ge \frac{2\lambda}{\tau}\right\}\subset A_0, \nonumber
\end{equation}
which completes building the concentration inequality for event $A_1$ since proposition \ref{prop:consistency} presents the concentration inequality for event $A_0$.  Then following the argument in proof of proposition \ref{prop:consistency}, we could complete the proof. 
\section{Appendix: Proofs for Linear Processes}\label{Appendix:proof_heavytail}
\subsection{Proof for Lemma \ref{lemma:heavy_tail_hanson}}
\begin{proof}
Proof for sub-Gaussian case is given by \cite{rudelson2013hanson} and proof for the  sub-exponential case is given by Lemma 8.3 in \cite{erdHos2012bulk}. We will show the proof for case \eqref{C3} based on Markov inequality. We will show tail bound for both  diagonal part and non-diagonal part for any $\eta > 0$ 
one by one. For diagonal part,  let $y_i = \varepsilon^2_{ii}-1$. Then 
$\mathbb{E}y_i = 0$ and $\mathbb{E}y^2_i = \mathbb{E}\varepsilon^4_{ii} - 2\mathbb{E}\varepsilon^2_{ii}+1 \le   K-1< K$. 
Therefore, noticing  $\mathbb{E}\varepsilon^\top A\varepsilon = \text{tr}(A)$ under this setting,
\begin{equation*}
\begin{aligned}
\mathbb{P}\left[\left|\sum_{i=1}^n \varepsilon^2_{ii} A_{ii} - \text{tr}(A)\right|\ge \eta\right] &= \mathbb{P}\left[\left|\sum_{i=1}^n y_iA_{ii}\right|\ge \eta\right]\\
&\le \frac{\mathbb{E}(\sum_{i=1}^n y_iA_{ii})^2}{\eta^2} \le \frac{K\sum_{i=1}^n A^2_{ii}}{\eta^2}, 
\end{aligned}
\end{equation*}
where the second last inequality follows from  $\mathbb{E} y_iy_j = 0$. 
For the non-diagonal part, note that 

\begin{equation*}
\begin{aligned}
\mathbb{P}\left[\left|\sum_{1 \le  i\neq j \le n} A_{ij}\varepsilon_i\varepsilon_j \right|\ge \eta\right] &\le \frac{\mathbb{E}\left|\sum_{1 \le  i\neq j \le n} A_{ij}\varepsilon_i\varepsilon_j \right|^2}{\eta^2} \\
& = \frac{\sum_{1\le i \neq j\le n} A^2_{ij}(\mathbb{E}\varepsilon^2_1)^2}{\eta^2} +\frac{\sum_{1\le i\neq j\le n}A_{ij}A_{ji}(\mathbb{E}\varepsilon^2_1)^2}{\eta^2}\\
&\le  \frac{2\sum_{1\le i\neq j\le n} A^2_{ij}}{\eta^2}.
\end{aligned}
\end{equation*}
Here the second line holds since  $\mathbb{E}\varepsilon_i\varepsilon_j\varepsilon_p\varepsilon_q\neq 0$ iff $i=p,j=q$ or $i=q, j=p$ and the third line comes from the simple fact that 
$A_{ij}A_{ji}\le \frac{1}{2}(A^2_{ij}+A^2_{ji})$. \par 
Then plugging $\frac{\eta}{2}$ into above two parts, we get \begin{equation*}
\begin{aligned}
    &\mathbb{P}\left[|\varepsilon^\top A\varepsilon - \mathbb{E} \varepsilon^\top A\varepsilon|\ge \eta\right] \\
    \leq& \mathbb{P}\left[\left|\sum_{i=1}^n \varepsilon ^2_{ii} A_{ii} - \text{tr}(A)\right|\ge \eta/2\right]+\mathbb{P}\left[\left|\sum_{1\le i \neq j\le n} A_{ij}\varepsilon_i\varepsilon_j \right|\ge \eta/2\right]\\
    \leq& \max\{4K,8\} \frac{\|A\|_F^2}{\eta^2},
\end{aligned}
\end{equation*}
where we can set $c_3=\max\{4K,8\}$ and use the fact $\|A\|^2_F\le \rank(A)\|A\|^2$ to complete our proof. 
\end{proof}
\subsection{Proof of Proposition \ref{lemma:heavy_tail_time_hanson}} 

\begin{proof}
The proofs of the above inequalities for these three cases follow a common structure. We work with fixed values of $n$ and $p$, and construct a limiting argument as $L \rightarrow \infty$.  In the first step, we apply inequality in Lemma  \ref{lemma:heavy_tail_hanson} to the truncated process $X_{(L),t} = \sum_{\ell=0}^L B_\ell \varepsilon_{t-\ell}$, for some $L>0$. Then we show that this inequality holds in the limit  $L\rightarrow \infty$. For the sake of brevity, we only present the proof for sub-Gaussian case here. 

Let $\mathcal{X}_{(L)}$ be a $n \times p$ data matrix with $n$ consecutive observations from process $\{X_{(L), t}\}_{t \in \mathbb{Z}}$.  We can write 
$vec(\mathcal{X}_{(L)}^\top) = \Pi_L E_n$ where 
\begin{equation*}
    \Pi_L = \begin{bmatrix}
    0 & 0 & \dots & 0 & B_0 & B_1 & \dots & B_{L-1} & B_L \\
    0 & 0 & \dots & B_0 & B_1 & B_2 & \dots & B_L & 0 \\
    \vdots & \vdots & \ddots & \vdots & \vdots & \vdots &\ddots & \vdots & \vdots \\
    B_0 & B_1 & \dots & \dots &  \dots & \dots & B_L & 0  & 0
    \end{bmatrix}
\end{equation*}
and $E_n = (\varepsilon_n^\top, \dots, \varepsilon^\top_{1-L})^\top$. Without loss of generality, we assume $L>n$ in our representation of $\Pi_L$ and $E_n$. It follows from Lemma  \ref{lemma:max-L2-norm} that 
$\|\cov(vec(\mathcal{X}_{(L)}^\top), vec(\mathcal{X}_{(L)}^\top))\| = \|\Pi_L\Pi_L^\top\| \le \vertiii{f_{(L)}}$, where  $f_{(L)}(\omega)$ is the spectral density of $X_{(L), t}$. Then using the same technique as in the proof of Lemma \ref{lemma: hason_bound_time_gauss} and inequality for sub-Gaussian i.i.d. case introduced in Lemma \ref{lemma:heavy_tail_hanson}, we get 
\begin{equation}
\label{eq:sub_gauss_time_hanson_wright_truncated}
\begin{aligned}
&\mathbb{P}\left(\left|vec(\mathcal{X}_{(L)}^\top)^\top A ~vec(\mathcal{X}_{(L)}^\top) - \mathbb{E} \left[~vec(\mathcal{X}_{(L)}^\top )^\top A ~vec(\mathcal{X}_{(L)}^\top)\right]\right| >2\pi \eta \vertiii{f_{(L)}} \right)\\
&\le 2\exp\left[-c\min\left\{\cfrac{\eta}{\|A\|}, \cfrac{\eta^2}{\rank(A)\|A\|^2}\right\}\right].
\end{aligned}
\end{equation} 
Next we note that by Lemma \ref{lemma:L2_convergence_truncate}, for any fixed $n, p$,  $vec(\mathcal{X}_{(L)}^\top)\overset{L_2}{\rightarrow}vec(\mathcal{X^\top})$ as $L \rightarrow \infty$. Since $L_2$ convergenece implies convergence in probability, by  continuous mapping theorem, we have   
\begin{equation}
vec(\mathcal{X}_{(L)}^\top)^\top A ~vec(\mathcal{X}_{(L)}^\top)\overset{\mathbb{P}}{\rightarrow}vec(\mathcal{X}^\top)^\top A ~vec(\mathcal{X}^\top)
\end{equation}
as $L \rightarrow \infty$. The $L_2$-norm  convergence also ensures $L_1$-norm convergence,  which implies   
\begin{equation}
     \mathbb{E} \left[~vec(\mathcal{X}_{(L)}^\top)^\top A ~vec(\mathcal{X}_{(L)}^\top)\right] \rightarrow  \mathbb{E} \left[~vec(\mathcal{X}^\top )^\top A ~vec(\mathcal{X}^\top)\right].
\end{equation}
A detailed derivation is outlined in the remarks after Lemma \ref{lemma:L2_convergence_truncate}. Together with Lemma \ref{lemma:spectral_convergence}, we obtain  $2\pi \eta\vertiii{f_{(L)}}  \rightarrow 2\pi \eta\vertiii{f}$. Putting pieces together, we have
\begin{equation*}
\begin{aligned}
&vec(\mathcal{X}_{(L)}^\top)^\top A ~vec(\mathcal{X}_{(L)}^\top )-\mathbb{E} \left[~vec(\mathcal{X}_{(L)}^\top)^\top A ~vec(\mathcal{X}_{(L)}^\top)\right]-2\pi \eta\vertiii{f_{(L)}} \\
\end{aligned}
\end{equation*}
converges in probability, and hence in distribution, to 
\begin{equation*}
\begin{aligned}
&
vec(\mathcal{X}^\top )^\top A ~vec(\mathcal{X}^\top)-\mathbb{E} \left[~vec(\mathcal{X}^\top)^\top A ~vec(\mathcal{X}^\top)\right]-2\pi \eta\vertiii{f}. 
\end{aligned}
\end{equation*}
Thus, if we take $L\rightarrow \infty$ from both sides in \eqref{eq:sub_gauss_time_hanson_wright_truncated}, we obtain the final bound. 
\end{proof}

\section{Appendix: Additional Proofs of Technical Results}\label{Appendix:proof_technical_lemmas}
\label{sec:proof_for_technical_lemmas}

\begin{lem}
\label{lemma:q_norm_eq}
For any matrix $A\in \mathbb{C}^{p \times p}$ and  $0\le q<1$, define $\|A\|_q:= \max_{\|x\|_q=1} \|Ax\|_q$, where $q$ norm for vector is defined as $\|x\|_q= (\sum_{i=1}^p |x_i|^q)^{1/q}$ for any vector x of length $p$(Again, it is indeed a norm iff $q\ge 1$). Then 
\begin{equation}
\max_{s=1}^p  \sum_{r=1}^p |A_{rs}|^q = \|A\|_q^q. \nonumber
\end{equation}
\begin{proof}
First, for two vectors $v_1, v_2\in \mathbb{C}^p$, $\|v_1+v_2\|_q^q\le \|v_1\|_q^q + \|v_2\|_q^q$ for $0\le q<1$, since for scalars $x,y\in \mathbb{C}$, $|x+y|^q \le |x|^q+|y|^q$. Then let $A_i$ be the $i^{th}$ column of $A$. Based on the definition of $\|A\|_q$, we have 
\begin{equation*}
\begin{aligned}
& \|A\|_q^q = \max_{\|x\|_q = 1} \|\sum_{i=1}^p A_ix_i\|_q^q \\
& \le  \max_{\|x\|_q = 1} \sum_{i=1}^p \|A_ix_i\|_q^q  = \sum_{i=1}^p |x_i|^q\|A_i\|_q^q\\
&\le (\max_{i=1}^p \|A_i\|_q^q) \sum_{i=1}^p\|x_i\|^q = \max_{i=1}^p \|A_i\|_q^q.
\end{aligned}
\end{equation*}
Noticing if we set $x$ above as the indicator vector $e_r$, where $r=\argmax_i \|A_i\|^q $, the equality holds, we finish the proof.  
\end{proof}
\end{lem}

\begin{lem}
\label{lemma:spectral_simple} 
For any matrix $A\in \mathbb{R}^{p\times p}$, and a positive constant $\epsilon$, we could find a matrix $E$ such that $A+E$ has distinct eigenvalues and $\|E\|\le \epsilon$. 
\begin{proof}
Consider the Schur decomposition(\citep{golub2012matrix}) of $A$ as $A=QUQ^\dag$ where $Q$ is an unitary matrix and $U$ is an upper triangular matrix. Construct a diagonal matrix
$D$ with each element less than $\epsilon$ and make $U_{i,i} + D_{i,i}$ distinct. Set $E = QDQ^\dag$, we have $A+QDQ^\dag = Q (E+D)Q^\dag$ with eigenvalues as $U_{i,i} + D_{i,i}, i=1,\dots p$ which are distinct. By setting $E= QDQ^\dag$ and noticing $\|E\| = \|QDQ^\dag\| = \|D\|\le \epsilon$ we complete the proof. 
\end{proof}
\begin{remark}
$\lambda_{\text{max}}(A)$ is continuous mapping from the set of $p\times p$ complex matrices to the set of real numbers. Thus, we can always find perturbation $\|E\|$ small enough to guarantee $\|A+E\|<1$. To quantify this, we can apply the result from \citet{bhatia1990bounds} for perturbation bound on potentially non-symmetric matrices
\begin{equation}
\label{eq:eigen_bound}
|\lambda_{\textup{max}}(A+E)-\lambda_{\textup{max}}(A)|\le 12\|A\|^{1-1/p}\|E\|^{1/p}.
\end{equation}
\end{remark}
\end{lem}

\begin{lem}
\label{lemma:orthogonal-cos-sin}
For any $j, k$ in $F_n$, the inner product between $C_j$ and $S_k$ can only have the  following forms:
\begin{enumerate}
\item[(a)] $C^{\top}_j S_k = 0$
\item[(b)] 
\begin{equation*}
\begin{aligned}
C_j^\top C_k = 0 ~~\text{if}~~ |j|\neq |k|; ~~~~  C_j^\top C_j = 
\begin{cases}
1 & ~\text{if}~ j\in \{0, \frac{n}{2}\}\\
\frac{1}{2} & \text{otherwise}
\end{cases}; ~~~C_j^\top C_{-j} = 
\begin{cases}
1 & ~\text{if}~ j=0\\
\frac{1}{2} & \text{otherwise}
\end{cases}
\end{aligned}
\end{equation*}
\item[(c)] 
\begin{equation*}
\begin{aligned}
S_j^\top S_k = 0 ~~\text{if}~~ |j|\neq |k|; ~~~~  S_j^\top S_j = 
\begin{cases}
0 & ~\text{if}~ j\in \{0, \frac{n}{2}\}\\
\frac{1}{2} & \text{otherwise}
\end{cases}; ~~~S_j^\top S_{-j} = 
\begin{cases}
0 & ~\text{if}~ j=0\\
-\frac{1}{2} & \text{otherwise}.
\end{cases}
\end{aligned}
\end{equation*}
\end{enumerate}

\begin{proof}
We first state Lagrange's trigonometric identities: 
\begin{equation}
\label{eq:cos_series}
\sum_{\ell=1}^n \cos(\ell\theta) = 
\begin{cases}
n & \theta = 2k\pi ~\text{for some integer}~ k \\
-\frac{1}{2}+\frac{\sin\left(n+\frac{1}{2}\right)\theta}{2\sin \frac{\theta}{2}}  & \text{otherwise}
\end{cases}
\end{equation}
and 
\begin{equation}
\label{eq:sin_series}
\sum_{\ell=1}^n \sin(\ell\theta) =
\begin{cases}
0 & \theta = 2k\pi ~\text{for some integer}~ k \\ 
\frac{\cos (\frac{1}{2}\theta)}{2\sin (\frac{1}{2}\theta)}-\frac{\cos\left(n+\frac{1}{2}\right)\theta}{2\sin \frac{\theta}{2}} & \text{otherwise}
\end{cases}
\end{equation}
Now we consider a special case where we set $\mathbf{\theta} = \omega_j = \frac{2j\pi}{n}, j\in \mathbb{Z}$. Here we relax $j\in F_n$ to all integers. After this relaxation, we can write $\omega_j+\omega_k = \omega_{j+k}$ and $\omega_j-\omega_k = \omega_{j-k}$. Using \eqref{eq:cos_series} and \eqref{eq:sin_series}, 
for any $\omega_j$, $ j \in \mathbb{Z}$, and fixed $n$, we have the following identities
\begin{equation}
\label{eq:omega_cos}
\sum_{\ell=1}^n \cos(\ell\omega_j) = 
\begin{cases}
n &  \mbox{if } j \equiv 0\pmod{n} \\
0  & \text{otherwise}
\end{cases},
\end{equation}
\begin{equation}
\label{eq:omega_sin}
\sum_{\ell=1}^n \sin(\ell\omega_j) = 0.
\end{equation}\par
Now we prove (a), (b) and (c).

\begin{enumerate}
\item[(a)] For any $j$ and $k$ in $F_n$, \eqref{eq:omega_sin} implies
\begin{equation}
\begin{aligned}
\label{eq:case_a_proof}
&C_j^\top S_k = \frac{1}{2n}\sum_{\ell=1}^n [\sin(\ell (\omega_j+\omega_k))-\sin(\ell (\omega_k-\omega_j))] \\
& = \frac{1}{2n}\left[\sum_{\ell=1}^n \sin(\ell \omega_{j+k}) - \sum_{\ell=1}^n \sin(\ell \omega_{k-j})\right] = 0
\end{aligned}
\end{equation}
\item[(b)] For any $j,k\in F_n$, 
\begin{equation}
\label{eq:case_b_proof}
C_j^\top C_k = \frac{1}{2n}\left(\sum_{\ell=1}^n \cos (\ell\omega_{j+k}) + \sum_{\ell=1}^n \cos (\ell\omega_{j-k})\right)
\end{equation}
For the case $j=k$ or $j=-k$, we have
\begin{equation}
\label{eq:j=k_cos}
C_j^\top C_k = \frac{1}{2n} \left(\sum_{\ell=1}^n\cos (\ell\omega_{2k}) + \sum_{\ell=1}^n \cos (\ell\omega_{0})\right).
\end{equation}
\eqref{eq:cos_series} implies that if $j \in \left\{ 0,\frac{n}{2}\right\}$, \eqref{eq:j=k_cos} is 1. In other cases, $0<2k<n$, \eqref{eq:omega_cos} implies that $\sum_{\ell=1}^n \cos(\ell\omega_{2k})=0$ which further shows that the right hand side in \eqref{eq:j=k_cos} is $1/2$. \par  

For the other cases, since $-n<j+k<n$ and $-n<j-k<n$, 
$j+k\not \equiv 0 \pmod{n}$ and $j-k \not \equiv 0 \pmod{n}$, the right hand side in equation \eqref{eq:case_b_proof} becomes 0. 
\item[(c)] for any $j,k\in F_n$, 
\begin{equation}
\label{eq:case_c_proof}
C_j^\top C_k + S_j^\top S_k = \frac{1}{n}\sum_{\ell=1}^n \cos (\ell\omega_j) \cos (\ell\omega_k) + \sin (\ell\omega_j)\sin (\ell\omega_k) = \frac{1}{n}\sum_{\ell=1}^n \cos (\ell\omega_{j-k})
\end{equation}
If $k=j$, the right hand side in \eqref{eq:case_c_proof} is 1 and in other cases, the right hand side is 0. Then plugging in  the value of $C_j^\top C_k$ listed in case (a), we complete  our proof for case (c). 
\end{enumerate}
\end{proof}
\end{lem}

\begin{lem}
\label{lemma:maximum_L2_Q}
$\|Q_{F_n}\|= 1$
where 
\begin{equation}
Q_{F_n} = \begin{bmatrix}
C_{-[\frac{n-1}{2}]}^\top\\ 
S_{-[\frac{n-1}{2}]}^\top \\
\vdots \\
C_{[\frac{n}{2}]}^\top\\ 
S_{[\frac{n}{2}]}^\top \\
\end{bmatrix}
\end{equation}
and each $C_j, S_j, j\in F_n$ follow the definition in \eqref{eq:cos_sin_coef} 
\begin{proof}
Since row permutation does not change the $L_2$ norm of a matrix, we can stack rows in $Q_{F_n}$ such that $S_j,C_j,S_{-j}, C_{-j}$ appear adjacently, if there exists such a pair $\{j, -j\}$. Then $\|Q_{F_n}\|=\|Q^\top_{F_n}\| = \sqrt{\lambda_{\max}(Q_{F_n}Q_{F_n}^\top)}$. Lemma \ref{lemma:orthogonal-cos-sin} implies that $Q_{F_n}Q_{F_n}^\top$ can only be block-wise diagonal with three possible blocks:
\begin{equation*}
B_1=\begin{bmatrix}
1&0\\
0&0
\end{bmatrix}, ~~B_2 = \begin{bmatrix}
\frac{1}{2} & 0 & \frac{1}{2} & 0\\
0 & \frac{1}{2} & 0 & -\frac{1}{2}\\
\frac{1}{2} & 0 & \frac{1}{2} & 0 \\
0 & -\frac{1}{2} & 0 & \frac{1}{2}
\end{bmatrix}, ~~
B_3 = \begin{bmatrix}
\frac{1}{2} & 0\\
0 & \frac{1}{2}
\end{bmatrix}.
\end{equation*}
Here $B_1$ corresponds to the block formed with $C_0, S_0$, $B_2$ corresponds to the block formed of $C_j, S_j, C_{-j}, S_{-j}, j\neq 0$ and $B_3$ corresponds to the block formed of single j: $C_j, S_j$. 
It can be checked that $\|B_i\| \le 1$ for $i=1,2,3$. 
It follows that $\|Q_{F_n}\|= \sqrt{\lambda_{\max}(Q_{F_n}Q_{F_n}^\top )} \le \max_{i=1}^3 \|B_i\|=1$, completing our proof. 
\end{proof}
\end{lem}

\begin{lem}
\label{lemma:max-L2-norm}
\[
\|\cov(vec(\mathcal{\mathcal{X}}^\top), vec(\mathcal{\mathcal{X}}^\top))\|  \le 2\pi \vertiii{f},
\]
where $\vertiii{f}=\esssup_{\omega \in [-\pi, \pi]}\|f(\omega)\|$. 
\begin{proof}
The proof follows from Proposition 2.3 in \citet{Basu2015}. 
\end{proof}
\end{lem}

\begin{lem}
\label{lemma:max-L2-norm-Y}
For any matrix $A_{p\times m}$, the time series $Y_t = A^\top {X}_t$ satisfies
\begin{equation}
\vertiii{f_Y} \le \|A\|^2 \vertiii{f}. \nonumber
\end{equation}
\begin{proof}
The autocovariance function of the time series  $Y_t$ can be written as 
\begin{equation}
    \Gamma_Y(\ell) = \cov(A^\top {X}_t, A^\top {X}_{t-\ell}) = A^\top \Gamma_{X}(\ell) A,
\end{equation}
which immediately leads to  
\begin{equation}
f_Y(\omega) = \sum_{\ell=-\infty}^\infty A^\top \Gamma_{X}(\ell)A e^{-\iu\omega\ell} = A^\top  f(\omega)A.
\end{equation}
Thus for any $\omega \in [-\pi, \pi]$, $\|f_Y(\omega)\| \le \|A\|^2\vertiii{f}$. Taking supremum over $\omega$ on the left side completes the proof. 
\end{proof}
\end{lem}

\begin{lem}
\label{lemma:linear_assumption}
For stationary linear processes $\Gamma(\ell)$ is well defined, and Assumption \ref{assumption:finite_auto} holds.
\begin{proof}
Since $(\sum_{i=1}^n |a_i|)^2 \ge \sum_{i=1}^n a_i^2$, we have 
\begin{equation}
\sum_{\ell=0}^\infty \|B_\ell\|_F \le \sum_{\ell=0}^\infty \sum_{1\le i,j\le p} |B_{\ell, (i,j)}|<\infty. \nonumber
\end{equation}
Then by equivalence of norms, it follows that
\begin{equation}
\sum_{\ell=0}^\infty \|B_\ell\| < \infty.
\end{equation}
Noticing for $h>0$, $\Gamma(h) = \Gamma^\top(-h)$, we have $\|\Gamma(h)\| = \|\Gamma(-h)\|$ for $h\ge 0$. Therefore, 
\begin{equation}
\begin{aligned}
\label{eq:finite_sum_auto}
&\sum_{\ell = -\infty}^\infty \|\Gamma(\ell)\| \le 2\sum_{\ell=0}^\infty \|\Gamma(\ell)\|
= 2\sum_{\ell = 0}^\infty \|\sum_{t=0}^\infty B_{t+\ell} B_{t}^\top \| \\
&<2\sum_{\ell = 0}^\infty \sum_{t=0}^\infty \|B_{\ell+t}\|\|B_\ell^\top\| = 
2\sum_{t_1=0}^\infty \sum_{t_2=0}^\infty \|B_{t_1}\|\|B_{t_2}\|=2\left[\sum_{t=0}^\infty \|B_t\|\right]^2<\infty.
\end{aligned}
\end{equation}
\end{proof}
\end{lem}

\begin{lem}
\label{lemma:L2_convergence_truncate}
\begin{equation}
\lim_{L\rightarrow \infty}\mathbb{E} \left[ \|vec(\mathcal{\mathcal{X}}_{(L)}^\top) - vec(\mathcal{\mathcal{X}}^\top)\|^2 \right]  = 0, \nonumber
\end{equation}
where $\mathcal{\mathcal{X}}_{n\times p} = [{X}_1: \ldots : {X}_n]^\top$ is a $n \times p$ data matrix with $n$ consecutive observations from a stationary linear process defined in \eqref{eq:infinite_ma}.
\begin{proof}
Since
\begin{equation}
\|vec(\mathcal{\mathcal{X}}_{(L)}^\top) - vec(\mathcal{\mathcal{X}}^\top)\|^2 = \sum_{t=1}^n \|{X}_t-{X}_{(L), t}\|^2, \nonumber
\end{equation}
it suffices to show that $\lim_{L\rightarrow \infty}\mathbb{E} \left[ \|{X}_{(L), t} - {X}_t\|^2 \right] = 0$ for any given $t\in \{1,\dots, n\}$. It follows that 
\begin{equation}
\label{eq:dct_dominant}
\|{X}_{(L), t}-{X}_t\|^2 = \sum_{\ell_1=L+1}^\infty \sum_{\ell_2=L+1}^\infty \varepsilon_{t-{\ell_1}}^\top B^\top_{\ell_1} B_{\ell_2}\varepsilon_{t-{\ell_2}} \le \sum_{\ell_1=0}^\infty \sum_{\ell_2=0}^\infty\|B_{\ell_1}\|\|B_{\ell_2}\|\|\varepsilon_{t-{\ell_1}}\|
\|\varepsilon_{t-{\ell_1}}\|.
\end{equation}
Since each coordinate of $\varepsilon_t$ has finite second moment ($1$ to be precise), we let 
$\mathbb{E}\|\varepsilon_{t-{\ell_1}}\| = c_\varepsilon <\infty $. Then the expected value of right part in \eqref{eq:dct_dominant} is
\begin{equation}
c_\varepsilon^2 \sum_{\ell_1=0}^{\infty}\sum_{\ell_2=0}^{\infty} \|B_{\ell_1}\| \|B_{\ell_2}\| = c_\varepsilon^2(\sum_{\ell=0}^\infty \|B_{\ell}\|)^2<\infty, \nonumber
\end{equation}
where the last inequality was established in the proof of lemma \ref{lemma:linear_assumption}. Then we apply dominated convergence theorem to show that 
\begin{equation*}
\begin{aligned}
&\mathbb{E} \left[\|{X}_{(L), t}-{X}_t\|^2 \right] = \sum_{\ell_1=L+1}^\infty \sum_{\ell_2=L+1}^\infty \mathbb{E}\left[ \varepsilon_{t-{\ell_1}}^\top B^\top_{\ell_1} B_{\ell_2}\varepsilon_{t-{\ell_2}} \right]\\
& = \sum_{\ell=L+1}^\infty \mathbb{E} \left[ \varepsilon_{t-\ell}^\top B^\top_{\ell} B_{\ell} \varepsilon_{t-\ell} \right] \le c_\varepsilon^2 (\sum_{\ell=L+1}^\infty \|B_\ell\|)^2,
\end{aligned}
\end{equation*}
because $\sum_{\ell=0}^\infty \|B_\ell\|<\infty$, above goes to zero when $L\rightarrow \infty$. 
\end{proof}
\end{lem}
\begin{remark}
The above convergence argument immediately implies several useful results,
\begin{enumerate}
    \item $vec(\mathcal{\mathcal{X}}_{(L)}^\top)\overset{\mathbb{P}}{\to} vec(\mathcal{\mathcal{X}}^\top)$
    \item For any real matrix $A_{np\times np}$, 
    \begin{equation*}
       \lim_{L\rightarrow \infty}\mathbb{E} \left[ vec(\mathcal{\mathcal{X}}_{(L)}^\top)^\top A ~ vec(\mathcal{\mathcal{X}}_{(L)}^\top)\right] =  \mathbb{E} \left[ vec(\mathcal{\mathcal{X}}^\top)^\top A ~ vec(\mathcal{\mathcal{X}}^\top)\right]
    \end{equation*}
This is because
\begin{equation}
\label{eq:cross_terms_bound}
\begin{aligned}
& \left|\mathbb{E} \left[ vec(\mathcal{X}_{(L)}^\top)^\top A ~ vec(\mathcal{X}_{(L)}^\top)\right] - \mathbb{E} \left[ vec(\mathcal{X}^\top)^\top A ~ vec(\mathcal{X}^\top)\right]\right| \\
\le & \left|\mathbb{E} \left[ vec(\mathcal{X}_{(L)}^\top)^\top A  \left(vec(\mathcal{X}_{(L)}^\top)-vec(\mathcal{X}^\top)\right)\right] \right|+ \left| \mathbb{E} \left[ \left(vec(\mathcal{X}_{(L)}^\top)-vec(\mathcal{X}^\top)\right)^\top A ~ vec(\mathcal{X}^\top )\right]\right|.
\end{aligned}
\end{equation}
Applying Cauchy-Schwarz inequality to the first part in second line of \eqref{eq:cross_terms_bound}, we get 
\begin{equation*}
\begin{aligned}
&\left|\mathbb{E} \left[ vec(\mathcal{\mathcal{X}}_{(L)}^\top)^\top A  \left(vec(\mathcal{X}_{(L)}^\top)-vec(\mathcal{\mathcal{X}}^\top)\right)\right]\right|^2 \\
&\le \|A\| \mathbb{E} \left[ \|vec(\mathcal{X}_{(L)}^\top)\|^2\right] \mathbb{E} \left[ \left\|\left(vec(\mathcal{X}_{(L)}^\top)-vec(\mathcal{\mathcal{X}}^\top)\right)\right\|^2\right].
\end{aligned}
\end{equation*}
In addition, from Lemma \ref{lemma:L2_convergence_truncate}, we have 
$\mathbb{E} \left[ \|vec(\mathcal{\mathcal{X}}_{(L)}^\top)\|^2\right]\rightarrow \mathbb{E} \left[ \|vec(\mathcal{\mathcal{X}}^\top)\|^2\right] $ and \\
 $ \mathbb{E} \left[ \|\left(vec(\mathcal{X}_{(L)}^\top)-vec(\mathcal{\mathcal{X}}^\top)\right)\|^2\right]\rightarrow 0$. This implies that the first part in second line of \eqref{eq:cross_terms_bound} converges to zero when $L$ goes to infinity. A similar argument ensures that the second part in second line of \eqref{eq:cross_terms_bound} goes to zero as well, completing our proof.
\end{enumerate}
\end{remark}

\begin{lem}
\label{lemma:spectral_convergence}
$\lim_{L\rightarrow \infty}\vertiii{f_{(L)}} = \vertiii{f}$.
\begin{proof}
Let $\Gamma_{(L)}(h)$ and $f_{(L)}(\omega)$ be the autocovariance function and spectral density of the truncated process $X_{(L),t}$. 
We list expressions for $\Gamma_{(L)}(h)$
and $\Gamma(h)$ in order to make a comparison later where we focus on the case $h>0$ (as pointed before, $\Gamma(h) = \Gamma^\top(-h)$ for $h>0$)
\begin{equation}
\begin{aligned}
&\Gamma(h) = \mathbb{E} X_t X_{t-h}^\top = \mathbb{E}\left(\sum_{\ell=0}^\infty B_\ell \varepsilon_{t-\ell}\right) 
\left(\sum_{\ell=0}^\infty B_\ell \varepsilon_{t-h-\ell}\right)^\top = \sum_{\ell=0}^\infty B_{\ell+h}B_{\ell}^\top, \nonumber 
\end{aligned}
\end{equation}
and 
\begin{equation}
\begin{aligned}
&\Gamma_L(h) = \mathbb{E} X_t X_{t-h}^\top = \mathbb{E}\left(\sum_{\ell=0}^L B_\ell \varepsilon_{t-\ell}\right) 
\left(\sum_{\ell=0}^L B_\ell \varepsilon_{t-h-\ell}\right)^\top = \sum_{\ell=0}^{L-h} B_{\ell+h}B_{\ell}^\top, \nonumber 
\end{aligned}
\end{equation}
which indicates $\Gamma_L(h)= 0$ if $h>L$. \par 

Now we show that $\|\Gamma_{(L)}(h) - \Gamma(h)\|$ goes to zero with $L\rightarrow \infty$. Since we consider $L$ goes to inftty, we assume $L>|h|$. Without losing generality, for any given positive integer $h$, 
\begin{equation*}
\begin{aligned}
\lim_{L\rightarrow \infty}\|\Gamma_{(L)}(h) - \Gamma(h)\|& = \lim_{L\rightarrow \infty} \left\|\sum_{\ell= L - h +1}^\infty B_{\ell+h} B_{\ell}^\top \right\| \\ 
&\le \lim_{L\rightarrow \infty}\sum_{\ell=L-h+1}^\infty \|B_\ell\|  \|B_{\ell+h}\|\\ 
&\le \lim_{L\rightarrow \infty} \left(\sum_{\ell=0}^\infty \|B_{\ell}\|\right) \left(\sum_{\ell=L+1}^\infty \|B_{\ell}\|\right)  =0.
\end{aligned}
\end{equation*}
The last equality comes from the fact that $\sum_{\ell=0}^\infty \|B_\ell\|<\infty$. 
Considering the relation that $\Gamma(h)=\Gamma^\top(-h)$, for $h<0$, following also holds: 
\begin{equation}
\lim_{L\rightarrow \infty} \|\Gamma_{L}(h)-\Gamma(h)\| = 0.
\end{equation}
Based on the expression of $\Gamma(h)$ and $\Gamma_L(h)$, we have
\begin{equation*}
\begin{aligned}
\max\left\{\sum_{h=-\infty}^\infty \|\Gamma(h)\|, \sum_{h=-\infty}^\infty\|\Gamma_{(L)}(h)\|\right\}\le 2(\sum_{\ell=0}^\infty \|B_\ell\|)^2<\infty,
\end{aligned}
\end{equation*}
which in turn implies 
\begin{equation}
\label{eq:autocovariance_dct}
\sum_{h=-\infty}^\infty \|\Gamma_{(L)}(h) - \Gamma(h)\| \le 2(\sum_{\ell=0}^\infty \|B_\ell\|)^2<\infty.
\end{equation}
Therefore, by dominant convergence theorem, 
\begin{equation*}
\begin{aligned}
&\lim_{L\rightarrow \infty}\esssup_{\omega\in [-\pi, \pi]}\|f_{(L)}(\omega)-f(\omega)\|\le \lim_{L\rightarrow \infty}\sum_{h=-\infty}^\infty \|\Gamma_{(L)}(h)-\Gamma(h)\|\\
& = \sum_{h=-\infty}^\infty  \lim_{L\rightarrow \infty}  \|\Gamma_{(L)}(h)-\Gamma(h)\| = 0,
\end{aligned}
\end{equation*}
Finally
\begin{equation*}
\begin{aligned}
\lim_{L\rightarrow \infty}\left|\vertiii{f_{(L)}} - \vertiii{f}\right| &=\lim_{L\rightarrow \infty} \left|\esssup_{\omega\in [-\pi, \pi] }\|f_{(L)}(\omega)\|-\esssup_{\omega\in [-\pi, \pi] }\|f(\omega)\| \right|\\
&\le \lim_{L\rightarrow \infty} \esssup_{\omega \in [-\pi, \pi]}\|f_{(L)}(\omega)-f(\omega)\| =0, 
\end{aligned}
\end{equation*}
which completes the proof.  
\end{proof}
\end{lem}
\section{Appendix: Additional Table and Graphs}\label{appendix:more_tables}
This section contains a table on precision, recall and F1 measures of the three different types of thresholding methods in selecting the non-zero entries of the spectral density matrices of VMA and VAR models of different dimension using different  sample sizes. The simulation settings are described in Section \ref{sec:simulation}. 

We also present enlarged images of the adjacency matrices of coherence networks obtained using adaptive lasso thresholding and shrinkage methods on the real data analysis in Section \ref{sec:realdata}. These images contain names of brain regions so that interesting strong connectivity patterns between regions can be identified easily.

\begin{sidewaystable}
\small
\centering
\def~{\hphantom{0}}
\begin{tabular}{l@{\hskip 0.4in} ccc ccc ccc}
& \multicolumn{3}{c}{Hard Thresholding}  & \multicolumn{3}{c}{Lasso} & \multicolumn{3}{c}{Adaptive Lasso}\\VMA & precision & recall & F1 & precision & recall & F1 & precision & recall & F1\\
 p = 12 & & & & & & & & & \\
\multicolumn{1}{r}{n = 100}&93.07(2.79)&55.95(2.58)&68.8(1.93)&80.1(4.87)&75.97(4.19)&72.7(2.54)&93.07(2.79)&61.04(3.62)&70.55(2.07)\\
\multicolumn{1}{r}{n = 200}&92.31(2.45)&68.04(2.84)&77.65(2.1)&75.53(4.11)&91.07(2.34)&79.25(2.54)&92.31(2.45)&75.92(2.88)&80.35(1.92)\\
\multicolumn{1}{r}{n = 400}&91.07(2.15)&83.72(1.96)&88.4(1.51)&70.37(3.39)&98.16(0.68)&79.83(2.52)&91.07(2.15)&91.32(1.61)&89.67(1.6)\\
\multicolumn{1}{r}{n = 600}&91.89(1.82)&90.02(1.59)&92.75(1.06)&69.73(3.74)&99.4(0.3)&80.13(2.77)&91.89(1.82)&95.68(1.02)&92.83(1.28)\\
 p = 24 & & & & & & & & & \\
\multicolumn{1}{r}{n = 100}&97.93(0.98)&45.58(0.94)&62.33(0.81)&88.59(3.35)&63.0(3.6)&70.4(1.89)&97.93(0.98)&50.07(2.04)&65.36(1.44)\\
\multicolumn{1}{r}{n = 200}&96.12(1.36)&52.79(1.6)&68.13(1.2)&80.36(4.06)&84.67(2.81)&79.53(2.22)&96.12(1.36)&66.14(2.77)&76.51(1.72)\\
\multicolumn{1}{r}{n = 400}&94.62(1.0)&71.14(2.26)&81.57(1.53)&74.21(2.63)&96.7(0.88)&82.47(1.74)&94.62(1.0)&86.86(1.81)&89.47(1.1)\\
\multicolumn{1}{r}{n = 600}&94.56(1.1)&81.6(1.73)&88.68(1.11)&71.44(2.66)&99.0(0.34)&81.79(1.9)&94.56(1.1)&93.81(0.95)&93.64(0.79)\\
 p = 48 & & & & & & & & & \\
\multicolumn{1}{r}{n = 100}&99.42(0.37)&43.4(0.23)&60.5(0.22)&94.44(1.85)&52.88(1.96)&66.51(1.14)&99.42(0.37)&45.31(0.75)&62.08(0.61)\\
\multicolumn{1}{r}{n = 200}&98.57(0.51)&45.51(0.56)&62.4(0.48)&87.23(2.05)&75.2(2.38)&78.7(1.31)&98.57(0.51)&55.86(1.86)&70.44(1.33)\\
\multicolumn{1}{r}{n = 400}&96.99(0.6)&58.26(1.35)&72.75(1.03)&79.44(2.01)&93.35(0.9)&84.74(1.17)&96.99(0.6)&79.2(1.65)&86.21(1.09)\\
\multicolumn{1}{r}{n = 600}&96.87(0.45)&70.21(1.46)&81.53(1.02)&77.86(1.56)&97.52(0.48)&85.89(1.01)&96.87(0.45)&89.23(1.1)&92.38(0.61)\\
 p = 96 & & & & & & & & & \\
\multicolumn{1}{r}{n = 100}&99.9(0.08)&42.95(0.09)&60.09(0.09)&98.48(0.57)&46.43(0.85)&62.87(0.68)&99.9(0.08)&43.5(0.3)&60.59(0.28)\\
\multicolumn{1}{r}{n = 200}&99.58(0.17)&43.47(0.22)&60.58(0.21)&93.6(0.91)&64.54(1.62)&75.21(1.02)&99.58(0.17)&49.24(0.99)&65.61(0.83)\\
\multicolumn{1}{r}{n = 400}&98.67(0.25)&49.43(0.7)&65.85(0.6)&86.03(1.31)&87.97(1.14)&86.08(0.64)&98.67(0.25)&70.45(1.18)&81.42(0.78)\\
\multicolumn{1}{r}{n = 600}&98.18(0.24)&59.63(0.96)&74.05(0.73)&82.83(1.08)&94.85(0.53)&87.92(0.57)&98.18(0.24)&82.83(0.92)&89.29(0.54)\\
VAR & precision & recall & F1 & precision & recall & F1 & precision & recall & F1\\
 p = 12 & & & & & & & & & \\
\multicolumn{1}{r}{n = 100}&90.08(2.96)&45.23(2.11)&57.52(1.41)&79.5(4.16)&60.65(4.06)&61.26(1.93)&90.08(2.96)&48.83(2.96)&58.42(1.74)\\
\multicolumn{1}{r}{n = 200}&88.99(2.48)&51.84(2.15)&62.68(1.25)&75.61(3.8)&73.26(3.52)&68.11(2.0)&88.99(2.48)&57.57(2.7)&64.86(1.83)\\
\multicolumn{1}{r}{n = 400}&88.99(1.96)&61.3(2.08)&71.14(1.49)&72.33(2.58)&86.34(2.19)&74.52(1.59)&88.99(1.96)&70.4(2.42)&74.7(1.69)\\
\multicolumn{1}{r}{n = 600}&86.88(0.11)&69.07(0.09)&76.14(0.41)&66.34(0.41)&93.65(0.53)&74.74(0.4)&86.88(0.11)&79.31(0.18)&79.61(0.05)\\
 p = 24 & & & & & & & & & \\
\multicolumn{1}{r}{n = 100}&94.85(1.6)&36.88(0.81)&52.6(0.55)&85.36(3.08)&47.49(2.25)&56.34(1.05)&94.85(1.6)&39.38(1.1)&53.8(0.76)\\
\multicolumn{1}{r}{n = 200}&94.43(1.39)&39.78(0.85)&55.71(0.74)&81.51(2.38)&59.89(2.32)&64.32(1.43)&94.43(1.39)&45.61(1.43)&59.44(1.08)\\
\multicolumn{1}{r}{n = 400}&92.9(1.21)&48.0(1.0)&63.48(0.86)&75.01(2.24)&77.5(1.81)&72.69(1.26)&92.9(1.21)&59.46(1.57)&70.39(1.19)\\
\multicolumn{1}{r}{n = 600}&92.74(0.46)&58.03(0.27)&72.01(0.18)&71.01(1.52)&88.3(0.12)&76.35(0.86)&92.74(0.46)&71.38(0.25)&79.11(0.05)\\
 p = 48 & & & & & & & & & \\
\multicolumn{1}{r}{n = 100}&97.63(0.95)&34.28(0.3)&50.8(0.23)&91.73(2.08)&39.63(1.17)&53.21(0.73)&97.63(0.95)&35.47(0.57)&51.44(0.39)\\
\multicolumn{1}{r}{n = 200}&97.72(0.51)&35.35(0.33)&52.02(0.31)&88.42(1.47)&48.61(1.34)&60.42(0.92)&97.72(0.51)&39.06(0.66)&55.16(0.55)\\
\multicolumn{1}{r}{n = 400}&96.34(0.59)&40.86(0.48)&57.54(0.47)&80.55(1.42)&66.78(1.11)&70.84(0.81)&96.34(0.59)&50.95(0.69)&65.76(0.63)\\
\multicolumn{1}{r}{n = 600}&95.71(0.12)&48.49(0.72)&64.74(0.77)&77.08(1.29)&79.01(0.31)&76.29(0.73)&95.71(0.12)&61.51(0.52)&74.1(0.44)\\
 p = 96 & & & & & & & & & \\
\multicolumn{1}{r}{n = 100}&99.02(0.52)&33.59(0.11)&50.21(0.08)&96.16(1.12)&35.66(0.52)&51.24(0.36)&99.02(0.52)&34.02(0.25)&50.51(0.19)\\
\multicolumn{1}{r}{n = 200}&99.15(0.31)&33.84(0.12)&50.54(0.12)&94.04(1.06)&41.37(0.77)&56.44(0.58)&99.15(0.31)&35.67(0.34)&52.29(0.31)\\
\multicolumn{1}{r}{n = 400}&98.25(0.29)&36.62(0.26)&53.44(0.26)&87.0(1.18)&57.22(0.96)&67.82(0.59)&98.25(0.29)&44.75(0.62)&61.02(0.58)\\
\multicolumn{1}{r}{n = 600}&97.74(0.12)&43.05(0.19)&59.87(0.18)&82.79(0.08)&69.24(0.59)&74.33(0.38)&97.74(0.12)&55.26(0.48)&70.27(0.33)\\
\end{tabular}
\label{table:precision-homogeneous-final}
\caption{Precision, Recall, F1 Score ( in $\%$) of three different thresholding methods: hard threshold, lasso and adaptive lasso. }%
\end{sidewaystable}

\begin{figure}[p]
    \centering
    \includegraphics[width=1.1\textwidth]{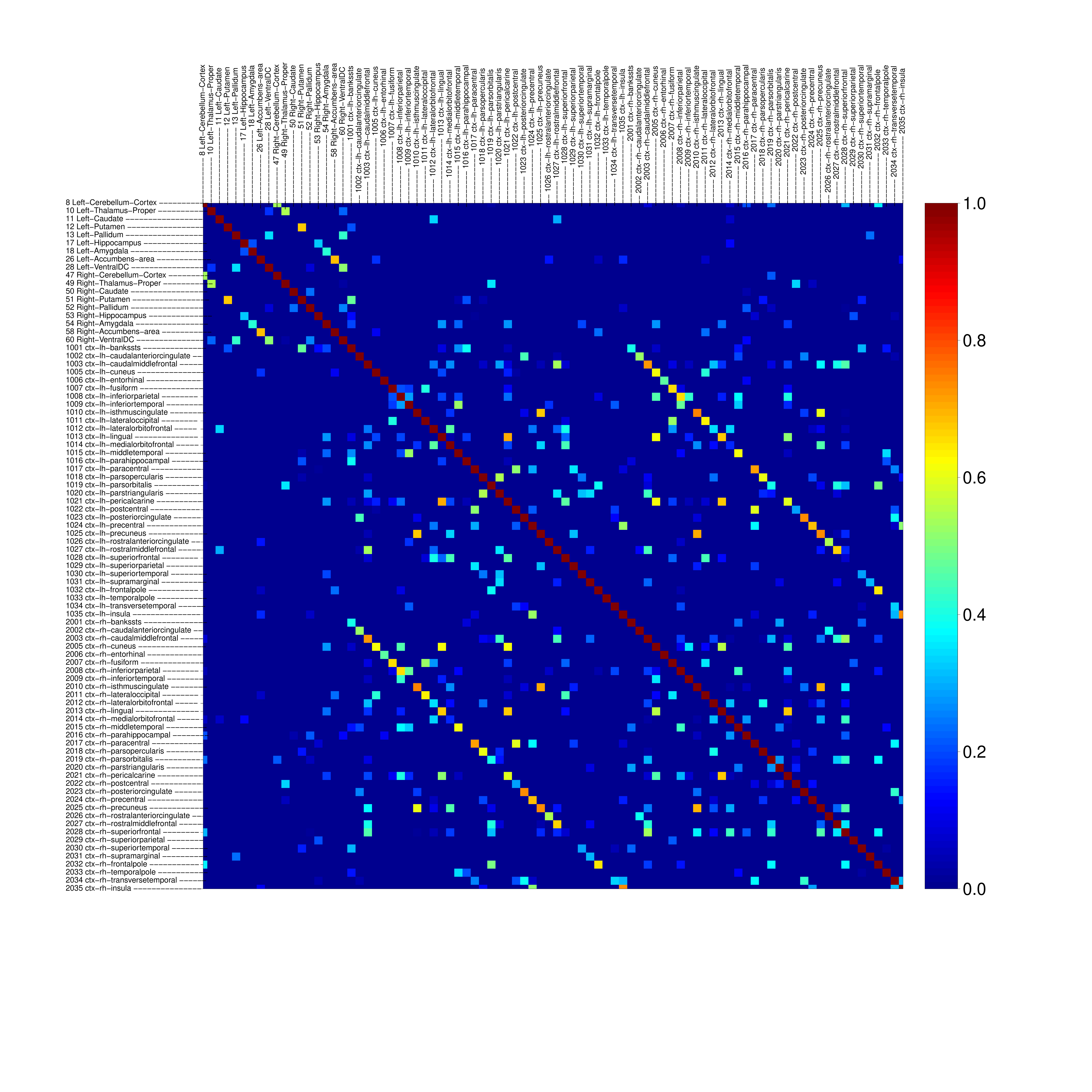}
    \caption{Heat map of absolute coherence matrix (at frequency $0$) estimated using adaptive lasso thresholding of averaged periodogram.}
    \label{fig:realdatafullalasso}
\end{figure}

\begin{figure}[p]
    \centering
    \includegraphics[width=1.1\textwidth]{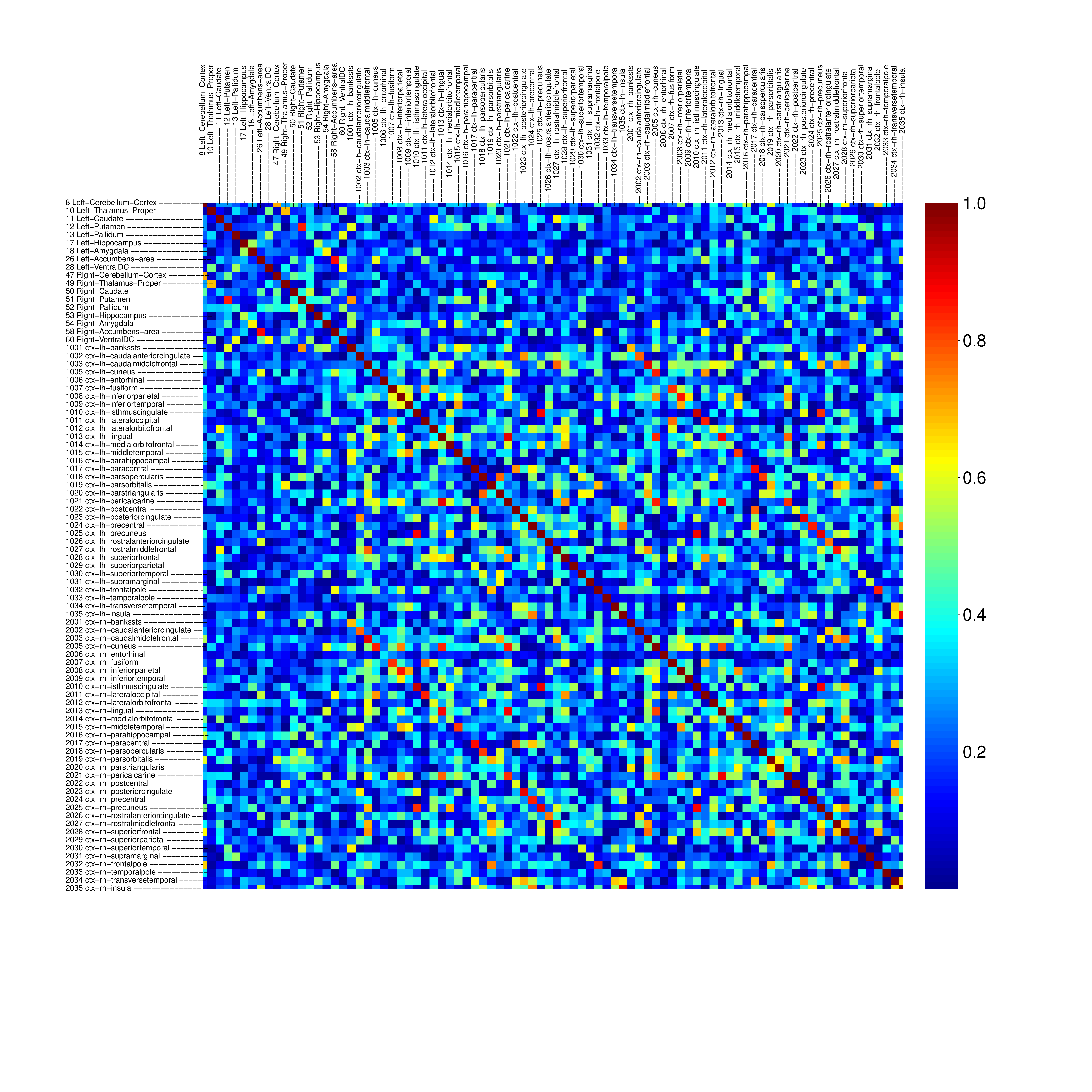}
    \caption{Heat map of absolute coherence matrix (at frequency $0$) estimated using diagonal shrinkage of averaged periodogram.}
    \label{fig:realdatafullshrinkage}
\end{figure}

\newpage 
\bibliographystyle{abbrvnat} 
\bibliography{biblio,paper-ref}

\end{document}